\pdfoutput=1
      [1999/12/01 v1.4c Il Nuovo Cimento]
\documentclass[varenna]{cimento}

\usepackage{graphicx}
\title{Nuclear physics aspects of double beta decay
}

\author{Petr Vogel}

\institute{Kellogg Radiation Laboratory 106-38 \\
California Institute of Technology \\
Pasadena, CA 91125, USA \\ \\
{\rm Lecture notes for course CLXX ``MEASUREMENTS OF NEUTRINO MASS" \\
Int. School of Physics ``Enrico Fermi", Varenna, June 2008 }}

\PACSes{\PACSit{00.00}{By the way, which PACS is it, the 00.00? GOK.}
\PACSit{---.---}{\ldots}}

\begin{document}

\maketitle

\PACSes{\PACSit{21.60.-n}{}
\PACSit{23.40.Bw}{}
\PACSit{23.40.Hc}{}}

\begin{abstract}

Comprehensive description of the phenomenology of the $\beta\beta$ decay is given,
with emphasis on the nuclear physics aspects. After a brief review of the neutrino oscillation
results and of motivation to test the lepton number conservation, the mechanism of the
$0\nu\beta\beta$ is discussed.  Its relation to the lepton flavor violation involving charged leptons
and its use as a diagnostic tool of the $0\nu\beta\beta$ mechanism is described.
Next the basic nuclear physics of both $\beta\beta$-decay modes is presented, and the
decay rate formulae derived. The nuclear physics methods used, the nuclear shell model
and the quasiparticle random phase approximation, are described next, and the choice of input 
parameters is discussed in the following section. Finally, the numerical values of the
nuclear matrix elements, and their uncertainty, are presented. In the appendix the relation
of the search for the neutrino magnetic moment to the Dirac versus Majorana nature
of neutrinos is described.

\end{abstract}

\section{Introduction  to  $\beta\beta$ decay}

In the last decade neutrino
oscillation experiments have convincingly and triumphantly
shown that neutrinos have a finite mass and that the lepton flavor is not
a conserved quantity. These results opened the door to what is often called
the ``Physics Beyond the
Standard Model". In other words, accommodating these findings into a consistent
scenario requires generalization of the Standard Model of electroweak 
interactions that postulates that neutrinos are massless and that consequently
lepton flavor, and naturally, also the total lepton number, are  conserved quantities.
 
In oscillation
experiments only the differences in squares of the neutrino masses, 
$\Delta m^2 \equiv | m_2^2 - m_1^2| $, is measured, and the results
do not depend on the charge conjugation properties of neutrinos, i.e. whether
they are Dirac or Majorana fermions.
Nevertheless, from oscillation experiments one can establish a lower limit on 
the absolute value  of the neutrino mass scale, 
$m_{scale} =  \sqrt{|\Delta m^2|}$. Thus, one or more neutrinos have
a mass of at least $\sqrt{|\Delta m^2_{atm}| \sim}$ 50 meV,
and another one has mass of at least  $\sqrt{|\Delta m^2_{sol}| \sim}$
10 meV. In addition, an upper limit on the masses of all
active neutrinos $\sim$ 2 - 3 eV can be derived from the combination of
analysis of the tritium beta-decay experiments and the neutrino
oscillation experiments. Combining these constraints, masses of at least two
(out of the total of three active) neutrinos are bracketted by 10 meV $\le m_{\nu} \le$
2 - 3 eV. 

Thus, neutrino masses are  six or more orders of
magnitude smaller than the masses of the other fermions. Moreover,
the pattern of masses, i.e. the mass ratios of neutrinos, is rather different
(even though it remains largely unknown) than the pattern of masses
of the up- or down-type quarks or charged leptons. All of these facts
suggest that, perhaps, the origin of the neutrino mass is different
than the origin (which is still not well understood) of the masses of the other fermions.    

The discoveries of neutrino oscillations, in turn, are causing a renaissance
of enthusiasm in the double beta decay community and a slew of
new experiments that are expected to reach,
within a near future, the sensitivity corresponding to the
neutrino mass scale. 
Below I review the current status of the double beta decay and the effort devoted
to reach the required sensitivity, as well as various issues 
in theory (or phenomenology) related to the relation
of the $0\nu\beta\beta$ decay rate to the absolute neutrino mass scale and to
the general problem of the Lepton Number Violation (LNV). 
And, naturally, substantial emphasis is devoted to the nuclear structure issues.
But before doing that I very briefly summarize
the achievements of the neutrino oscillation searches and the role that the search
for the neutrinoless double beta decay plays in the elucidation of the pattern
of neutrino masses and mixing.

There is a consensus that the measurement of atmospheric
neutrinos by the SuperKamiokande collaboration\cite{SKatm01}
can be only interpreted as a consequence
of the nearly maximum mixing between  $\nu_{\mu}$ and $\nu_{\tau}$ neutrinos,
with the corresponding mass squared difference  
$|\Delta m_{atm}^2| \sim  2.4\times10^{-3}{\rm eV}^2$.
This finding was confirmed  by the K2K experiment
\cite{K2K01} that uses accelerator  
$\nu_{\mu}$ beam pointing towards the SuperKamiokande detector
250 km away as well as  by the very recent result of the  MINOS 
experiment located at the Sudan mine in Minnesota 735 km away from 
Fermilab\cite{MINOS}. 
Several large long-baseline experiments are being build to further elucidate
this discovery, and determine the corresponding parameters even
more accurately. 

\begin{figure}
\centerline{\includegraphics[width=9.0cm]{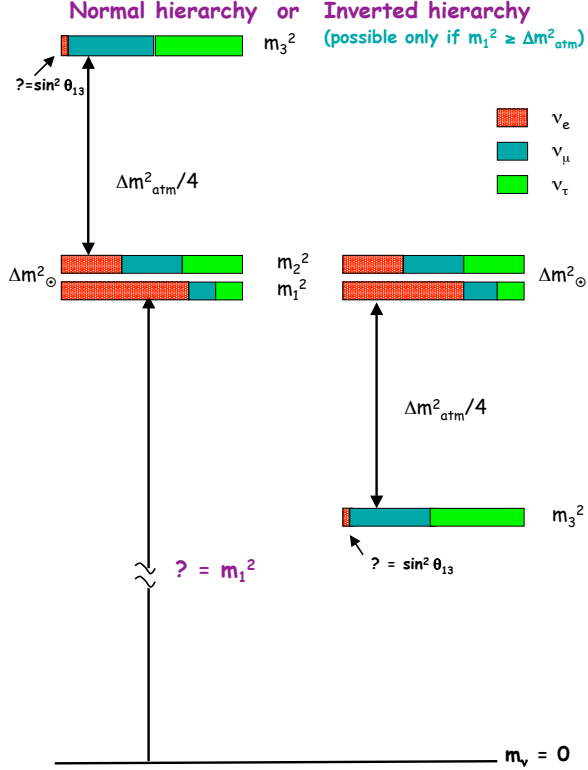}}
\caption{ Schematic illustration  of the decomposition
of the neutrino mass eigenstates $\nu_i$ in terms of the flavor eigenstates.
The two hierarchies cannot be, at this time, distinguished. The small admixture
of $\nu_e$ into $\nu_3$ is an upper limit, and the mass square of the neutrino 
$\nu_1$, the quantity $m_1^2$, remains unknown. }
\label{fig_osc}
\end{figure}

At the same time the "solar neutrino puzzle", which has been with us for over thirty years
since the pioneering chlorine experiment of Davis\cite{chlorine}, also reached
the stage where the interpretation of the measurements in terms of  oscillations
between the $\nu_e$ and some combination of active, 
i.e., $\nu_{\mu}$ and $\nu_{\tau}$ neutrinos, is inescapable. In particular, the
juxtaposition of the results of the
SNO experiment\cite{SNO01} and SuperKamiokande\cite{SKsol01},
together with the earlier solar neutrino flux determination in
the gallium experiments\cite{Gallex,Sage} and, of course chlorine\cite{chlorine},
leads to that conclusion.
The value of the corresponding oscillation parameters,
however, remained uncertain, with several
"solutions" possible, although the so-called Large Mixing Angle (LMA) solution
with $\sin^2 2\theta_{sol} \sim 0.8$ and 
$\Delta m_{sol}^2 \sim 10^{-4}{\rm eV}^2$ was preferred. 
A decisive confirmation of the "solar" oscillations was provided by
the nuclear reactor experiment KamLAND \cite{Kaml1,Kaml2,Kaml3} that
demonstrated that the flux of the reactor $\bar{\nu}_e$ is reduced
and its spectrum distorted at the distance $L_0 \sim$ 180 km from 
nuclear reactors. The most recent KamLAND results \cite{Kaml3},
combined with the existing solar neutrino data launched the era
of precision neutrino measurements, with the corresponding parameters
$\Delta m_{21}^2 = 7.59^{+0.21}_{-0.21} \times 10^{-5}{\rm eV}^2$ and
$\tan^2 \theta_{12} = 0.47^{+0.06}_{-0.05}$ determined with an unprecedented
accuracy.
Analysis of that experiment, moreover, clearly shows the oscillatory
behavior of the detection probability as a function of $L_0/E_{\nu}$.
That behavior can be traced in ref. \cite{Kaml3} over two
full periods.

The pattern of neutrino mixing is further simplified by the constraint due to
the Chooz and Palo Verde reactor neutrino experiments\cite{Chooz99,Palo01}
which lead to the
conclusion that the third mixing angle, $\theta_{13}$, is
small,  $\sin^2 2\theta_{13} \le 0.1$.  The two remaining possible 
neutrino mass patterns
are illustrated in Fig.\ref{fig_osc}.

As already stated, oscillation experiments cannot determine the absolute
magnitude of the masses and, in particular, cannot at this stage separate
two rather different scenarios, the hierarchical pattern of neutrino
masses in which $m \sim \sqrt{\Delta m^2}$ and the degenerate pattern
in which  $m \gg \sqrt{\Delta m^2}$. It is hoped that the 
search for the neutrinoless double beta
decay, reviewed here, will help in foreseeable future
 in determining or at least narrowing down the absolute neutrino mass
scale, and in deciding which of these two possibilities is applicable.

Moreover, even more important is the fact that the oscillation results 
do not tell us anything about the properties of neutrinos
under charge conjugation. While the charged leptons are Dirac particles, distinct from
their antiparticles, neutrinos may be the ultimate neutral particles, as
envisioned by Majorana, that are identical to their antiparticles. That fundamental
distinction becomes important only for massive particles.
Neutrinoless double beta decay proceeds only when neutrinos are massive 
Majorana particles, hence its observation would resolve the question.

The argument for the ``Majorana nature" of the neutrinos can be traced to the
observation by Weinberg \cite{Wei79} who pointed out almost thirty year ago 
that there exists only one
lowest order (dimension 5, suppressed by only one inverse power of the
corresponding high energy scale $\Lambda$) gauge-invariant operator given the content
of the standard model
\begin{equation}
{\cal L}^{(5)} = C^{(5)}/\Lambda (\bar{L}^c \epsilon H)(H^T \epsilon L) ~,
\end{equation}
where $L$ is the lepton doublet, $\bar{L}^c = L^T C$ with $C$ the charge conjugation operator,
$\epsilon = -i \tau_2$, and $H$ represent the Higgs boson. After the spontaneous symmetry
breaking the Higgs acquires vacuum expectation value and the above operator
represents the neutrino Majorana mass that violates the total lepton number
conservation law by two units
\begin{equation}
{\cal L}^{(M)} = \frac{C^{(5)}}{\Lambda} \frac{ v^2}{2} (\bar{\nu}^c \nu) + h.c. ~~,
\end{equation}
where $v \sim$ 250 GeV and the neutrinos are naturally light because their mass
is suppressed by the large value of the new physics scale $\Lambda$ in the denominator.

The most popular explanation of the smallness of neutrino mass is the see-saw mechanism,
 which is also roughly thirty years old \cite{seesaw}. In it, the existence of heavy right-handed
 neutrinos $N_R$ is postulated, and by diagonalizing the corresponding mass matrix
 one arrives at the formula
 \begin{equation}
 m_{\nu} = \frac{m_D^2}{M_N}
 \end{equation}
 where the Dirac mass $m_D$ is expected to be a typical charged fermion mass and $M_N$ is
 the Majorana mass of the heavy neutrinos $N_R$. Again, the small mass of the standard neutrino
 is related to the large mass of the heavy right-handed partner. Requiring that $m_{\nu}$ is of the
 order of 0.1 eV means that $M_N$ (or $\Lambda$) is $\sim 10^{14-15}$ GeV, i.e. near the GUT
 scale. That makes this template scenario particularly attractive.
 
 Clearly, one cannot reach such high energy scale experimentally. But, these scenarios imply that 
 neutrinos are Majorana particles, and consequently that the total lepton number should not
 be conserved. Hence the tests of the lepton number conservation acquires a fundamental
 importance.

Double beta decay ($\beta\beta$) is a nuclear transition 
$(Z,A) \rightarrow (Z+2,A)$ in which two neutrons
bound in a nucleus  are  simultaneously transformed into two protons plus two
electrons (and possibly other light neutral particles). This transition is
possible and potentially observable because 
nuclei with even $Z$ and $N$ are more bound than the odd-odd nuclei with
the same $A = N + Z$. 
Analogous transition of two protons into two neutrons
are also, in principle, possible
in several nuclei, but phase space considerations give preference to the former
mode.

\begin{figure}
\centerline{\includegraphics[width=9.0cm]{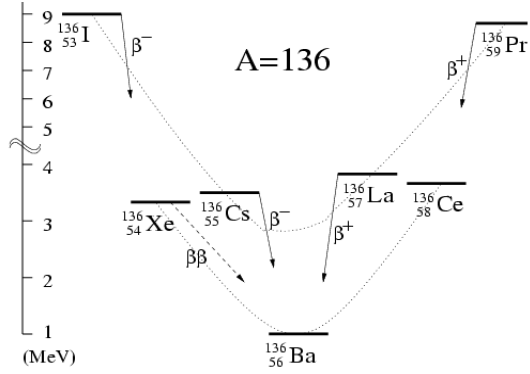}}
\caption{ Atomic masses of the isotopes with $A$ = 136. Nuclei $^{136}$Xe, $^{136}$Ba and
$^{136}$Ce are stable against the ordinary $\beta$ decay; hence they exist in nature. However,
energy conservation alone allows the transition
$^{136}$Xe $\rightarrow$ $^{136}$Ba + $2e^-$ (+ possibly
other neutral light particles) and the analogous decay of $^{136}$Ce with the positron emission. }
\label{fig_bb}
\end{figure}

An example is shown in Fig. \ref{fig_bb}. The situation shown there is not really exceptional. 
There are eleven analogous cases (candidate nuclei) with the $Q$-value (i.e. the kinetic energy
available to leptons) in excess of 2 MeV.  

There are two basic modes of the $\beta\beta$ decay. In the two-neutrino mode ($2\nu\beta\beta$)
there are 2 $\bar{\nu}_e$ emitted together with the 2 $e^-$. 
It is just an ordinary beta decay of two bound neutrons occurring simultaneously since
the sequential decays are forbidden by the energy conservation law. 
For this mode, clearly,
the lepton number is conserved and
this mode of decay is allowed in the standard model of electroweak interaction.
It has been repeatedly observed in a number of cases and proceeds with a typical half-life
of $\sim 10^{19-20}$years for the nuclei with $Q$-values above 2 MeV.
In contrast, in the neutrinoless
mode ($0\nu\beta\beta$) only the 2$e^-$ are emitted and nothing else. 
That mode clearly violates the law
of lepton number conservation and is forbidded in the standard model. Hence, its observation
would be a signal of a "new physics".

The two modes of the $\beta\beta$ decay have some common and some distinct features.
The common features are:
\begin{itemize}
\item  The leptons carry essentially all available energy. The nuclear recoil is negligible,
$Q/Am_p \ll 1$.
\item The transition involves the $0^+$ ground state of the initial nucleus and
(in almost all cases) the $0^+$ ground state of the final nucleus. In few cases the
transition to an excited $0^+$ or $2^+$ state in the final nucleus is energetically possible, but
suppressed by the smaller phase space available. (But the $2\nu\beta\beta$ decay to
the excited $0^+$ state has been observed in few cases.)
\item Both processes are of second order of weak interactions, $\sim G_F^4$, hence inherently
slow.  The phase space consideration alone (for the $2\nu\beta\beta$ mode $\sim Q^{11}$ and
for the $0\nu\beta\beta$ mode $\sim Q^5$) give preference to the $0\nu\beta\beta$ which is,
however, forbidden by the lepton number conservation.
\end{itemize}
The distinct features are:
\begin{itemize}
\item In the $2\nu\beta\beta$ mode the two neutrons undergoing the transition are uncorrelated
(but decay simultaneously) while in the $0\nu\beta\beta$ the two neutrons are correlated.
\item In the $2\nu\beta\beta$ mode the sum electron kinetic energy $T_1 + T_2$
spectrum is continuous and peaked below $Q/2$. 
This is due to the electron masses and the Coulomb attraction.
As $T_1 + T_2 \rightarrow Q$ the
spectrum approaches zero approximately like $(\Delta E/Q)^6$.
 \item On the other hand 
 in the $0\nu\beta\beta$ mode the sum of the electron kinetic
 energies is fixed, $T_1 + T_2 = Q$, smeared only by the detector resolution.
 \end{itemize}
 
 \begin{figure}
\centerline{\includegraphics[width=7.0cm]{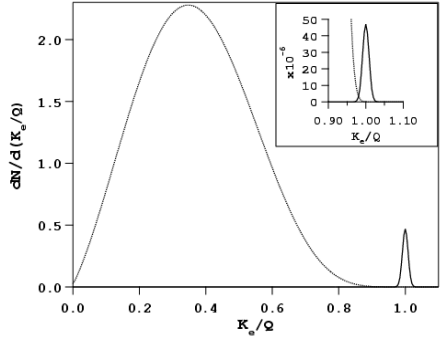}}
\caption{ Separating the $0\nu\beta\beta$ mode from the $2\nu\beta\beta$ by the shape of the
sum electron spectrum, including the effect of the 2\% resolution smearing. The figure is for the
rate ratio 1/100 and the insert for 1/$10^6$.}
\label{fig_2nu}
\end{figure}

These last distinct features allow one to separate the two modes experimentally
by measuring the sum energy of the emitted electrons with a good energy
resolution, even if the decay rate for the $0\nu\beta\beta$ mode is much smaller
than for the   $2\nu\beta\beta$ mode.
This is illustrated in Fig.\ref{fig_2nu} where the insert shows the situation for the rate
ratio of $1:10^6$ corresponding to the most sensitive current experiments.

Various aspects, both theoretical and experimental, 
of the $\beta\beta$ decay have been reviewed many times. Here I quote
just some of the review articles\cite{FS98,SuhCiv98,Ver02,EV02,EE04,AEE07}, earlier references
can be found there. 

\begin{figure}
\centerline{\includegraphics[width=7.0cm]{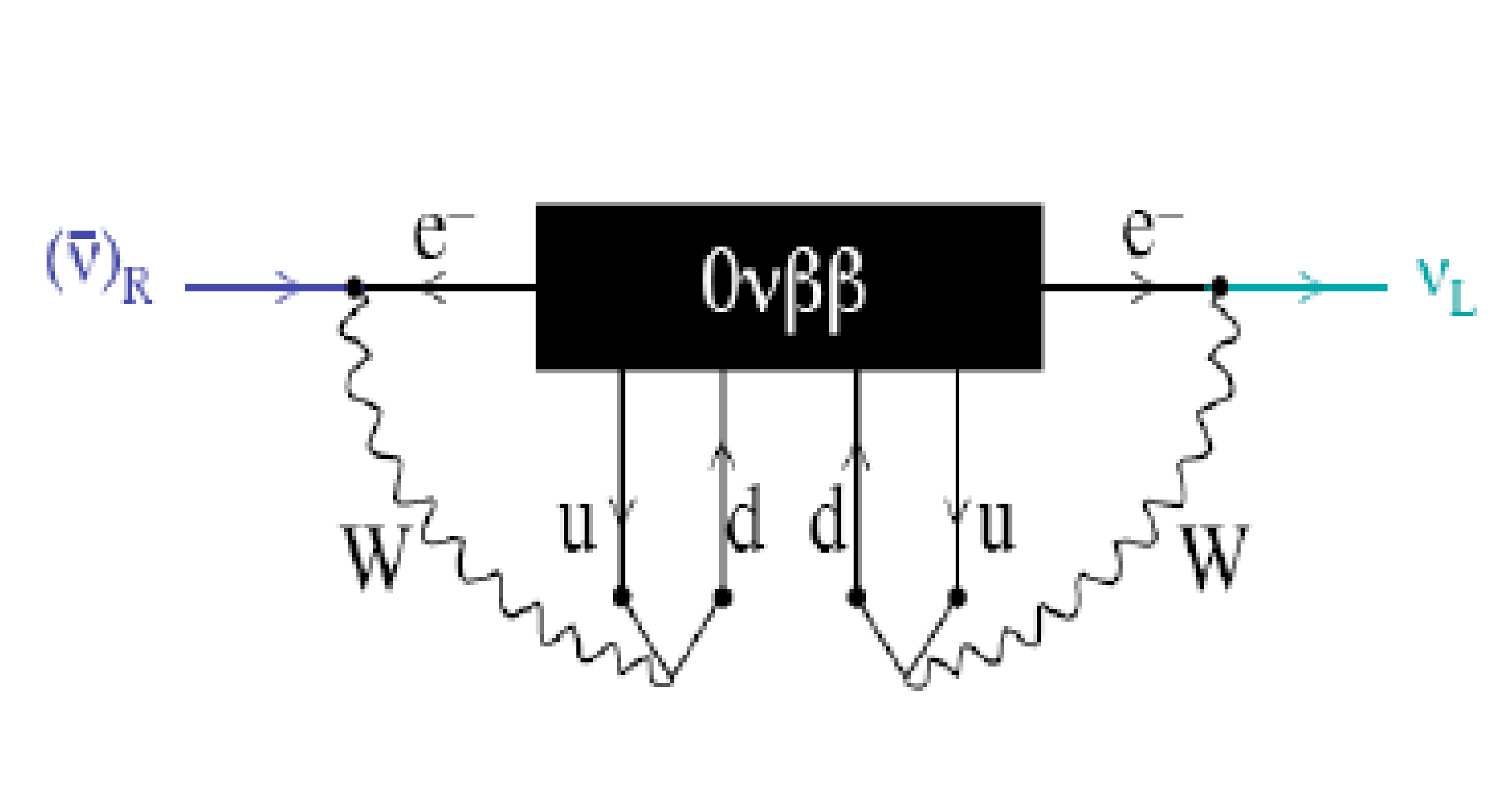}}
\caption{ By adding loops involving only standard weak interaction processes the
$\beta\beta$ decay amplitude (the black box)
implies the existence of the Majorana neutrino mass.  }
\label{fig_SV}
\end{figure}

In this introductory section let me make only few general remarks. 
The existence of the $0\nu\beta\beta$ decay
would mean that on the elementary particle level a six fermion 
lepton number violating amplitude
transforming two $d$ quarks into two $u$ quarks and two electrons
is nonvanishing.  As was first pointed out by Schechter and Valle\cite{SV82} 
more than twenty years ago,
this fact alone would guarantee that neutrinos are massive 
Majorana fermions (see Fig. \ref{fig_SV}). This qualitative statement (or theorem),
however, does not in general allow us to deduce the magnitude of the neutrino mass
once the rate of the $0\nu\beta\beta$ decay have been determined.
It is important to stress, however, that quite generally an observation of {\bf any} total
lepton number violating process, not only of the $0\nu\beta\beta$ decay, would necessarily 
imply that neutrinos are massive Majorana fermions.

There is no indication at the present time that neutrinos have nonstandard interactions,
i.e. they seem to have only interactions carried by the $W$ and $Z$ bosons
that are contained in the Standard Electroweak Model.
All observed oscillation phenomena can be understood if one assumes
that neutrinos interact exactly the way the Standard Model prescribes, but are massive
fermions forcing a generalization of the model. If we accept this, but in addition assume that
neutrinos are Majorana particles, we can in fact relate the   $0\nu\beta\beta$ decay  rate
to a quantity containing information about  the absolute neutrino mass. With these caveats that relation 
can be expressed as 
\begin{equation}
\frac{1}{T_{1/2}^{0\nu}} = G^{0\nu}(Q,Z) |M^{0\nu}|^2 \langle m_{\beta\beta} \rangle^2 ~,
\label{eq_rate}
\end{equation}
where $G^{0\nu}(Q,Z)$ is a phase space factor that depends on the transition $Q$ value and through
the Coulomb effect on the emitted electrons on the nuclear charge and that can be
easily and accurately calculated
(a complete list of the phase space factors $G^{0\nu}(Q,Z)$ and $G^{2\nu}(Q,Z)$
can be found, e.g. in Ref. \cite{BV92}), 
$M^{0\nu}$ is the nuclear matrix element that can be
evaluated in principle, although with a considerable uncertainty
and is discussed in detail later, and finally the quantity
$\langle m_{\beta\beta} \rangle$ is the effective neutrino Majorana mass, representing
the important particle physics ingredient of the process.

In turn, the effective mass $\langle m_{\beta\beta} \rangle$ is related 
to the mixing angles $\theta_{ij}$ (or to the matrix elements $|U_{e,i}|$ of the
neutrino mixing matrix)
that are determined or constrained by the oscillation experiments, to the absolute neutrino
masses $m_i$ of the mass eigenstates $\nu_i$ and to the totally unknown additional
parameters, as fundamental as the mixing angles $\theta_{ij}$, 
the so-called Majorana phases $\alpha(i)$,
\begin{equation}
 \langle m_{\beta\beta} \rangle = | \Sigma_i |U_{ei}|^2 e^{i \alpha (i)} m_i | ~.
 \label{eq_mbb}
 \end{equation}
Here $U_{ei}$ are the matrix elements of the first row of the neutrino mixing matrix.

\begin{figure}
\centerline{\includegraphics[width=11.0cm]{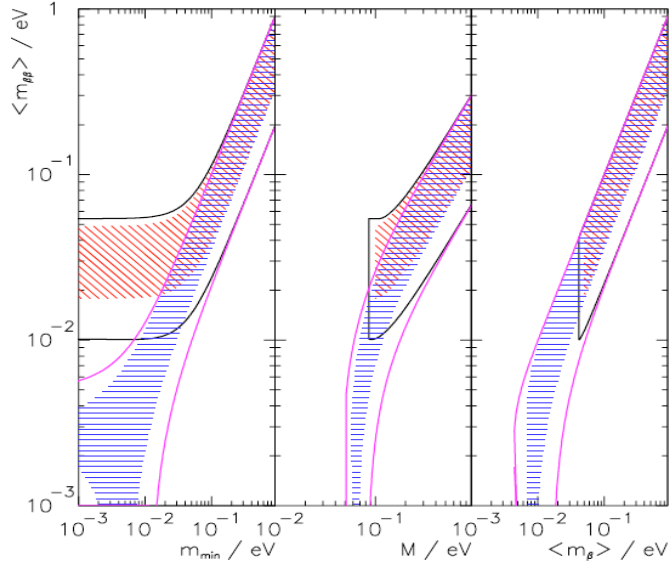}}
\caption{The left panel shows the dependence of $\langle m_{\beta\beta} \rangle$ on the absolute 
mass of the lightest neutrino mass eigenstate $m_{min}$, the middle one shows the relation between
 $\langle m_{\beta\beta} \rangle$ and the sum of   neutrino masses $M = \Sigma m_i$
 determined or costrained by the "observational cosmology", and the right
 one depicts the relation between $\langle m_{\beta\beta} \rangle$ and the effective mass 
 $\langle m_{\beta} \rangle$ determined or contrained by the ordinary $\beta$ decay. In all panels the
 width of the hatched area is due to the unknown Majorana phases and therefore irreducible. 
 The solid lines indicate the allowed regions by taking into account the current uncertainties in the
 oscillation parameters; they will shrink as the accuracy improves. The two sets of curves  
 correspond to the normal and inverted hierarchies, they merge above about 
 $\langle m_{\beta\beta} \rangle \sim$ 0.1 eV, where the degenerate mass pattern begins.  }
\label{fig_5}
\end{figure}

It is straightforward to use eq.(\ref{eq_mbb}) and the known 
neutrino oscillation results in order to compare $\langle m_{\beta\beta} \rangle$
with other neutrino mass related quantities. This is illustrated in Fig.\ref{fig_5}.
Traditionally such plot is made as in the left panel. However, the lightest
neutrino mass $m_{min}$ is not an observable quantity. For that reason
the other two panels show the relation of $\langle m_{\beta\beta} \rangle$
to the sum of the neutrino masses $M$
that is constrained and perhaps one day will be determined
by the ``observational cosmology", and also to $\langle m_{\beta} \rangle$
that represent the parameter that can be determined or constrained in ordinary
$\beta$ decay,
\begin{equation}
\langle m_{\beta} \rangle^2 = \Sigma_i |U_{ei}|^2 m_i^2 ~.
\end{equation}

Several remarks are in order. First, the observation of the $0\nu\beta\beta$ decay 
and determination of $\langle m_{\beta\beta} \rangle$, even when combined
with the knowledge of $M$ and/or $\langle m_{\beta} \rangle$ does not allow,
in general, to distinguish between the normal and inverted mass orderings. This is
a consequence of the fact that the Majorana phases are unknown. In regions in
Fig. \ref{fig_5} where the two hatched bands overlap it is clear that two solutions
with the same  $\langle m_{\beta\beta} \rangle$ and the same $M$ 
(or the same $\langle m_{\beta} \rangle$) 
always exist and cannot be distinguished.

On the other hand, obviously, if one can determine that
$\langle m_{\beta\beta} \rangle \ge$ 0.1 eV we would conclude that the 
mass pattern is degenerate. And in the so far hypothetical case
that one could show that $\langle m_{\beta\beta} \rangle \le  $ 0.01 eV, but nonvanishing
nevertheless, the normal hierarchy would be established\footnote{In that case
also the $\langle m_{\beta} \rangle$ in the right panel would not represent the quatity
directly related to the ordinary $\beta$ decay. There are no ideas, however, how to reach 
the corresponding sensitivity in ordinary $\beta$ decay at the present time.}.

It is worthwhile noting that if the inverted mass ordering is realized in nature,
(and neutrinos are Majorana particles) the quantity $\langle m_{\beta\beta} \rangle$
is constrained from below by $\sim$0.01 eV. This is within the reach of the
next generation of experiments. Also, at least in principle, in the
case of the normal hierarchy while all neutrinos could be massive
Majorana particles it would be still possible that 
$\langle m_{\beta\beta} \rangle$ = 0. Such a situation, however, requires "fine tuning"
or reflects a symmetry of some kind. 
For example, if $\theta_{13}=0$ the relation $m_1/m_2 = \tan^2 \theta_{12}$
must be realized in order that $\langle m_{\beta\beta} \rangle$ = 0. This implies,
therefore, a definite relation between the neutrino masses and mixing angles.
There is only one value of $m_1$ = 4.6 meV for which this condition is valid.

Let us finally remark that the $0\nu\beta\beta$ decay is not the only LNV process for which 
important experimental constraints exist. Examples of the other LNV
processes with important limits are
\begin{eqnarray}
&& \mu^- + (Z,A)  \rightarrow  e^+ + (Z-2,A); {~\rm exp.~ branching~ ratio} \le 10^{-12} ~,
\nonumber \\
&&  K^+  \rightarrow   \mu^+ \mu^+ \pi^-;  {~\rm exp. ~branching ~ratio} \le 3 \times 10^{-9} ~,
  \nonumber \\
&&  \bar{\nu}_e {\rm ~emission~ from~ the~ Sun};  {~\rm exp.~ branching~ ratio} \le 10^{-4} ~.
\end{eqnarray}  
However, detailed analysis suggests that the study of the $0\nu\beta\beta$ decay is by far the
most sensitive test of LNV. In simple terms, this is caused by the amount of tries one can make.
A 100 kg $0\nu\beta\beta$ decay source contains $\sim 10^{27}$ nuclei
that can be observed for a long time (several years). This can be contrasted
with the possibilities of first producing muons or kaons, and then searching for the unusual
decay channels. The Fermilab accelerators, for example, produce $\sim 10^{20}$ protons on target
per year in their beams and thus correspondingly smaller numbers of muons or kaons.  

\section{Mechanism of the $0\nu\beta\beta$ decay}

\begin{figure}
\centerline{\includegraphics[width=9.0cm]{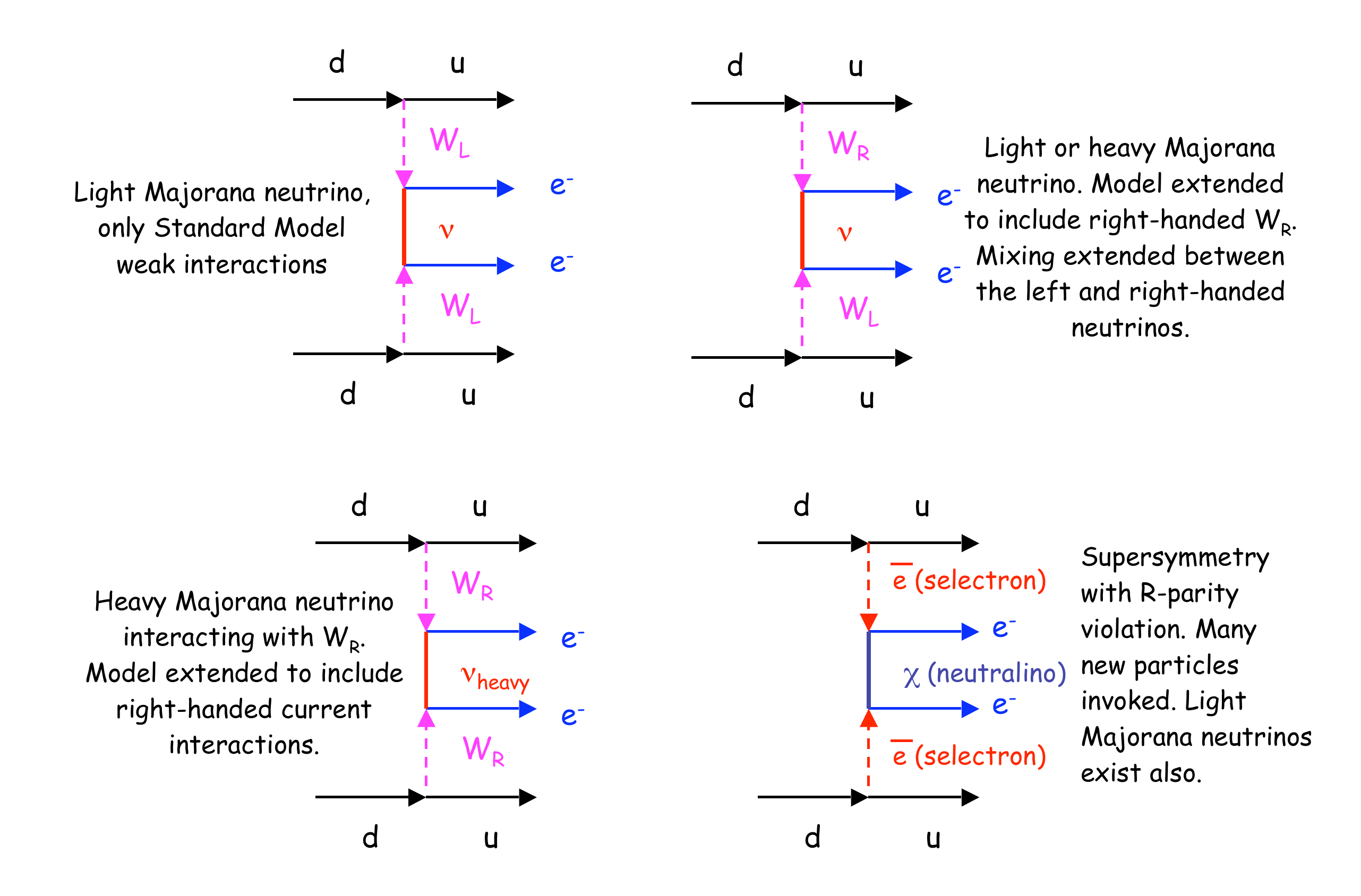}}
\caption{All these symbolic Feynman 
graphs potentially contribute to the  $0\nu\beta\beta$-decay
amplitude}
\label{fig:lnv}
\end{figure}

It has been recognized long time ago that the relation between the $0\nu\beta\beta$-decay
rate and the effective Majorana mass $\langle m_{\beta\beta} \rangle$
is to some extent problematic. The rather conservative
assumption leading to eq.(\ref{eq_rate}) is that the only possible way
the $0\nu\beta\beta$ decay can occur is through the exchange
of a virtual light, but massive, 
Majorana neutrino between the two nucleons undergoing the transition,
and that these neutrinos interact by the standard left-handed weak currents. But that is not
the only possible mechanism. LNV interactions involving so far unobserved 
much heavier ($\sim$ TeV) particles
can lead to a comparable $0\nu\beta\beta$ decay rate. 
Some of the possible mechanisms of the elementary $dd \rightarrow uu+e^-e^-$
transition (the ``black box in Fig. \ref{fig_SV})
are indicated in Fig. \ref{fig:lnv}.  Only the graph in the upper left
corner would lead to eq. (\ref{eq_rate}). Thus, in the absence of additional
information about the mechanism responsible for the $0\nu\beta\beta$ decay
one could not unambiguously infer the magnitude of  $\langle m_{\beta\beta} \rangle$  
from the $0\nu\beta\beta$-decay rate.

In general $0\nu\beta\beta$ decay can be generated by (i) light massive Majorana 
neutrino exchange or (ii) heavy particle exchange (see, e.g. Refs.\cite{heavy,Pre03}),
resulting from LNV dynamics at some scale $\Lambda$ above the electroweak one.
The relative size of heavy ($A_H$) versus light 
particle ($A_L$) exchange contributions to the decay amplitude 
can be crudely estimated as follows~\cite{Mohapatra:1998ye}: 
\begin{equation}
A_L \sim G_F^2  \frac{\langle m_{\beta \beta} \rangle}{\langle k^2 \rangle}  ,~ 
 A_H \sim G_F^2  \frac{M_W^4}{\Lambda^5}  ,~
\frac{A_H}{A_L} \sim \frac{M_W^4 \langle k^2 \rangle } 
{\Lambda^5  \langle m_{\beta \beta} \rangle }  \ , 
\label{eq_estimate}
\end{equation}
where $\langle m_{\beta \beta} \rangle$ is the effective neutrino
Majorana mass, 
$\langle k^2 \rangle \sim ( 100 \ {\rm MeV} )^2 $ is the
typical light neutrino virtuality, and $\Lambda$ is the heavy
scale relevant to the LNV dynamics. 
Therefore,  $A_H/A_L \sim O(1)$ for  $\langle m_{\beta \beta} \rangle \sim 0.1-0.5$ 
eV and $\Lambda \sim 1$ TeV, and  thus the LNV dynamics at the TeV
scale leads to similar $0 \nu \beta \beta$-decay rate as the
exchange of light Majorana neutrinos with the effective mass 
$\langle m_{\beta \beta} \rangle \sim 0.1-0.5$ eV. 

Obviously, the lifetime measurement by itself
does not provide the means for determining the underlying mechanism.
The spin-flip and non-flip exchange can be, in principle,
distinguished by the measurement of the single-electron spectra or
polarization (see e.g. \cite{Doi}).  However, in most cases the
mechanism of light Majorana neutrino exchange, and of
heavy particle exchange cannot be separated by the observation
of the emitted electrons. Thus one must look for other phenomenological
consequences of the different mechanisms. Here I discuss the
suggestion\cite{LNVus} that under natural assumptions the presence of low scale
LNV interactions, and therefore the absence of proportionality between
$\langle m_{\beta \beta} \rangle^2$ and the $0\nu\beta\beta$-decay
rate also affects muon lepton flavor violating (LFV)
processes, and in  particular enhances the $\mu \to e$ conversion 
compared to the $\mu \to e \gamma$ decay.

The discussion is
concerned mainly with the branching ratios $B_{\mu \rightarrow e \gamma} = \Gamma
(\mu \rightarrow e \gamma)/ \Gamma_\mu^{(0)}$ and $B_{\mu \to e} =
\Gamma_{\rm conv}/\Gamma_{\rm capt} $, where $\mu \to e \gamma$ is
normalized to the standard muon decay rate $\Gamma_\mu^{(0)} = (G_F^2
m_\mu^5)/(192 \pi^3)$, while $\mu \to e$ conversion is normalized to
the capture rate $\Gamma_{\rm capt}$. The main diagnostic tool in our
analysis is the ratio 
\begin{equation}
{\cal R} = B_{\mu \to e}/B_{\mu \rightarrow e \gamma} ~,
\end{equation}
and the relevance of our observation relies on the potential
for LFV discovery in the forthcoming experiments  MEG~\cite{MEG}
($\mu \to e \gamma$) and MECO~\cite{MECO} 
($\mu \to e$ conversion)\footnote{Even though the MECO 
experiment, that aimed at substantial incerase in
sensitivity of the $\mu \rightarrow e$ conversion,  was recently cancelled, proposals
for experiments with similar sensitivity exist elsewhere.}.

At present, the most stringent limit on the branching ratio $B_{\mu \rightarrow e \gamma}$
is \cite{mega} $1.2 \times 10^{-11}$ and the MEG experiment aims at the sensitivity
about two orders of magnitude better. For the muon conversion the best experimental limit
\cite{conv} used gold nuclei and reached $B_{\mu \to e} < 8 \times 10^{-13}$. The
various proposals aim at reaching sensitivity of  about $10^{-17}$.

It is useful to formulate the problem in terms of effective low energy
interactions obtained after integrating out the heavy degrees of
freedom that induce LNV and LFV dynamics.
Thus, we will be dealing only with the Standard Model particles
and all symmetry relations will be obeyed. However, operators of dimension $>$ 4 will
be suppressed by $1/\Lambda^{d-4}$, where $\Lambda$ is the
scale of new physics. If the scales for both 
LNV and LFV are well above the weak scale, then one would not 
expect to observe any signal in the forthcoming LFV experiments, nor would 
the effects of heavy particle exchange enter $0\nu\beta\beta$ 
at an appreciable level. In this case, the only origin of a signal in 
$0\nu\beta\beta$ at the level of prospective experimental sensitivity 
would be the exchange of a light Majorana neutrino, leading to eq.(\ref{eq_rate}),
and allowing one to extract  $\langle m_{\beta \beta} \rangle$ from the decay rate.

In general, however, the two scales may be distinct, as in
SUSY-GUT~\cite{Barbieri:1995tw} or SUSY see-saw~\cite{Borzumati:1986qx} models. 
In these scenarios, both the Majorana neutrino mass as well as LFV effects are 
generated at the GUT scale.
The effects of heavy Majorana neutrino exchange in $0\nu\beta\beta$ are, thus, 
highly suppressed. In contrast,  the effects of GUT-scale LFV are transmitted 
to the TeV-scale by a soft SUSY-breaking sector without mass suppression 
via renormalization group running of the high-scale LFV couplings. 
Consequently, such scenarios could lead to observable effects 
in the upcoming LFV experiments but with an ${\cal O}(\alpha)$ 
suppression of the branching ratio 
$B_{\mu\to e} $ relative to $B_{\mu\to e\gamma}$ 
due to the exchange of a virtual photon in the conversion process 
rather than the emission of a real one.

As a specific example let us quote the SUSY SU(5) scenario where \cite{Barbieri1994}
\begin{equation}
B_{\mu\to e\gamma} = 2.4 \times 10^{-12} \left( \frac{|V_{ts}|}{0.04} \frac{|V_{td}|}{0.01}\right)^2
\left(\frac{100 \rm{GeV}}{m_{\bar{\mu}}}\right)^4 ~,
\end{equation}
\begin{equation}
B_{\mu\to e} =  5.8 \times 10^{-12} \alpha \left( \frac{|V_{ts}|}{0.04} \frac{|V_{td}|}{0.01}\right)^2
\left(\frac{100 \rm{GeV}}{m_{\bar{\mu}}}\right)^4 ~,
\end{equation}
where the gaugino masses were neglected.

Another example is the evaluation of the ratio 
${\cal R} = B_{\mu \to e}/B_{\mu \rightarrow e \gamma} $
in the constrained seesaw minimal supersymmetric model \cite{arganda}
with very high scale LNV and ${\cal R} \sim 1/200$ for a variety
of input parameters. There are, however, exceptions, like the
recent evaluation of ${\cal R}$ in a variety of SUSY SO(10)
models \cite{carl} with high scale LNV but with ${\cal R}$ as large as 0.3
in one case.

The case where the
scales of LNV and LFV are both relatively low ($\sim$ TeV)
is more subtle and requires more detailed analysis.
This is the scenario which might lead to observable signals
in LFV searches and at the same time generate ambiguities in
interpreting a  positive signal in $0 \nu \beta \beta$.
This is the case where one needs to develop some
discriminating criteria.

Denoting the new physics scale by $\Lambda$, one has a LNV
effective lagrangian (dimension $d=9$ operators) of the form
\begin{equation}
{\cal L}_{0 \nu \beta \beta} = \displaystyle\sum_i \
\frac{\tilde{c}_i}{\Lambda^5}  \  \tilde{O}_i
\qquad  \tilde{O}_{i} =  \bar{q} \Gamma_1 q \,  \
\bar{q} \Gamma_2 q \,   \bar{e} \Gamma_3 e^c   \ ,
\label{eq:lag1}
\end{equation}
where we have suppressed the flavor and Dirac structures
(a complete list of the dimension nine operators
$\tilde{O}_i$ can be found in Ref.~\cite{Pre03}).

For the LFV interactions (dimension $d=6$ operators), one has
\begin{equation}
{\cal L}_{\rm LFV} = \displaystyle\sum_i \
\frac{c_i}{\Lambda^2}  \  O_i  \  ,
\label{eq:lag2}
\end{equation}
and a complete operator basis can be found in
Refs.\cite{Raidal:1997hq,Kitano:2002mt}.  The LFV operators relevant to
our analysis are of the following type (along with their analogues
with $L \leftrightarrow R$):
\begin{eqnarray}
O_{\sigma L} & = &   \displaystyle\frac{e}{(4 \pi)^2}
 \overline{\ell_{iL}} \, \sigma_{\mu \nu} i
/ \hspace{-0.23cm}D \, \ell_{jL}  \  F^{\mu \nu}  + {\rm h.c.}
\nonumber \\
O_{\ell L} & = &   \overline{\ell_{iL}} \, \ell^c_{jL} \
\overline{\ell^c_{kL}} \, \ell_{mL}
\nonumber \\
O_{\ell q} & = &   \overline{\ell_{i}} \Gamma_\ell \ell_{j} \
\overline{q} \Gamma_q  q  \  .
\end{eqnarray}

Operators of the type $O_{\sigma}$ are typically generated at one-loop level,
hence our choice to explicitly display the loop factor
$1/(4 \pi)^2$.   On the
other hand, in a large class of models, operators of the type $O_{\ell}$ or
$O_{\ell q}$ are generated by tree level exchange of heavy degrees of
freedom. With the above choices,
all non-zero $c_i$ are nominally of the same size,
typically the product of two Yukawa-like couplings or gauge couplings
(times flavor mixing matrices). 

With the notation established above, the ratio ${\cal R}$
of the branching ratios $\mu \to e$ to $\mu \to e + \gamma$  can be 
written schematically as follows (neglecting flavor indices in the
effective couplings and the term with $L \leftrightarrow R$)
\cite{LNVus}:
\begin{eqnarray}  
{\cal R} &=& 
\displaystyle\frac{\Phi}{48 \pi^2} \,
\Big| \lambda_1  \, e^2  c_{\sigma L}  + e^2 \left( 
\lambda_2   c_{\ell L} + \lambda_3 c_{\ell q} \right)   
\log \displaystyle\frac{\Lambda^2}{m_\mu^2} 
\nonumber \\
&+&   \lambda_4 (4 \pi)^2  c_{\ell q} \ + \dots 
\Big|^2 / 
\left[ e^2 \left( |c_{\sigma L}|^2  + |c_{\sigma R}|^2 \right) \right] \, .
\label{eq:main1}  
\end{eqnarray}

\begin{figure}
\centerline{\includegraphics[width=4.5cm]{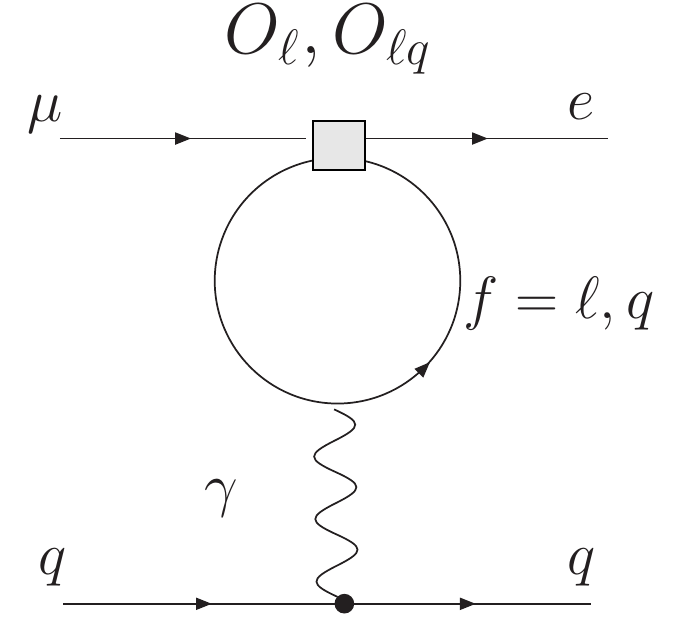}}
\caption{Loop contributions to $\mu \to e$ conversion through insertion of 
operators $O_{\ell}$ or $O_{\ell q}$, generating the large logarithm.}
\label{fig_6}
\end{figure}

In the above formula $\lambda_{1,2,3,4}$ are numerical factors of
$O(1)$, while the overall factor $\frac{\Phi}{48 \pi^2}$ 
arises from phase space and overlap
integrals of electron and muon wavefunctions in the nuclear field. For
light nuclei $\Phi = (Z F_p^2)/(g_V^2 + 3 g_A^2) \sim O(1)$ ($g_{V,A}$
are the vector and axial nucleon form factors at zero momentum
transfer, while $F_p$ is the nuclear form factor at
$q^2 = -m_\mu^2$~\cite{Kitano:2002mt}). The dots indicate 
subleading terms, not
relevant for our discussion, such as loop-induced
contributions to $c_{\ell}$ and $c_{\ell q}$ that are analytic in
external masses and momenta.  In contrast the 
logarithmically-enhanced loop contribution given by the second term in
the numerator of ${\cal R}$ plays an essential role. This term arises 
whenever the operators $O_{\ell L,R}$ and/or $O_{\ell q}$ appear at
tree-level in the effective theory and generate one-loop
renormalization of $O_{\ell q}$~\cite{Raidal:1997hq} (see
Fig.~\ref{fig_6}).

The ingredients in eq.~(\ref{eq:main1}) lead to several observations:
(i) In absence of tree-level $c_{\ell L }$ and $c_{\ell q}$, one
obtains ${\cal R} \sim (\Phi \, \lambda_1^2 \, \alpha)/(12 \pi) \sim
10^{-3}-10^{-2}$, due to gauge coupling and phase space
suppression. 
(ii) When present, the logarithmically enhanced contributions,
i.e. when either  $c_{\ell L }$ or  $c_{\ell q}$ or both are nonvanishing,
compensate for the gauge coupling and phase space suppression, leading
to ${\cal R} \sim O(1)$. 
(iii) If present, the tree-level coupling $c_{\ell q}$ dominates the
$\mu \to e$ rate leading to ${\cal R} \gg 1$. 

Thus, we can formulate our main conclusions regarding the discriminating
power of the ratio ${\cal R}$:
\begin{enumerate} 
\item 
Observation of both the LFV muon processes
$\mu \to e$ and $\mu \to e \gamma$ with relative ratio ${\cal R} \sim
10^{-2}$ implies, under generic conditions, that $\Gamma_{0 \nu \beta
\beta} \sim \langle m_{\beta \beta} \rangle^2$. Hence the relation
of the $0\nu\beta\beta$ lifetime to the absolute neutrino mass scale
is straightforward.
\item 
On the other hand, observation of LFV muon processes with
relative ratio ${\cal R} \gg 10^{-2}$ could signal non-trivial LNV
dynamics at the TeV scale, whose effect on $0 \nu \beta \beta$ has to
be analyzed on a case by case basis. Therefore, in this scenario no
definite conclusion can be drawn based on LFV rates.
\item
Non-observation of LFV in muon processes in forthcoming 
experiments would imply either that the scale of non-trivial LFV and
LNV is  above a few TeV, and thus 
$\Gamma_{0 \nu \beta\beta} \sim \langle m_{\beta \beta} \rangle^2$, 
or that any TeV-scale LNV is
approximately flavor diagonal (this is an important caveat).
\end{enumerate}

The above statements are illustrated using two explicit cases\cite{LNVus}: the
minimal supersymmetric standard model (MSSM) with R-parity violation
(RPV-SUSY) and the Left-Right Symmetric Model (LRSM). Limits on
the rate of the $0\nu\beta\beta$ decay were used in the past 
to constrain parameters of these two models \cite{heavy}.

\begin{figure}
\centerline{\includegraphics[width=6.0cm]{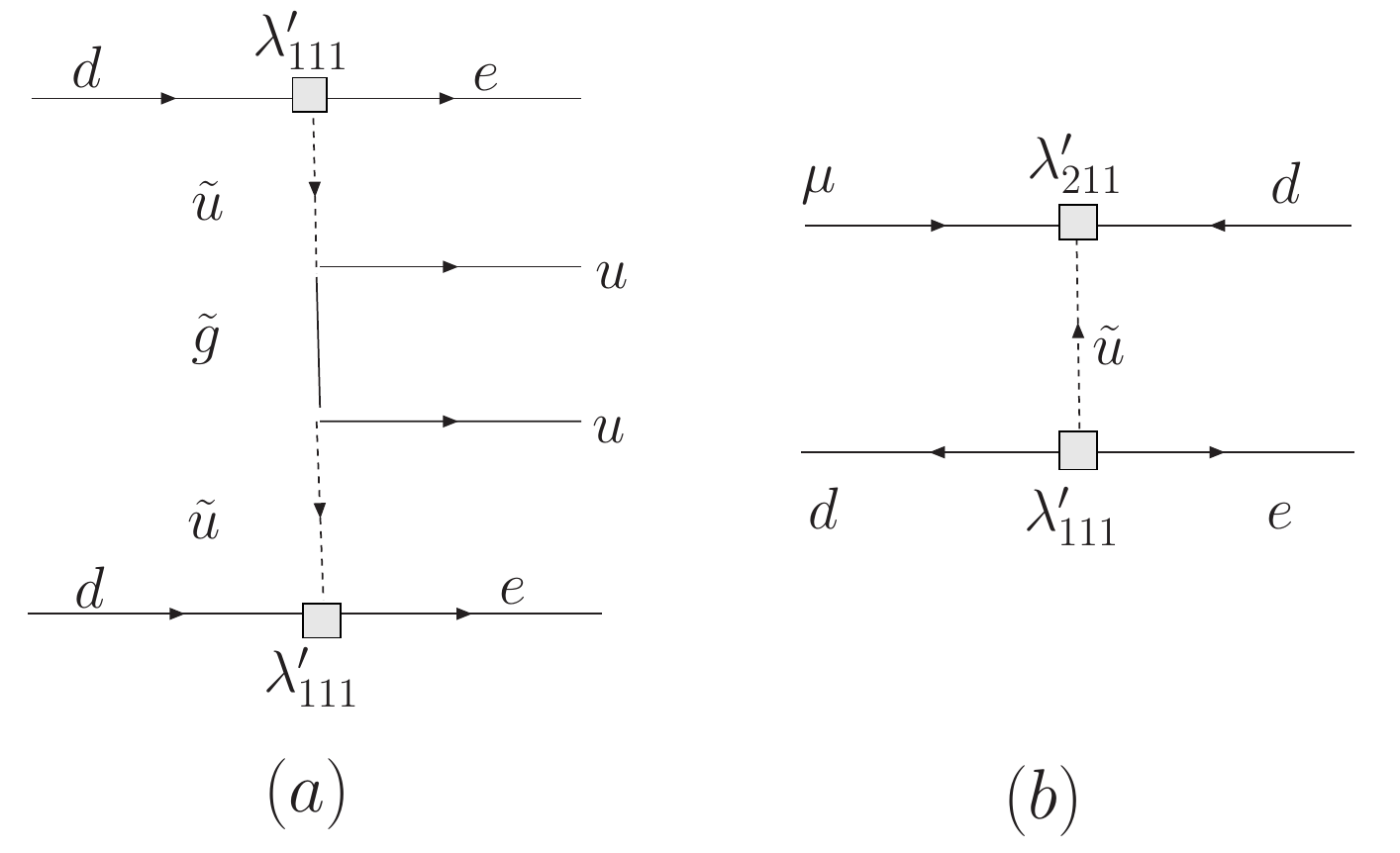}}
\caption{Gluino exchange contribution to $0 \nu \beta \beta$ $(a)$, 
and typical tree-level contribution to $O_{\ell q}$ $(b)$ in RPV SUSY.}
\label{fig_7}
\end{figure}

{\em RPV SUSY ---}  If one does not impose R-parity conservation 
[$R = (-1)^{3 (B-L) + 2 s}$], the MSSM superpotential includes, in addition
to the standard Yukawa terms, lepton and baryon number violating
interactions, compactly written as (see, e.g.~\cite{Dreiner:1997uz})  
\begin{eqnarray}
W_{RPV} &=& \lambda_{ijk} L_i L_j E_k^c + 
\lambda_{ijk}' L_i Q_j D_k^c + 
\lambda_{ijk}''  U_i^c D_j^c D_k^c 
\nonumber \\
& & +  \mu_{i}' L_i  H_u   \ , 
\end{eqnarray}
where $L$ and $Q$ represent lepton and quark doublet superfields,
while $E^c$, $U^c$, $D^c$ are lepton and quark singlet superfields.
The simultaneous presence of $\lambda'$ and $\lambda ''$ couplings
would lead to an unacceptably large proton decay rate (for SUSY mass
scale $\Lambda_{SUSY} \sim$ TeV), so we focus on the case of
$\lambda'' = 0$ and set $\mu'=0$ without loss of
generality. In such case, lepton number is
violated by the remaining terms in $W_{RPV}$, leading to short
distance contributions to $0 \nu \beta \beta$
[see Fig.~\ref{fig_7}(a)], with typical
coefficients [cf. eq.~(\ref{eq:lag1})]
\begin{equation} 
\frac{\tilde{c_i}}{\Lambda^5} \sim 
\frac{\pi
\alpha_s}{m_{\tilde{g}}} \frac{\lambda_{111}'^2}{m_{\tilde{f}}^4} \, ; \, 
\frac{\pi
\alpha_2}{m_\chi} \frac{\lambda_{111}'^2}{m_{\tilde{f}}^4} \ ,  
\end{equation}
where $\alpha_s, \alpha_2$ represent the strong and weak gauge
coupling constants, respectively.  The RPV interactions also lead to
lepton number conserving but lepton flavor violating operators [see
 Fig.~\ref{fig_7}(b)], with coefficients
[cf. eq.~(\ref{eq:lag2})]
\begin{eqnarray}
\frac{c_{\ell}}{\Lambda^2} &\sim &
\frac{ \lambda_{i11} \lambda_{i21}^*}{m_{\tilde{\nu}_i}^2} , 
\frac{ \lambda_{i11}^* \lambda_{i12}}{m_{\tilde{\nu}_i}^2}  \ , 
\nonumber \\
\frac{c_{\ell q}}{\Lambda^2} & \sim & 
\frac{ \lambda_{11i}'^* \lambda_{21i}'}{m_{\tilde{d}_i}^2}, 
\frac{ \lambda_{1i1}'^* \lambda_{2i1}'}{m_{\tilde{u}_i}^2} \ , 
\nonumber \\
\frac{c_\sigma}{\Lambda^2} 
& \sim &  \frac{ \lambda \lambda^* }{m_{\tilde{\ell}}^2}, 
\frac{\lambda'  \lambda'^* }{m_{\tilde{q}}^2}  \ ,
\end{eqnarray}
where the flavor combinations contributing to $c_{\sigma}$ can be
found in Ref.~\cite{deGouvea:2000cf}.  Hence, for generic flavor
structure of the couplings $\lambda$ and $\lambda'$ 
the underlying LNV dynamics generate both short distance
contributions to $0 \nu \beta \beta$ and LFV contributions that lead
to ${\cal R} \gg 10^{-2}$.  

Existing limits on rare processes strongly constrain combinations of
RPV couplings, assuming $\Lambda_{SUSY}$ is 
between a few hundred GeV and $\sim$ 1 TeV.  Non-observation of LFV at future
experiments MEG and MECO could be attributed either to a larger 
$\Lambda_{SUSY}$ ($>$ few TeV) or to suppression of couplings that
involve mixing among first and second generations.  In the former
scenario, the short distance contribution to $0\nu \beta 
\beta$ does not compete with the long distance one
[see eq.~(\ref{eq_estimate})], so that $\Gamma_{0 \nu \beta \beta}
\sim \langle m_{\beta \beta} \rangle^2$.  On the other hand, there is an exception to this
"diagnostic tool". If the $\lambda$ and
$\lambda'$ matrices are nearly flavor diagonal, the exchange of
superpartners may still make non-negligible contributions to $0\nu
\beta \beta$ without enhancing the ratio ${\cal R}$ .

{\em LRSM ---} The LRSM provides a natural scenario for introducing
non-sterile, right-handed neutrinos and Majorana
masses~\cite{Mohapatra:1979ia}.  The corresponding electroweak gauge
group $SU(2)_L \times SU(2)_R \times U(1)_{B-L}$, breaks down to
$SU(2)_L \times U(1)_Y$ at the scale $\Lambda \ge {\cal O}({\rm
TeV})$.  The symmetry breaking is implemented through an extended
Higgs sector, containing a bi-doublet $\Phi$ and two triplets
$\Delta_{L,R}$, whose leptonic couplings generate both Majorana
neutrino masses and LFV involving charged leptons:

\begin{eqnarray}
{\cal L}_Y^{\rm lept} &=& - \  
\overline{L_L}\, ^{i} \, 
\left( 
 y_D^{ij}  \,  \Phi \ + \   \tilde{y}_D^{ij} \,  \tilde{\Phi} 
\right) \, 
L_{R}^j  \\
&-&  
\overline{(L_{L})^c}\, ^i \    y_M^{ij} \, \tilde{\Delta}_L \  L_{L}^j
\ - \ 
 \overline{(L_{R})^c}\, ^i \  y_M^{ij}  \, \tilde{\Delta}_R  \ L_{R}^j \ . 
\nonumber 
\end{eqnarray}
Here $ \tilde{\Phi} = \sigma_2 \Phi^* \sigma_2$,
$\tilde{\Delta}_{L,R} = i \sigma_2 \Delta_{L,R}$, and leptons belong 
to two isospin doublets $L_{L,R}^i = (\nu_{L,R}^i,
\ell_{L,R}^i)$.  The gauge symmetry is broken through the VEVs
$\langle \Delta^0_R \rangle = v_R$, $\langle \Delta^0_L \rangle = 0$,
$ \langle \Phi \rangle = {\rm diag}(\kappa_1, \kappa_2) $. 
After diagonalization of the lepton mass matrices, LFV arises from
both non-diagonal gauge interactions and the Higgs Yukawa
couplings. In particular, the $\Delta_{L,R}$-lepton interactions are
not suppressed by lepton masses and have the structure ${\cal L} \sim
\Delta_{L,R}^{++} \, \overline{\ell_i^c} \, h_{ij} \, (1 \pm \gamma_5)
\ell_j + {\rm h.c.}$. The couplings $h_{ij}$ are in general
non-diagonal and related to the heavy neutrino mixing
matrix~\cite{Cirigliano:2004mv}.

\begin{figure}
\centerline{\includegraphics[width=7.0cm]{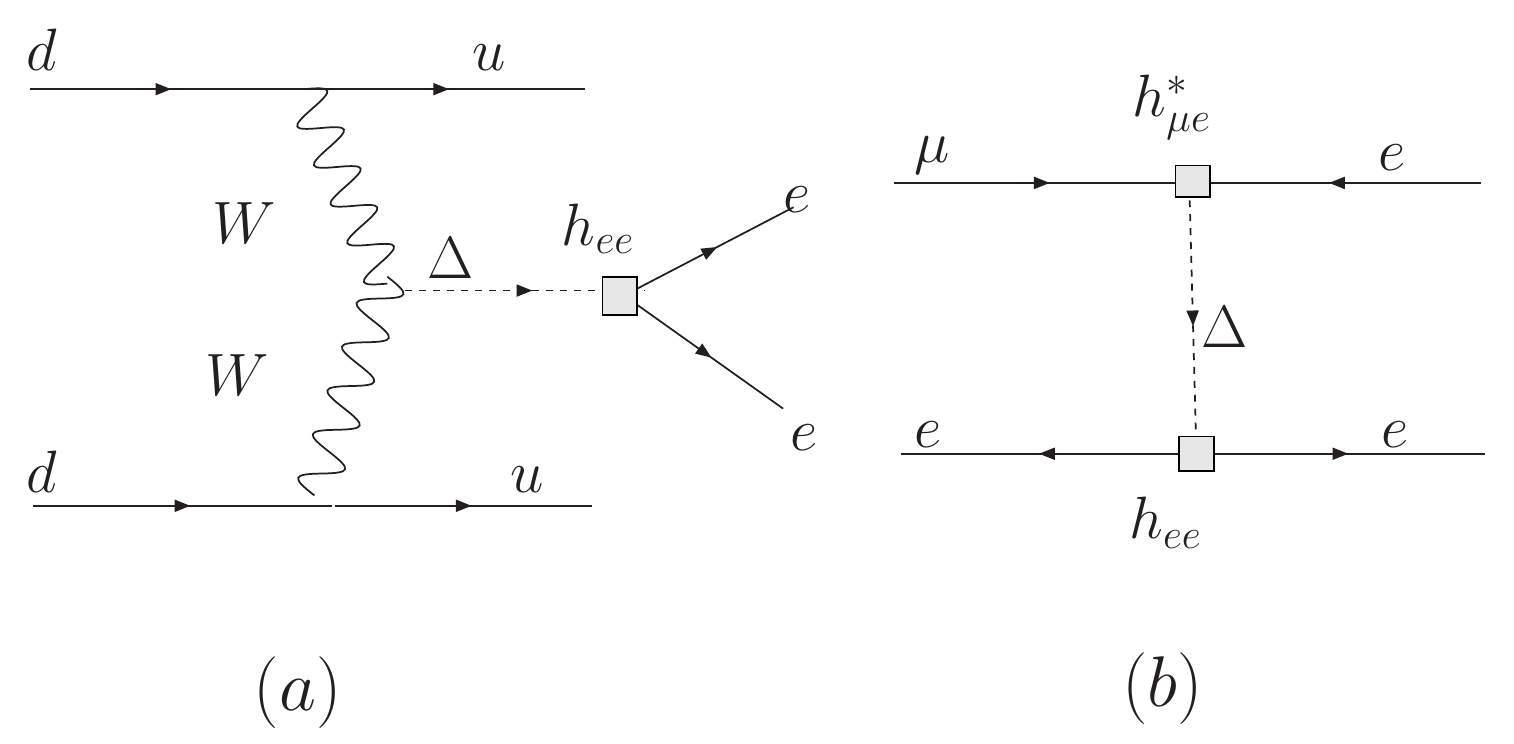}}
\caption{Typical doubly charged Higgs contribution to $0 \nu \beta \beta$ $(a)$ 
and to $O_{\ell}$ $(b)$ in the LRSM.}
\label{fig_8}
\end{figure}

Short distance contributions to $0\nu \beta \beta$ arise from the
exchange of both heavy $\nu$s and $\Delta_{L,R}$ 
(see Fig.~\ref{fig_8})(a), with
\begin{equation}
\frac{\tilde{c}_i}{\Lambda^5} \sim  
\frac{g_2^4}{M_{W_R}^4} \frac{1}{M_{\nu_R}} 
\,  ; \, \frac{g_2^3}{M_{W_R}^3} \frac{h_{ee}}{M_\Delta^2} \ , 
\end{equation}
where $g_2$ is the weak gauge coupling. LFV operators are also generated 
through non-diagonal gauge and Higgs vertices, with~\cite{Cirigliano:2004mv}
(see Fig.~\ref{fig_8}(b))
\begin{equation}
\frac{c_{\ell}}{\Lambda^2} \sim \frac{h_{\mu i} h_{ie}^*}{m_{\Delta}^2} \qquad 
\frac{c_{\sigma}}{\Lambda^2} \sim 
\frac{(h^\dagger h)_{e \mu}}{M_{W_R}^2}  \quad   i=e, \mu, \tau \ . 
\end{equation}
Note that the Yukawa interactions needed for the Majorana neutrino
mass necessarily imply the presence of LNV and LFV couplings $h_{ij}$
and the corresponding LFV operator coefficients $c_{\ell}$, 
leading to ${\cal R} \sim O(1)$. 
Again, non-observation of LFV in the next generation of
experiments would typically push $\Lambda$ into the multi-TeV range,
thus implying a negligible short distance contribution to $0 \nu \beta
\beta$.  As with RPV-SUSY, this conclusion can be evaded by assuming a  
specific flavor structure, namely $y_M$ approximately diagonal 
or a nearly degenerate heavy neutrino spectrum.

In both of these phenomenologically viable models
that incorporate LNV and LFV at low scale ($\sim$ TeV), one finds
${\cal R} \gg 10^{-2}$~\cite{Raidal:1997hq,deGouvea:2000cf,Cirigliano:2004mv}.
It is likely that the basic mechanism at work in 
these illustrative cases is  generic: low scale LNV interactions
($\Delta L = \pm 1$ and/or $\Delta L= \pm 2$), which in general
contribute to $0 \nu \beta \beta$, also generate sizable contributions
to $\mu \to e$ conversion, thus enhancing this process over $\mu \to e
\gamma$.

In conclusion of this section, 
the above considerations suggest that the ratio ${\cal R} = B_{\mu \to
e}/B_{\mu \rightarrow e \gamma}$ of muon LFV processes will provide
important insight about the mechanism of neutrinoless double beta
decay and the use of this process to determine the absolute scale of
neutrino mass.  Assuming observation of LFV processes in forthcoming
experiments, if ${\cal R} \sim 10^{-2}$ the mechanism of $0 \nu
\beta \beta$ is light Majorana neutrino exchange
and, therefore, $1/T_{1/2} \sim \langle m_{\beta \beta} \rangle^2$; 
on the other hand, if ${\cal R}
\gg 10^{-2}$, there might be TeV scale LNV dynamics, and no definite
conclusion on the mechanism of $0 \nu \beta \beta$ decay can be drawn based
only on LFV processes.

\section{Overview of the experimental status of the search for $\beta\beta$ decay}

Before embarking on the discussion of the nuclear structure aspects of the 
$\beta\beta$ decay let us briefly describe the experimental status of the field
(more detailed information on this topics are in the lectures by A. Giuliani).
The topic has a venerable history. The rate of the $2\nu\beta\beta$ decay was
first estimated by Maria Goeppert-Meyer already in 1937 in her thesis work
suggested by E. Wigner, basically correctly. 
Yet, first experimental observation in a laboratory
experiment was achieved only in 1987, fifty years later \cite{Moe}.
(Note that this is not really exceptional in neutrino physics. It took more
than twenty years since the original suggestion of Pauli to show
that neutrinos are real particles in the pioneering experiment by
Raines and Cowan. And it took another almost fifty years since that
time to show that neutrinos are massive fermions.)
Why it took so long in the case of the $\beta\beta$ decay?  
As pointed out above, the typical half-life
of the $2\nu\beta\beta$ decay is $\sim 10^{20}$ years. Yet, its 
``signature" is very similar to natural radioactivity, present to some extent
everywhere, and governed by the half-life of $\sim 10^{10}$ years
or much less for most of  the man-made or cosmogenic radioactivities.
So, background suppression is the main problem to overcome
when one wants to study either of the $\beta\beta$ decay modes.

During the last two decades the $2\nu\beta\beta$ decay has been observed
in ``live" laboratory experiments
in many nuclei, often by different groups and using different 
methods. That shows not only the ingenuity of the experimentalists who
were able to overcome the background nemesis, but makes it possible
at the same time to extract the corresponding $2\nu$ nuclear matrix element
from the measured decay rate. In the $2\nu$ mode the half-life is given by
\begin{equation}
1/T_{1/2} = G^{2\nu}(Q,Z) |M^{2\nu}|^2 ~,
\label{eq_m2nu}
\end{equation}
where $G^{2\nu}(Q,Z) $ is an easily and accurately calculable phase space factor.

\begin{figure}
\centerline{\includegraphics[width=9.0cm]{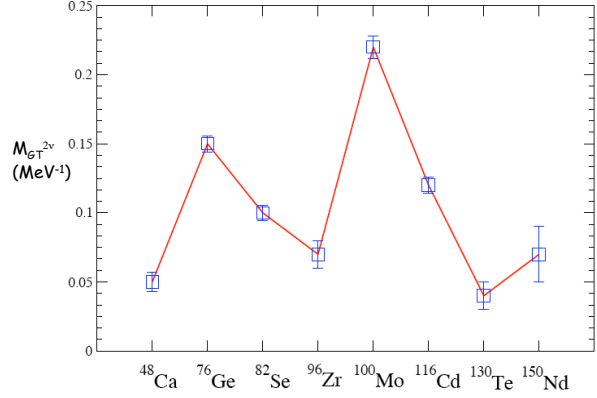}}
\caption{Nuclear matrix elements for the $2\nu\beta\beta$ decay extracted from the
measured half-lives.}
\label{fig_m2nu}
\end{figure}

The resulting nuclear matrix elements $M^{2\nu}$, which have the dimension energy$^{-1}$,
are plotted in Fig.\ref{fig_m2nu}. Note the pronounced shell dependence; the matrix element
for $^{100}$Mo is almost ten times larger than the one for $^{130}$Te. Evaluation of these
matrix elements, to be discussed below,
 is an important test for the nuclear theory models that aim at the determination
of the analogous but different quantities for the $0\nu$ neutrinoless mode.

The challenge of detecting the $0\nu\beta\beta$ decay is, at first blush, easier. Unlike the
continuous $2\nu\beta\beta$ decay spectrum with a broad
maximum at rather low energy where the background
suppression is harder, the $0\nu\beta\beta$ decay spectrum is 
sharply peaked at the known $Q$ value (see Fig.\ref{fig_2nu}),
at energies that are not immune to the background, but a bit less difficult to manage.
However, as also indicated in Fig.\ref{fig_2nu}, to obtain interesting results at the
present time means to reach sensitivity to the $0\nu$ half-lives that are $\sim10^6$ times
longer than the $2\nu$ decay half-life of the same nucleus.

\begin{figure}
\centerline{\includegraphics[width=9.0cm]{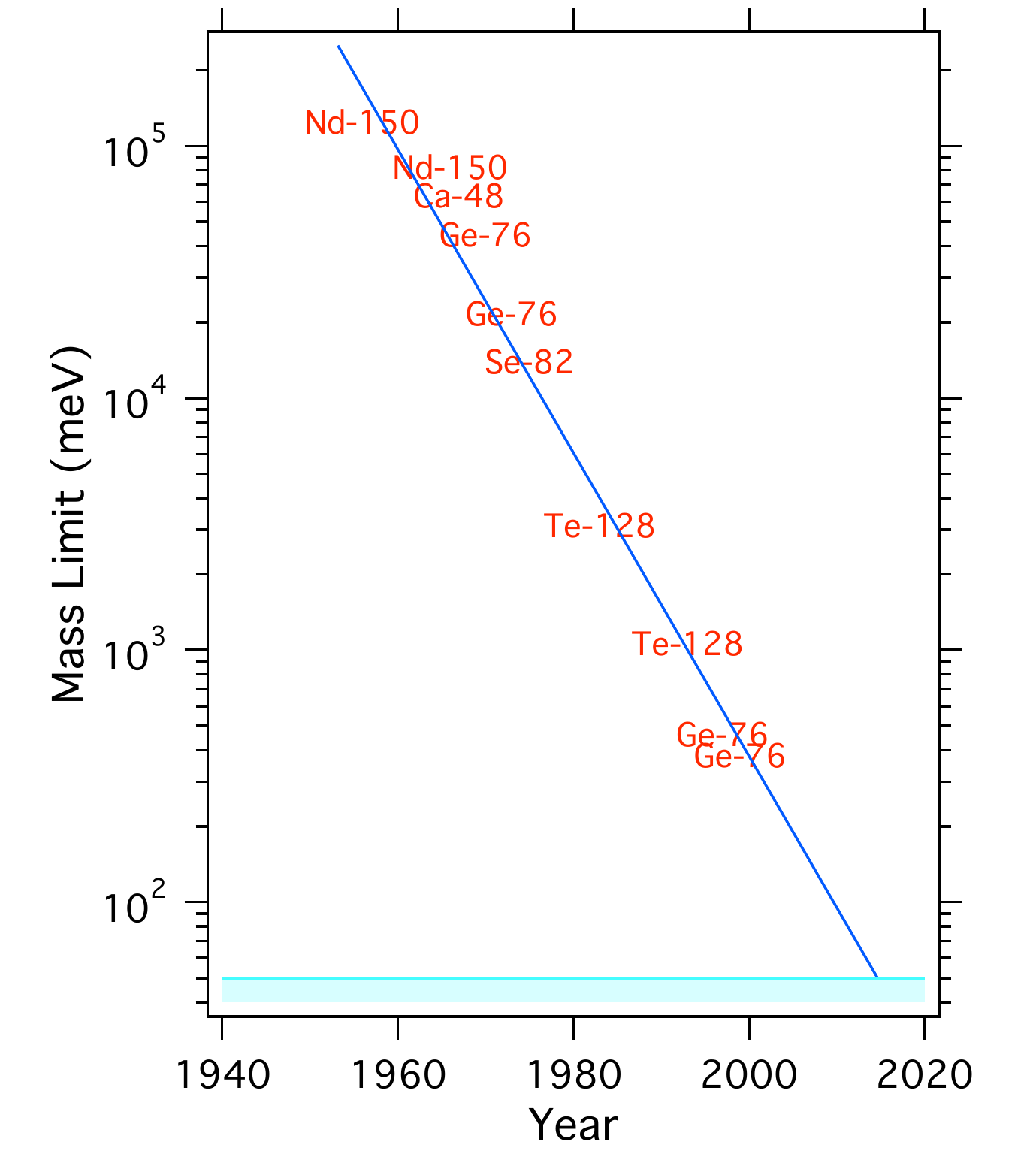}}
\caption{The limit of the effective mass $\langle m_{\beta \beta} \rangle$ 
extracted from the experimental lower limits on the $0\nu\beta\beta$ decay half-life
versus the corresponding year. The gray
band near bottom indicates the $\sqrt{\Delta m^2_{atm}}$
value. Figure originally made by S. Elliott.}
\label{fig_history}
\end{figure}

The historical lessons are illustrated in Fig.\ref{fig_history} where the past limits on the
$0\nu\beta\beta$ decay half-lives of various candidate nuclei are translated using 
eq.(\ref{eq_rate})   into the limits on the effective mass $\langle m_{\beta \beta} \rangle$.
When plotted in the semi-log plot this figure represent ``Moore's law" of double beta decay,
and indicates that, provided that the past trend will continue, the mass scale corresponding to 
$\Delta m^2_{atm}$ will be reached in about 7 years. 
This is also the time scale of significant experiments these days. Note that the figure was made using
some assumed values of the corresponding nuclear matrix elements, without including
their uncertainty. For such illustrative purposes they are, naturally, irrelevant.  

The past search for the neutrinoless double beta decay, illustrated in Fig.\ref{fig_history},
was driven by the then current technology and the resources of the individual experiments.
The goal has been simply to reach sensitivity to longer and longer half-lives.
The situation is different, however, now. The experimentalists at the 
present time can, and do, use the knowledge
summarized in Fig. \ref{fig_5} to gauge the aim of their proposals. Based on that figure,
the range of the mass parameter $\langle m_{\beta \beta} \rangle$  can be divided
into three regions of interest.
\begin{itemize}
\item The degenerate mass region where all $m_i \gg \sqrt{\Delta m^2_{atm}}$. In that
region $\langle m_{\beta \beta} \rangle \ge$ 0.1 eV, corresponding crudely to the $0\nu$
half-lives of $10^{26-27}$ years. To explore it (in a realistic time frame), 
$\sim$ 100 kg of the decaying nucleus is needed. Several experiments aiming at such 
sensitivity are being built and should run very soon and give results within the next 
$\sim$ 3 years.
Moreover, this mass region (or a substantial part of it) will be explored, in a similar
time frame, by the study of ordinary $\beta$ decay (in particular of tritium, 
see the lectures by C. Weinheimer) and by
the observational cosmology (see the lectures by S. Pastor). 
These techniques are independent on the Majorana
nature of neutrinos. It is easy, but perhaps premature, 
to envision various possible scenarios depending on the possible
outcome of these measurements.
\item The so-called inverted hierarchy mass region where 
$20 < \langle m_{\beta \beta} \rangle < 100$ meV and the $0\nu\beta\beta$ half-lives
are about $10^{27-28}$ years. (The name is to some extent a misnomer. In that interval
one could encounter not only the inverted hierarchy but also a quasi-degenerate but
normal neutrino mass ordering. Successfull observation of the $0\nu\beta\beta$ decay
will not be able to distinguish these possibilities, as I argued above.
This is so not only due to the anticipated experimental accuracy, but more fundamentally
due to the unknown Majorana phases.) To explore this
mass region, $\sim$ ton size sources would be required. Proposals for the corresponding
experiments exist, but none has been funded as yet, and presumably the real work will
begin depending on the experience with the various $\sim$ 100 kg size sources. Timeline
for exploring this mass region is $\sim$ 10 years.
\item Normal mass hierarchy region where $\langle m_{\beta \beta} \rangle \le$ 10-20 meV.
To explore this mass region, $\sim$ 100 ton sources would be required. There are no realistic
proposals for experiments of this size at present.
\end{itemize}

Over the last two decades, the methodology for double beta decay experiments has improved
considerably.
Larger amounts of high-purity enriched parent isotopes, combined with careful selection
of all surrounding materials and using deep-underground sites have lowered backgrounds 
and increased sensitivity. The most sensitive experiments to date use $^{76}$Ge, $^{100}$Mo,
$^{116}$Cd, $^{130}$Te, and $^{136}$Xe. For $^{76}$Ge the lifetime limit reached impressive 
values exceeding $10^{25}$years\cite{HM,IGEX}. The experimental lifetime limits have
been interpreted to yield effective neutrino mass limits typically a few eV and in $^{76}$Ge
as low as 0.3 - 1.0 eV (the spread reflects an 
estimate of the uncertainty in the nuclear matrix elements). Similar sensitivity to the neutrino
mass was reached recently also in the CUORICINO experiment with $^{130}$Te \cite{Cuoricino}.

While all these experiments reported lower limits on the $0\nu\beta\beta$-decay 
half-lives, a subset of members of the Heidelberg-Moscow collaboration 
\cite{Klapdor} reanalyzed the
data (and used additional information, e.g. the pulse-shape analysis and a different
algorithm in the peak search) and claimed to observe
a positive signal corresponding, in the latest publication, to the half-life
$T_{1/2} = 2.23_{-0.31}^{+0.44} \times 10^{25}$ years. That report has been followed by a lively discussion.  Clearly, such an extraordinary claim
with its profound implications, requires extraordinary evidence. It is fair to say that
a confirmation, both for the same $^{76}$Ge parent nucleus, and better yet also in another
nucleus with a different $Q$ value, would be required for a consensus. In any case, if that
claim is eventually confirmed
(and the mechanism of the $0\nu\beta\beta$ decay determined, i.e. the validity of eq. (\ref{eq_rate})
assured), the degenerate mass scenario will be implicated, and eventual
positive signal in the analysis of the tritium $\beta$ decay and/or observational cosmology
should be forthcoming. For the neutrinoless $\beta\beta$ decay the next generation of experiments,
which will use $\sim$ 100 kg of decaying isotopes will, among other things, test this
recent claim. 

\section{Basic nuclear physics of $\beta\beta$ decay}

\begin{figure}
\centerline{\includegraphics[width=8.0cm]{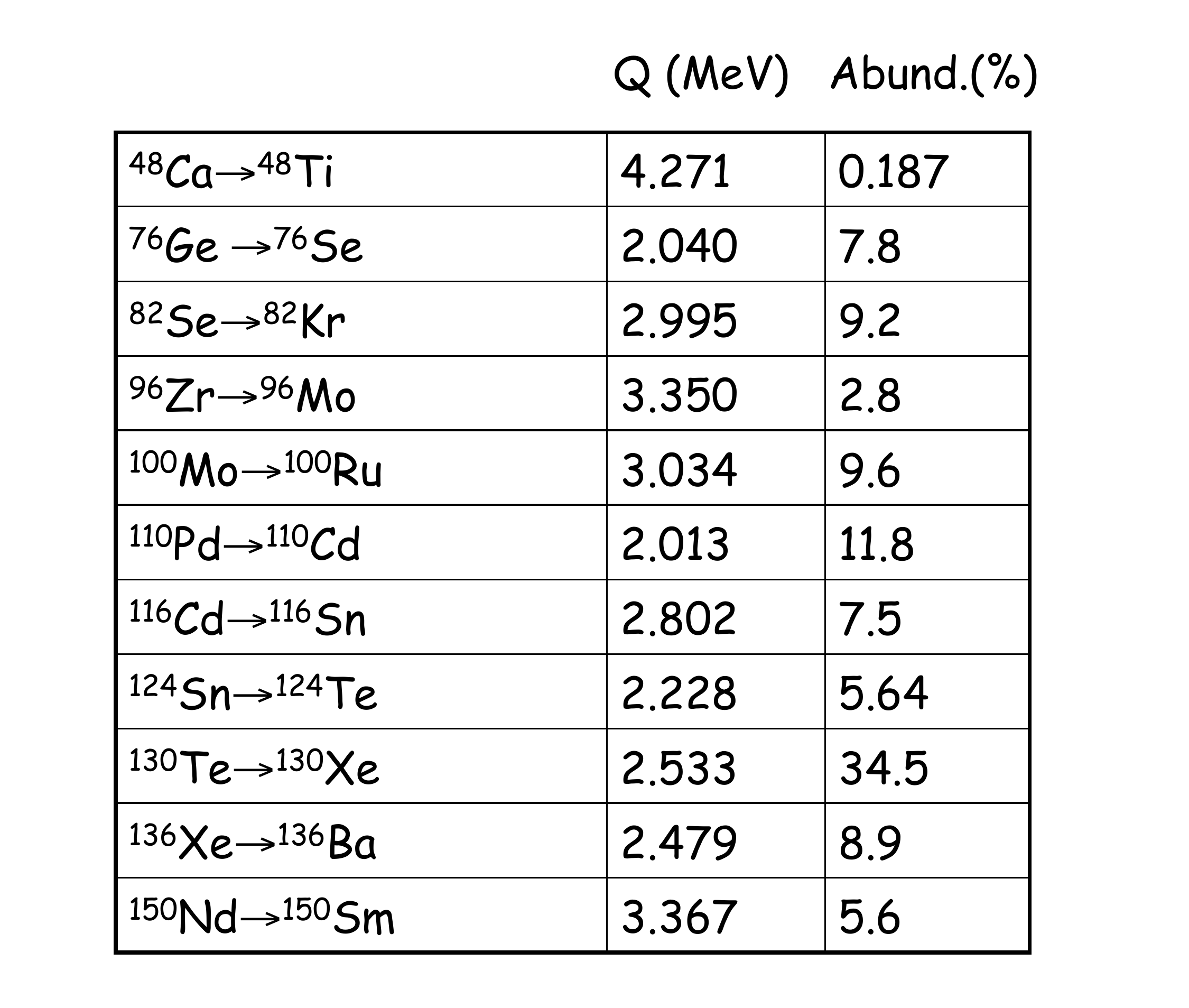}}
\caption{Candidate nuclei for $\beta\beta$ decay
with 2$e^-$ emission and  with the $Q$ value $>$2 MeV. The corresponding abundances
are also shown.}
\label{fig_tablebb}
\end{figure}

Whether a nucleus is stable or undergoes weak decay is determined by
the dependence of the atomic mass $M_A$ of the isotope $(Z,A)$ 
on the nuclear charge $Z$.
This functional dependence near its minimum can be approximated by a parabola
\begin{equation}
M_A (Z,A) ~ ~= ~ const ~ + ~ 2 b_{sym} \frac{ ( A/2 - Z )^2 }{ A^2 } ~ + ~ 
b_{Coul}\frac { Z^2 }{ A^{1/3} } ~ + ~ m_e Z ~ + ~ \delta ~,
\end{equation}
where the symmetry energy coefficient is $b_{sym} \sim 50$ MeV  
and the Coulomb energy coefficient is $b_{Coul} \sim 0.7$ MeV. 
The $m_e Z$ term represents the mass of the bound electrons; their
binding energy, for our purposes, is small enough to be neglected. The last
term $\delta$, decisive for the application to the $\beta\beta$ decay, describes nuclear 
pairing, the increase in binding as pairs of
like nucleons couple to angular momentum zero.
It is a small correction term and is given
in a crude approximation by
$\delta ~ \sim ~ \pm 12/A^{1/2}$ MeV for odd $N$ and odd $Z$, 
or even $N$ and even $Z$, respectively,
while $\delta = 0$ for odd $A$.
Thus, for odd $A$ nuclei, typically only one isotope is stable; nuclei 
with charge $Z$ smaller than the stable nucleus decay by electron emission,
while those with larger $Z$ decay by electron capture or positron
emission or by both these modes simultaneously.
For even $A$ nuclei the situation is different. Due 
to the pairing term $\delta$,
the even-even nuclei form one parabola while the odd-odd nuclei form
another one, at larger mass, as shown in Fig. \ref{fig_bb}, using $A$ = 136 as an
example. Consequently, in a typical case there exist two (or three as in Fig. \ref{fig_bb})
even-even nuclei
for a given $A$ which are stable against both electron and positron
(or EC) decays.
As these nuclei usually do not have the same mass, 
the heavier may decay into the lighter through a second order weak process
in which the nuclear charge changes by two units.
This process is double beta decay.
In Fig. \ref{fig_tablebb} all nuclei that are practical candidates for the search for 
the $0\nu\beta\beta$ decay are listed. All of them exist in nature since their lifetime is 
longer than the age of the solar system. However, with a single exception ($^{130}$Te), all
of them are relatively rare so that a large scale experiments also require a large
scale and costly (and sometimes technically difficult) isotope enrichment.

Double beta decay, therefore,  proceeds between two even-even nuclei.
All ground states of even-even nuclei have spin and parity $0^+$
and thus transitions $0^+  \rightarrow 0^+$ are expected in all cases.
Occasionally, population of the low-lying excited states of the daughter nucleus
is energetically possible, giving rise to $0^+ \rightarrow 2^+$ transitions
or to transitions to the excited $0^+$ states.

Double beta decay with the electron emission (both the $2\nu$ and $0\nu$ modes)
in which the nuclear charge increases by two units
is subject to the obvious condition
\begin{equation}
M_A (Z,A)   >  M_A (Z+2,A)~, ~~{\rm while} ~~M_A (Z,A)  <  M_A (Z + 1,A) ~,
\end{equation} 
where $M_A$ is the {\it atomic mass},
with the supplementary practical requirement that single beta decay is
absent, 
or that it is so much hindered (e.g., by the angular
momentum selection rules) that it does not compete with 
double beta decay. 

Double beta decay with the positron emission and/or
electron capture (both the $2\nu$ and $0\nu$ modes)
in which the nuclear charge decreases by two units
can proceed in three different ways:
\begin{itemize}
\item Two positron emission when $M_A (Z,A)   >  M_A (Z-2,A)  + 4m_e~ $,
\item One positron emission and one electron capture when 
$M_A (Z,A)   >  M_A (Z-2,A) + 2m_e + B_e ~ $, and
\item Two electron captures (only the two neutrino mode, see below for the comment
on the $0\nu$ mode) when $M_A (Z,A)   >  M_A (Z-2,A) + B_e(1) + B_e(2) ~ $,
\end{itemize}
where $B_e$ is the positive binding energy of the captured electron. 
As mentioned above, a complete list of all candidate nuclei (except the double electron 
capture) with the corresponding phase space factors for both $\beta\beta$-decay
modes is given in Ref.\cite{BV92}. 

Since the relevant masses are atomic masses, the processes with emission of positrons
have reduced phase space (terms with $m_e$ above). While from the point of view
of experimental observation
the positron emission seems advantageous (possibility to observe the annihilation
radiation), the reduction of phase space means that the corresponding lifetimes are quite
long. In fact, not a single such decay has been observed so far.

The two electron capture decay without neutrino emission requires a special
comment. Clearly, when the initial and final states have different energies, the process
cannot proceed since energy is not conserved. The radiative process, 
with bremsstrahlung photon emission, however,
can proceed and its rate, unlike all the other neutrinoless processes, increases
with decreasing $Q$ value \cite{SW04}. (However, the estimated decay rates are
quite small and lifetimes long). 
In the extreme case of essentially perfect degeneracy, a resonance
enhancement can occur \cite{Ber83}.

The case of resonance, though probably unrealistic, perhaps deserves some explanations.
The initial state is the atom $(Z,A)$, stable against ordinary $\beta$ decay. The final state
is the ion $(Z-2,A)$ with electron vacancies $H,H'$ and, in general, with the nucleus
in some excited state of energy $E^*$. The resonance occurs if the final energy
\begin{equation}
E = E^* + E_H + E_{H'}
\end{equation}
is close to the decay $Q$ value, 
i.e. the difference of the initial and final atomic masses,
and a perfect resonance occurs when $Q-E$ is less than
the width of the final state which is dominated by the electron hole widths $\Gamma_H, \Gamma_{H'}$.
The decay rate near resonance is given by the Breit-Wigner type formula
\begin{equation}
\frac{1}{\tau} = \frac{(\Delta M)^2}{(Q-E)^2 + \Gamma^2/4}\Gamma ~,
\end{equation}
where $\Delta M$ is the matrix element of weak interaction between the two degenerate
atomic states.

The states of definite energy, the eigenstates of the total hamiltonian, are superpositions
of the initial and final states, mixed by  $\Delta M$. But in reality, the initial state is pure,
and not a state of definite energy, since the final state decays essentially  immediately.

The mixing matrix element is \cite{Ber83}
\begin{equation}
\Delta M \sim  \frac{G_F^2 \cos^2 \theta_C}{4 \pi}  \langle m_{\beta \beta} \rangle
|\psi(0)|^2 g_A^2 M^{0\nu} ~,
\end{equation}
where $\psi(0)$ is the amplitude at the origin of the wave function of the
captured electrons and $M^{0\nu}$ is the nuclear matrix element discussed
later. Clearly, if the resonance can be approached, the decay rate would be
enhanced by the factor $4/\Gamma$
compared to $\Gamma/(E-Q)^2$, where 
the width $\Gamma$ is typically tens of eV. 
Estimates suggest that in such a case the decay lifetime for 
$\langle m_{\beta \beta} \rangle \sim$ 1 eV could be of the order of
$10^{24-25}$ years, competitive to the rate with 2$e^-$ emission.
However, chances of finding
a case of a perfect (eV size) resonance when $E$ is of order of MeV are
very unlikely. 

\section{Decay rate formulae}

\subsection{$2\nu$ decay}

Even though the $2\nu\beta\beta$-decay mode is unrelated to the fundamental
particle physics issues, it is worthwhile to discuss it in some detail. This is so
because it is the mode that is actually seen experimentally; it is also
the inevitable background for the $0\nu\beta\beta$-decay mode. In nuclear
structure theory the corresponding rate is used as a test of the adequacy of the 
corresponding nuclear models. And various auxiliary experiments can be
performed to facilitate the evaluation of the $M^{2\nu}$ nuclear matrix elements. 

The derivation of the $2\nu\beta\beta$-decay rate formula
is analogous to the treatment of ordinary beta decay. It
begins with the Fermi golden rule for second order weak decay
\begin{equation}
\frac{1}{\tau} = 2\pi \delta(E_0 - \Sigma_f E_f) 
\left[ \Sigma_{m,\beta} \frac{\langle f| H_{\beta} | m \rangle \langle m| H^{\beta} | i \rangle}
{E_i - E_m - p_{\nu} - E_e} \right]^2 ~,
\end{equation}
where the sum over $m$ includes all relevant virtual states in the intermediate odd-odd
nucleus and $\beta$ labels the different Dirac structures
of the weak interaction Hamiltonian. 
Next we take into account that the weak Hamiltonian is the product of the
nuclear and lepton currents; the corresponding formula will include the summation
over the indeces of the emitted leptons. Then the summation over the lepton polarizations
is performed, 
taking into account the indistinguishability of the final lepton pairs.
Because we are interested in the rate formula, we 
neglect terms linear in $\vec{p}_e$ and $\vec{p}_\nu$
that disappear after integration over 
angles. (The angular distribution of the electrons is of the
form (1 - $\vec{\beta}_1 \cdot \vec{\beta}_2$)
for the $0^+ \rightarrow 0^+ $ transitions, where $ \vec{\beta}_i $ is the velocity of the electron $i$).

The energy denominators are of the following form
\begin{equation}
K_m (M_m) = \frac{1}{E_m - E_i + p_{\nu1} + E_{e1}} \pm  \frac{1}{E_m - E_i + p_{\nu2} + E_{e2}} ~,
\end{equation}
where the $+$ sign belongs to $K_m$ and the $-$ sign to $M_m$ and
\begin{equation}
L_m (N_m) = \frac{1}{E_m - E_i + p_{\nu2} + E_{e1}} \pm  \frac{1}{E_m - E_i + p_{\nu1} + E_{e2}} ~.
\end{equation}
The energy denominators in the factors $K, ~ M, ~ L,~N $ 
contain contributions
of the nuclear energies $E _m - E _i$, as well as the lepton
energies $E_e + p_{\nu} $ . 
When calculating the $0^+ \rightarrow 0^+ $ transitions,
it is generally a very good approximation
to replace these lepton energies by the corresponding average value, i.e.,
$E_e  +  p_{\nu}  \sim E_0 /2$, where the $E_0 = M_i - M_f$ is the total decay 
energy including electron masses. In that case 
$M_m = N_n = 0$ and
\begin{equation}
K_m = L_m \sim \frac{1}{E_m - E_i + E_0/2} = \frac{1}{E_m - (M_i + M_f)/2} ~.
\label{eq_denom}
\end{equation}

The lepton momenta, for both electrons and neutrinos are all $q < Q$ and thus
$qR \ll 1$, where R is the nuclear radius. Hence the so-called long wavelength
approximation is valid and the rate formula  (with eq. (\ref{eq_denom})) separates
into a product of the nuclear and lepton parts, where the lepton part contains
just the phase space integral
\begin{equation}
~~~~\int_{m_e}^{E_0 - m_e} F(Z,E_{e1}) p_{e1} E_{e1} dE_{e1} \int_{m_e}^{E_0 - E_1}
F(Z,E_{e2}) p_{e2} E_{e2} dE_{e2} (E_0 - E_{e1} - E_{e2})^5/30 ~,
\end{equation}
where the integration over the neutrino momentum was already performed.
The single electron spectrum is obtained by performing
integration over $dE_{e2}$, while the 
spectrum of summed electron energies
is obtained by changing the variables 
to $E_{e1} + E_{e2}$ and  $E_{e1} - E_{e2}$ and 
performing the integration over the second variable. If an accurate
result is required, the relativistic form of the function $F(Z,E)$ must be used
and numerical evaluation is necessary.

For a qualitative and intuitive picture, one can use the 
simplified nonrelativistic Coulomb
expression, so-called Primakoff-Rosen approximation \cite{PR59} 
\begin{equation}
F(Z,E) = \frac {E}{p} \frac { 2 \pi Z \alpha }{ 1 - e^{ - 2 \pi Z \alpha }} ~.
\label{eq_PR}
\end{equation}
This approximation allows us to perform the required integrals analytically.

For example, the sum electron spectrum, which is of primary interest from 
the experimental point of view is then independent of $Z$,
\begin{equation}
\frac{dN}{dK} \sim K(T_0- K)^5  \left( 1  +  2K  +  \frac{4K^2}{3}  +  \frac{K^3}{3}
 + \frac{ K^4 }{30} \right) ~,
\end{equation}
where $K$ is the sum of the kinetic energies of both electrons, in units
of electron mass. The Coulomb effects result in shifting the maximum
of $dN/dK$ towards lower energy and in making the approach of 
$dN/dK$ to zero when $K \rightarrow T_0$ steeper.

The nuclear structure information is contained in the nuclear matrix element;
only the Gamow-Teller $\sigma \tau$ part contributes in the long wavelength
approximation
\begin{equation}
M^{2\nu} = \Sigma_m \frac { \langle 0^+_f | \vec{\sigma}_i \tau^+_i | m \rangle
\langle m | \vec{\sigma}_k \tau^+_k | 0^+_i \rangle}{E_m - (M_i + M_f)/2} ~.
\label{eq_m2nu}
\end{equation}
The individual terms in the eq. (\ref{eq_m2nu}) have a well defined meaning, in particular
for the most relevant ground state to ground state transitions. The terms
$\langle m | \vec{\sigma}_k \tau^+_k | 0^+_i \rangle$ represent the $\beta^-$
strength in the initial nucleus and can be explored in the nucleon exchange
reactions such as $(p,n)$ and $(^3He,t)$. On the other hand the terms
$\langle 0^+_f | \vec{\sigma}_i \tau^+_i | m \rangle$ represent the $\beta^+$
strength in the final nucleus and can be explored in the nucleon charge exchange
reactions such as $(n,p)$ and $(d,^2He)$. In this way one can (up to the sign)
explore the contribution of several low lying states to the $M^{2\nu}$ matrix element.

It turns out that in several nuclei the lowest (or few lowest) $1^+$ states give a 
dominant contribution to $M^{2\nu}$. (This is so-called ``single state dominance".)
In those cases the above mentioned experiments allow one to determine the $M^{2\nu}$
indirectly, independently of the actual $2\nu\beta\beta$ decay. Such data are, naturally,
a valuable testing ground of nuclear theory. On the other hand, it is not a priori
clear and easy to decide in which nuclei the sum over the $1^+$ states in eq. (\ref{eq_m2nu})
converges very fast and in which nuclei many states contribute. We will return to this
issue later.

Since, as stated in Section 3, the half-lives of many $2\nu\beta\beta$ decays was experimentally
determined. One can then extract the values of the nuclear matrix elements $M^{2\nu}$
using eq. (\ref{eq_m2nu}) . They are depicted in Fig. \ref{fig_m2nu}.
For completeness, we add a table of the most recent half-life measurements of the
$2\nu\beta\beta$ decay.

\begin{table}[htb]
\caption{ Summary of experimentally measured 
$2\nu\beta\beta$ half-lives and matrix elements, mostly from the
NEMO experiment \cite{NEMO}
($^{136}$Xe is an important exception where a limit is quoted).}

\begin{center}
\begin{tabular}{lll}  \hline\hline
Isotope                & T$_{1/2}^{2\nu}$ (y)           & $M_{GT}^{2\nu}$ (MeV$^{-1}$) \\ \hline
$^{48}$Ca              & $(3.9 \pm 0.7 \pm 0.6)\times 10^{19}$   & $0.05 \pm 0.01$      \\
$^{76}$Ge              & $(1.7 \pm 0.2)\times 10^{21}$               & $0.13 \pm 0.01$   \\
$^{82}$Se              & $(9.6 \pm 0.3 \pm 1.0) \times 10^{19}$      & $0.10 \pm 0.01$    \\
$^{96}$Zr  & $(2.0 \pm 0.3 \pm 0.2)\times 10^{19}$                   & $0.12 \pm 0.02$    \\
$^{100}$Mo             & $(7.11 \pm 0.02 \pm 0.54)\times 10^{18}$   & $0.23 \pm 0.01$   \\
$^{116}$Cd             & $(2.8 \pm 0.1 \pm 0.3)\times 10^{19}$         &  $0.13 \pm 0.01$  \\
$^{128}$Te$^{(1)}$     & $(2.0 \pm 0.1)\times 10^{24}$                 & $0.05 \pm 0.005$ \\
$^{130}$Te     & $(7.6 \pm 1.5 \pm 0.8)\times 10^{20}$         & $0.032 \pm 0.003$\\
$^{136}$Xe             & $>1.0 \times 10^{22}$ (90\% CL)                & $<$0.01 \\
$^{150}$Nd     & ($9.2 \pm 0.25 \pm 0.73) \times$ 10$^{18}$               & $0.06 \pm 0.003$ \\
$^{238}$U$^{(2)}$      & $(2.0\pm 0.6)\times 10^{21}$        &  $0.05 \pm 0.01$ \\ \hline
\end{tabular}
\end{center}

$^{(1)}$deduced from the geochemically determined half-life ratio $^{128}$Te/$^{130}$Te \\
$^{(2)}$radiochemical result for all decay modes
\end{table}

\subsection{$0\nu$ rate} We shall now indicate the
derivation of the electron spectra and decay rates 
associated with the nonvanishing value of $m_{\nu}$. The decay rate is of the
general form
\begin{equation}
\omega_{0\nu}  =  2 \pi \Sigma_{spin} | R_{0\nu} |^2 
\delta (E_{e1}  +  E_{e2}  +  E_f - M_i ) d^3 p_{e1} d^3 p_{e2} ~,
\end{equation}
where $E_f $ is the energy of the final nucleus and $R_{0\nu}$ is the
transition amplitude including both the lepton and nuclear parts.

The lepton part of the amplitude is written as
a product of two left-handed currents
\begin{equation}
\bar{e} (x) \gamma_{\rho} \frac{1}{2} (1 - \gamma_5 )  \nu_j (x)
\bar{e} (y) \gamma_{\sigma}  \frac{1}{2} (1 - \gamma_5 )  \nu_ k (y) ~,
\end{equation}
where, $\nu_j,  \nu_k$ represent neutrino mass eigenstates $j$ and $k$, and
there is a contraction over the two neutrino operators. 
The contraction above is allowed only if the neutrinos are Majorana particles.

After substitution for the neutrino propagator
and integration over the virtual neutrino momentum, 
the lepton amplitude acquires the form
\begin{equation}
-i \delta_{jk} \int \frac{d^4 q}{ (2 \pi )^4 }
\frac{e^{ -iq(x-y) }}{ q^2 - m_j^2 }
\bar{e} (x) \gamma_{\rho} \frac{1}{2} (1 - \gamma_5 )  
( q^{\mu} \gamma_{\mu} + m_j ) \frac{1}{2} (1 - \gamma_5 )  
\gamma_{\sigma}  e^C (y) ~.
\end{equation}

From the commutation properties of the gamma matrices it follows that
\begin{equation}
\frac{1}{2} (1- \gamma_5 )  
( q^{\mu} \gamma_{\mu} + m_j ) \frac{1}{2} (1 - \gamma_5 )  
 =  m_j \frac{1}{2} (1 - \gamma_5 ) ~.
\end{equation}
Thus the decay amplitude for purely left-handed lepton currents is proportional
to the neutrino Majorana mass $m_j$.

Integration over the virtual neutrino energy leads to the replacement 
of the propagator
$(q^2 - m_j^2 )^{-1}$ by the 
residue $\pi/ \omega_ j$  with $ \omega_ j = ( \vec{q}^{~2} + m_j^2 )^{1/2}$. For the 
remaining integration
over the space part $d \vec{q}$ 
we have to consider, besides this denominator $\omega_j$,
the energy denominators of the second order
perturbation expression. Denoting
\begin{equation}
A_n = E _n -  E_i  +  E _e ~,
\end{equation}
we find that integration over $d \vec{q}$ leads to 
an expression representing the effect of the neutrino propagation
between the two nucleons. This expression has the form of a "neutrino
potential" and appears in the corresponding nuclear matrix elements,
introducing dependence of the transition operator on the coordinates
of the two nucleons, as well as a weak dependence on the excitation
energy $E _n  -  E_i$ of the virtual state in the odd-odd intermediate nucleus.

There are several neutrino potentials as explained in the next subsection.
The main one is of the form
\begin{equation}
H(r,E_m )  =  \frac{R}{ 2 \pi^2 g_A^2}  \int  \frac{ d \vec{q}}{ \omega} 
\frac {1}{ \omega + A_m }  e^{ i \vec{q} \cdot \vec{r}}  = 
\frac{2R}{ \pi r g_A^2}  \int_0^ {\infty} dq \frac{ q \sin (qr) }{ \omega( \omega + A_m ) } ~.
\label{eq_nupot}
\end{equation}
Here we added the nuclear radius $R = 1.2A^{1/3}$ fm as an auxiliary factor
so that $H$ becomes dimensionless. A corresponding $1/R^2$ compensates
for this auxiliary quantity in the phase space formula. Note that a consistency
between the definitions of the nuclear radius $R$ is required. )
The first factor $\omega$ in the denominator of eq. (\ref{eq_nupot}) is the residue, while
the factor $\omega + A_m$ is the energy denominator of perturbation theory.
To obtain the final result one has to treat properly the antisymmetry
between the identical outgoing electrons (see \cite{Doi85}).

The momentum of the virtual neutrino is 
determined by the uncertainty relation
$q \sim 1/r$, where $r \le R$ is a typical spacing between two nucleons.
We will show later that in fact the relevant values of $r$ are only $r \le$ 2-3 fm,
so that the momentum transfer $q \sim$ 100-200 MeV. For the light neutrinos
the neutrino mass $m_j$ can then be safely neglected in the potential $H(r)$.
(Obviously, for heavy neutrinos, with masses $M_j \gg$ 1 GeV a different procedure
is necessary.) Also, given the large value of $q$ the dependence on the 
difference of nuclear energies $E_m - E_i$ is expected to be rather weak
and the summation of the intermediate states can be performed in closure
for convenience. (We will discuss the validity of that approximation later.)

Altogether, we can rewrite the expression for the neutrino potential as
\begin{equation}
H(r) = \frac{R}{r} \Phi(\omega r) ~,
\label{eq:numotp}
\end{equation}  
where $ \Phi(\omega r) \le 1$ is a relatively slowly varying function of $r$.
From that it follows that a typical value of $H(r)$ is larger than unity, but less than 5-10.
In the next subsection we present the exact expressions for the neutrino potentials
and the $0\nu\beta\beta$ transition operator for the most interesting case of
small but finite neutrino Majorana masses $m_j$.

For now we use the relation (\ref{eq_rate}) and evaluate the phase space function $G^{0\nu}$
\begin{equation}
~~~~~G^{0\nu}(Q,Z)  \sim \int F(Z,E_{e1})F(Z,E_{e2}) p_{e1}p_{e2}E_{e1}E_{e2} 
\delta(E_0 - E_{e1} - E_{e2}) dE_{e1}dE_{e2} ~.
\end{equation}
The constant factor in front of this expression is 
\begin{equation}
(G_F \cos \theta_C g_A )^4 \left( \frac{\hbar c}{R} \right)^2 
\frac{1}{\hbar} \frac{1}{\ln(2) 32 \pi^5} ~,
\end{equation}
and the values of $1/G^{0\nu}$ listed in Ref. \cite{BV92} are in years provided that the
neutrino masses are in eV.

Again, in the Primakoff-Rosen approximation, eq. (\ref{eq_PR}), $G^{0\nu}$ is independent 
of $Z$ and (only the $E_0$ dependence is shown)
\begin{equation}
G^{0\nu}_{PR} \sim  \left( \frac{E_0^5}{30} - \frac{2E_0^2}{3} + E_0 - \frac{2}{5} \right) ~,
\end{equation}
where $E_0$ is expressed in units of electron mass. Each of the two electrons observed 
separately will have energy spectrum determined by the phase space integral. In the
Primakoff-Rosen approximation its shape is
\begin{equation}
\frac{dN}{dT_e} \sim (T_e + 1)^2(T_0 - T_e + 1)^2 ~,
\end{equation}
where again the kinetic energies are in the units of electron mass.

It is of interest to contrast the $0\nu$ and $2\nu$ decay modes
from the point of view of the phase space integrals. The $0\nu$
mode has the advantage of the two-lepton final state, with the
characteristic $E_0^5$ dependence  compared
to the four-lepton final state with $E_0^{11}$ dependence
for the $2 \nu$ mode.
In addition, the large average momentum of the virtual neutrino,
compared with the typical nuclear excitation energy also makes the
$0\nu$ decay faster. Thus, if $\langle m_{\beta\beta} \rangle$ were to be of the order
of $m_e$, the $0\nu$ decay would be $\sim  10^5$ times
faster than the $2\nu$ decay. It is this phase space advantage
which makes the $0\nu\beta\beta$ decay a sensitive 
probe for Majorana neutrino mass.

Finally, let us remark that it is possible that, in
addition to the two electrons,
a light boson, the so called majoron,
is emitted. This transition would have three-body phase space,
giving rise to a continuous spectrum peaked
at approximately three quarters of the decay energy $T_0$.
We shall not discuss this topic here, but refer to Doi et al. \cite{Doi85} for a discussion 
of this issue.

Also, since we concentrate on the mechanism involving the exchange of light Majorana
neutrinos, we will not discuss in any detail the case of hypothetical right-handed currents.

\subsection{Exact expressions for the transition operator}

After the qualitative discussion in the preceding subsection, here we derive the
exact expressions for the transition operator.

The hadronic current, expressed in terms of nucleon fields $\Psi$, is 
\begin{equation}
J^{\rho \dagger}
=  \overline{\Psi} \tau^+ \left[ g_V(q^2) \gamma^\rho
+ i g_M (q^2) \frac{\sigma^{\rho \nu}}{2 m_p} q_\nu 
 - g_A(q^2) \gamma^\rho\gamma_5 - g_P(q^2) q^\rho \gamma_5 \right] \Psi~,
\label{a1}
\end{equation}
where $m_p$ is the nucleon mass and $q^\mu$ is the momentum transfer,
i.e.\ the momentum of the virtual neutrino.
Since in the $0\nu\beta\beta$ decay $\vec{q}^{~2} \gg q_0^2$ we take
$q^2 \simeq - \vec{q}^{~2}$.

For the vector and axial vector form factors we adopt the usual dipole approximation
\begin{equation}
g_V({\vec q}^{~2}) = {g_V}/{(1+{\vec q}^{~2}/{M_V^2})^2},~~~
g_A({\vec q}^{~2}) = {g_A}/{(1+{\vec q}^{~2}/{M_A^2})^2}~,
\end{equation}
with $g_V$ = 1, $g_A$ = 1.254, $M_V$ = 850 MeV, and $M_A$ = 1086 MeV.
These form factors are a consequence of the composite nature of nucleons.
With high momentum transfer the ``elastic" transitions, in which a nucleon remains
nucleon and no other hadrons are produced, is reduced.

We use the usual form for the weak magnetism,
and the Goldberger-Treiman relation for the induced pseudoscalar term: 
\begin{equation}
g_M({\vec q}^{~2})= (\mu_p-\mu_n) g_V({\vec q}^{~2})~,~~~
g_P({\vec q}^{~2}) = {2 m_p g_A({\vec q}^{~2})}/({{\vec q}^{~2} + m^2_\pi}) ~.
\end{equation}

Reducing the nucleon
current to the non-relativistic form yields (see Ref.\cite{erwe}):
\begin{equation}
J^{\rho\dagger}(\vec{x})=\sum_{n=1}^A \tau^+_n [g^{\rho 0} J^0({\vec q}^{~2}) +
\sum_k g^{\rho k}  J^k_n({\vec q}^{~2})] \delta(\vec{x}-{\vec{r}}_n),
\label{a3}
\end{equation}
where $J^0({\vec q}^{~2}) = g_V(q^2)$ and
\begin{equation}
{\vec J}_n({\vec q}^{~2}) =  g_M({\vec q}^{~2})
i \frac{{\vec{\sigma}}_n \times \vec{q}}{2 m_p}
+ g_A({\vec q}^{~2})\vec{\sigma}
-g_P({\vec q}^{~2})\frac{\vec{q}~ {\vec{\sigma}}_n \cdot \vec{q}}{2 m_p},
\label{a4}
\end{equation}
${\vec r}_n$ is the coordinate of the $n$th nucleon,  $k=1,2,3$, and $g^{\rho,\alpha}$
is the metric tensor.

This allows us to derive the effective two-body transition operator in the momentum
representation:
\begin{equation}
~~~~\Omega = \tau^+\tau^+ \frac{(- h_F + h_{GT} \sigma_{12} - h_T S_{12})}{q(q+E_m - (M_i + M_f)/2)}
~,~
\sigma_{12} = \vec{\sigma}_1 \cdot \vec{\sigma}_2
~,~
S_{12} = 3 \vec{\sigma}_1 \cdot \hat{q}  \vec{\sigma}_2 \cdot \hat{q} - \sigma_{12} ~.
\label{eq_oper}
\end{equation}
Here $h_F = g_V^2$ and
\begin{equation}
h_{GT} = g_A^2 \left[ 1 - \frac{2}{3} \frac{{\vec q}^{~2}}{{\vec q}^{~2} + m_{\pi}^2}
+ \frac{1}{3} \left( \frac{{\vec q}^{~2}}{{\vec q}^{~2} + m_{\pi}^2} \right)^2 \right] ,~
h_T = g_A^2 \left[ \frac{2}{3} \frac{{\vec q}^{~2}}{{\vec q}^{~2} + m_{\pi}^2} -
 \frac{1}{3} \left( \frac{{\vec q}^{~2}}{{\vec q}^{~2} + m_{\pi}^2} \right)^2 \right] .
\end{equation}
For simplicity the $q^2$ dependence of the form factors $g_V$ and $g_A$ is not indicated
and the terms containing $1/m_p^2$ are omitted. The parts containing
${\vec q}^{~2} + m_{\pi}^2$ come from the induced pseudoscalar form factor $g_P$ for which the
partially conserved axial-vector hypothesis (PCAC) is used.

In order to calculate the nuclear matrix element in the coordinate space, one has to evaluate
first the ``neutrino potentials" that, at least in principle, depend explicitly on the energy
$E_m$ of the virtual intermediate state
\begin{equation}
H_K (r_{12}) = \frac{2}{\pi g_A^2} R \int_0^{\infty} f_K (qr_{12})
\frac{ h_K(q^2) q dq}{q + E_m - (M_i + M-f)/2} ~,
\label{eq_poten}
\end{equation}
where $K = F,GT,T$ and $f_{F,GT} = j_0(qr_{12})$ and $f_T = -j_2(qr_{12})$,
and $r_{12}$ is the internucleon distance.

With these ``neutrino potentials" and the spin dependence given in eq. (\ref{eq_oper})
where, naturally, in the tensor operator the unit vector $\hat{q}$ is replaced by $\hat{r}_{12}$,
the nuclear matrix element is now written (see \cite{us06})
\begin{equation}
M'^{0\nu} = \left( \frac{g_A}{1.25}\right)^2 
\langle f | \frac{M^{0\nu}}{g_A^2} + M^{0\nu}_{GT} + M^{0\nu}_T | i \rangle ~,
\label{eq_me0nu}
\end{equation}
where $| f \rangle$ and $| i \rangle$ are the ground state wave functions of the final and, 
respectively, initial nuclei. The somewhat awkward definition 
of $M'^{0\nu}$ is used so that, if needed,
one can use an ``effective" value of the axial coupling constant $g_A$ but still use
the tabulated values of the phase space integral $G^{0\nu}$ that were evaluated with
$g_A = 1,25$.

\section{Nuclear structure issues}

We will discuss now the procedures to evaluate the ground state wave functions 
of the initial and final nuclei
$| f \rangle$ and $| i \rangle$ and the nuclear matrix element, eq. (\ref{eq_me0nu}).
There are two complementary methods to accomplish this task, the nuclear shell
model (NSM) and the quasiparticle random phase approximation (QRPA). Since
my own work deals with the QRPA method and the NSM will be covered
by Prof. Poves in his seminar, 
the NSM will be described only superficially and reference to its
results will be made mainly in comparison with the QRPA.

\begin{figure}
\centerline{\includegraphics[width=9.0cm]{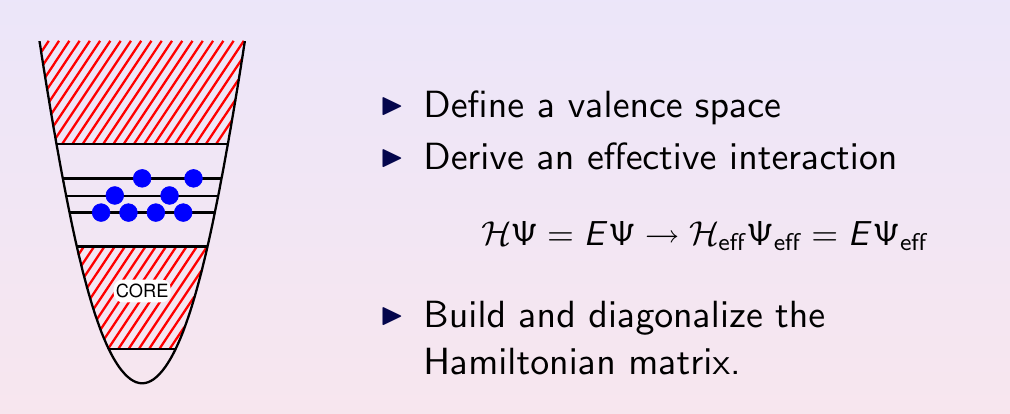}}
\caption{Schematic illustration of the basic procedures in the nuclear shell model.}
\label{fig_sm1}
\end{figure}

\subsection{Nuclear shell model}

The basic idea is schematically indicated in Fig. \ref{fig_sm1}. The procedure
should describe, at the same time, the energies and transition probabilities involving
the low-lying nuclear states as well as the $\beta$ and $\beta\beta$ decay nuclear
matrix elements. The hamiltonian matrix is diagonalized in one of two possible bases, either
\begin{equation}
{\rm m-scheme}  ~~| \Phi_{\alpha} \rangle = \Pi_{nljm\tau} a^+_i | 0 \rangle =
a^+_{i1} \cdots a^+_{iA} | 0 \rangle ~,
D \sim \left( \begin{array}{c}
d_{\pi} \\ p
\end{array}
\right)
\cdot
 \left( \begin{array}{c}
d_{\nu} \\ n
\end{array}
\right) ~,
\end{equation}
which is simple, based on the Slater determinants,
and the corresponding hamiltonian matrix is sparse, but it has a huge dimension
($d_{\pi}, d_{\nu}$ are the dimensions of the proton and neutron included subshells
and $p,n$ are the numbers of valence protons and neutrons). The other possible basis
uses states coupled to a given total angular momentum $J$ and isospin $T$. The dimensions
are smaller in that case but the evaluation of the hamiltonian matrix element is more complicated
and the corresponding hamiltonian matrix has fewer zero entries.
Due to the high dimensionality of the problem one cannot include in the valence space
too many single-particle orbits. Even the most advanced evaluations include just one
oscillator shell, in most $\beta\beta$ decay candidate nuclei the valence space usually
omits important spin-orbit partners. (For $^{76}$Ge for example, the valence space consists
of $p_{1/2}, p_{3/2}, f_{5/2}$ and $g_{9/2}$ orbits for protons and neutrons, while
omitting the essentially occupied $f_{7/2}$ and empty $g_{7/2}$ spin-orbit partners.)

In order to evaluate $M^{0\nu}$ the closure approximation is used and thus one evaluates
\begin{equation}
\langle f || O^K || i \rangle>  ~~{\rm with} ~~ O^K = \Sigma_{ijkl} W^{\lambda, K}_{ijkl}
\left[ (a^+_i a^+_j)^{\lambda}  (\tilde{a}_k \tilde{a}_l)^{\lambda} \right]^0 ~,
\end{equation}
where the creation operators create two protons and the annihilation operators annihilate
two neutrons. In this way the problem is reduced to a standard nuclear structure problem 
where the many-body problem is reduced to the
of evaluation of two-body transition densities. The matrix elements
$W^{\lambda, K}_{ijkl}$ involve only the mean field (usually harmonic oscillator)
one-body wave functions, and the transition operator defined above.

The hamiltonian $H_{eff}$ is an effective operator. It is not entirely based on first principle
reduction of the free nucleon-nucleon potential. Instead it relies on empirical data,
in particular on the energies of states in semi-magic nuclei. At the same time
effective operators should be used, in principle. Again, there is no well defined
and well tested procedure to obtain their form. Instead, one uses empirical data
and effective couplings (effective charges for the electromagnetic transitions,
effective $g_A$ values for the weak transitions).

\begin{figure}
\centerline{\includegraphics[width=11.0cm]{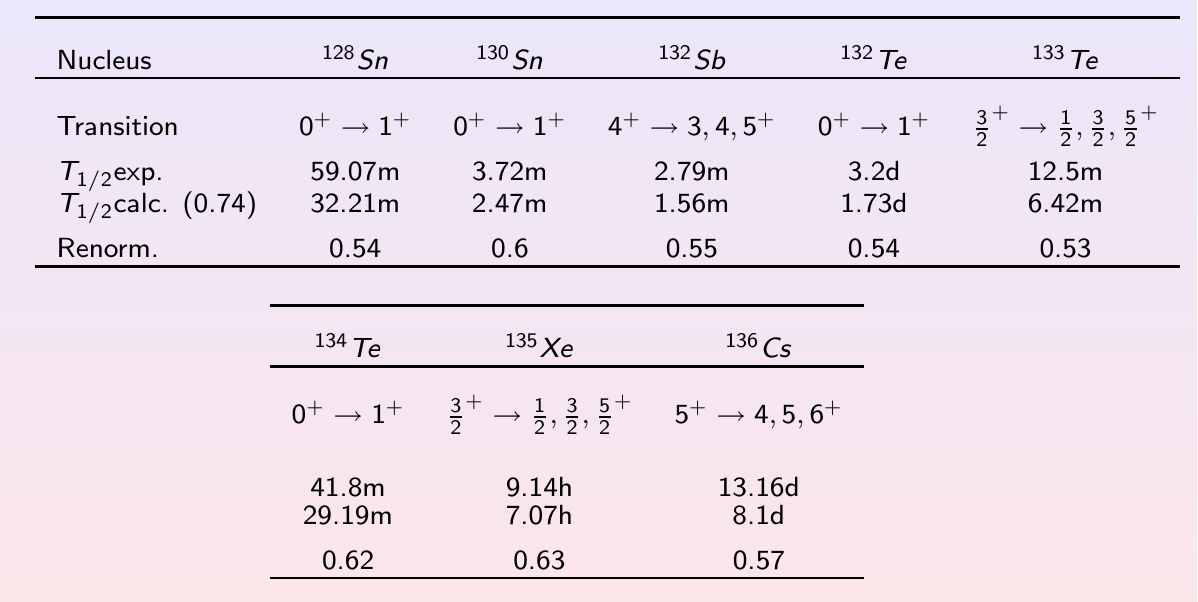}}
\caption{Experimental and calculated $\beta$-decay half-lives for several nuclei
with $A$ = 128-136 (adopted from ref. \cite{Now04}).}
\label{fig_sm2}
\end{figure}

\begin{figure}
\centerline{\includegraphics[width=11.0cm]{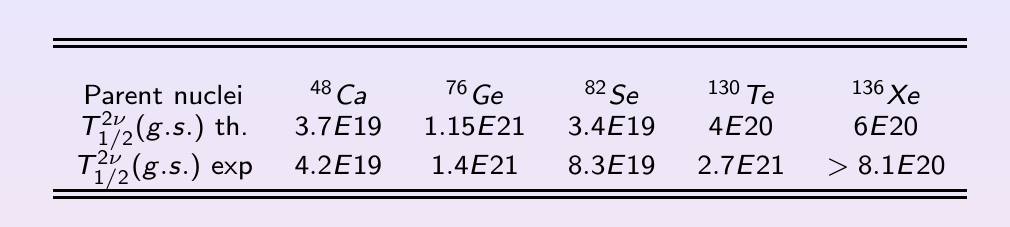}}
\caption{Experimental and calculated $2\nu\beta\beta$-decay half-lives for several nuclei
(adopted from ref. \cite{Now04}).}
\label{fig_sm2nu}
\end{figure}

How this procedure works is illustrated in Fig. \ref{fig_sm2} where the 
experimental $\beta$ decay
half-lives for a number of nuclei are compared with calculation and a quenching factor,
i.e.  the reduction factor of $g_A^2$ of $\sim$0.57 is obtained. And in Fig. \ref{fig_sm2nu}
we show the same comparison for the evaluation of the $2\nu\beta\beta$ half-lives.
(Note that this table is a bit obsolete as far as the experimental data are concerned.
In the meantime, the $T_{1/2}$ for $^{130}$Te was directly determined as 7.6$\times 10^{20}$
years, and the limit for $T_{1/2}$ in $^{136}$Xe is longer than indicated, 
$T_{1/2} > 1.0\times10^{22}$ years, see the Table in the preceding section.)

Various aspects of the application of the NSM to the $\beta\beta$ decay were reported
in a number of publications \cite{sm95,sm96,sm99,sm05,sm08a,sm08b,sm08c}. We will
return to these results later when we compare the NSM and QRPA methods.

\begin{figure}
\centerline{\includegraphics[width=7.0cm]{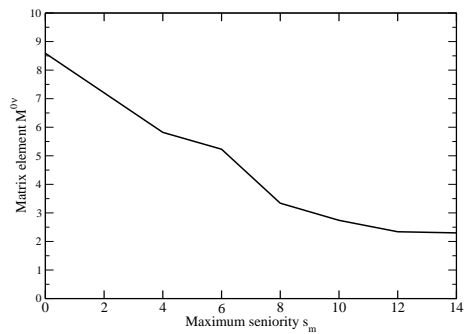}}
\caption{The full matrix element $M^{0\nu}$ for the $^{76}$Ge $\rightarrow ^{76}$Se
transition evaluated with the indicated maximum seniority $s_m$ included 
(adopted from ref. \cite{sm08c}).}
\label{fig_sen}
\end{figure}

Here we wish to stress one general result of NSM evaluation of $M^{0\nu}$.
In the NSM one can classify states by their ``seniority", i.e. by the number of valence
nucleons that  do not form Cooper-like pair and are therefore not coupled to $I^{\pi} = 0^+$.
The dimensionality of the problem increases fast with seniority and thus it is
of interest to see whether a truncation in seniority is possible and in general how the
magnitude of  $M^{0\nu}$ behaves as a function of the included seniority.
This is illustrated in Fig. \ref{fig_sen} from Ref. \cite{sm08c} for the case
of  the $^{76}$Ge decay. Note that the result does not saturate until $s_m = 12$
which is nearly the maximum seniority possible.
That behavior is observed in other cases as well
(the saturation in $s_m$ is faster in cases when one of the involved nuclei
is semimagic) , and is relevant when the
comparison to QRPA is made.

\begin{figure}
\centerline{\includegraphics[width=7.0cm]{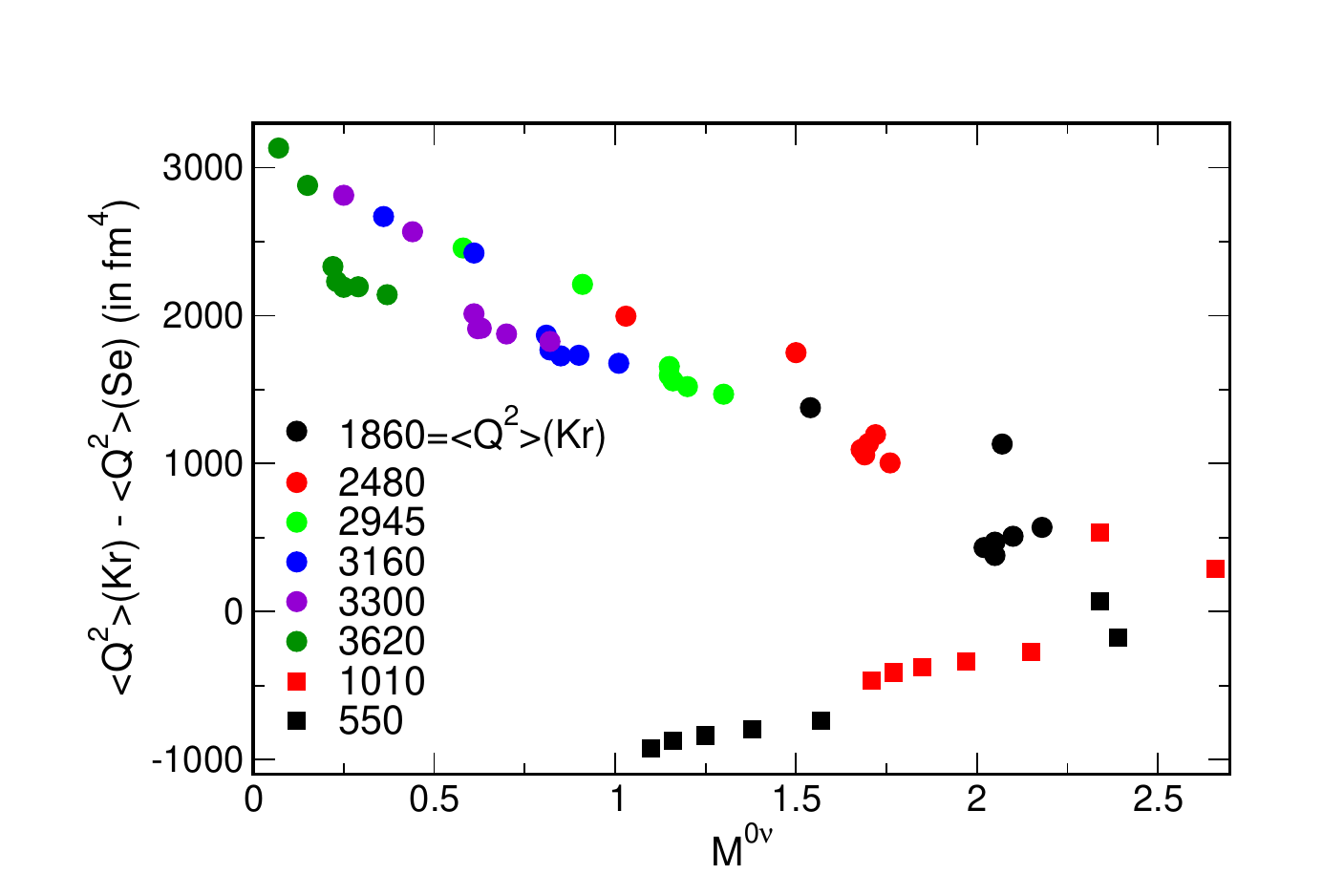}}
\caption{The matrix element $M^{0\nu}$ for the $^{82}$Se $\rightarrow ^{82}$Kr
transition for a large number of added quadrupole-quadrupole interaction strengths
that induces increased deformation in $^{82}$Se. The magnitude of $M^{0\nu}$ is ploted
against the difference in the squared quadrupole moments of the invloved nuclei. 
(adopted from ref. \cite{sm08c}).}
\label{fig_deform}
\end{figure}

NSM can also successfully describe nuclear deformation, unlike the usual application
of QRPA. Many of the considered $\beta\beta$ decay candidate nuclei are spherical
or nearly so. However, one seemingly attractive candidate nucleus, $^{150}$Nd,   is strongly
deformed. Moreover, the final nucleus, $^{150}$Sm, is considerably less deformed than  $^{150}$Nd.
Unfortunately, the NSM is unable to describe this system, the dimension is just too large.
However, to see qualitatively what the effect of deformation is, or the difference in deformation,
might be, in Ref. \cite{sm08c} the transition $^{82}$Se $\rightarrow ^{82}$Kr was considered
and the initial nucleus $^{82}$Se was artificially deformed by including in the hamiltonian
an additional (unrealistic) quadrupole-quadrupole interaction of varying strength. The result of that
exercise are shown in Fig. \ref{fig_deform}. One can see that as the difference in deformation
increases that magnitude of $M^{0\nu}$ decreases considerably. 

What are, then, the outstanding issues for NSM vis-a-vis the evaluation of the $M^{0\nu}$
matrix elements? The most important one, in my opinion, is the limited size of the valence
space. It is not clear how large (or small) the effect of the additional orbits might be. They
cannot be, at present, included directly. Perhaps a perturbative method of including them 
can be developed, either in the shell model codes or in the definition of the effective
operator. The other issue, that perhaps could be overcome, is the fact that the $H_{eff}$
has not been determined for nuclei with $A = 90-110$
and thus the $M^{0\nu}$ are not available. Among them are important $\beta\beta$
candidate nuclei $^{100}$Mo and $^{96}$Zr (a definitive 
calculations for $^{116}$Cd were not reported as yet either).
 
\subsection{QRPA basics}

The QRPA method was first applied to the charge changing modes by Halbleib and Sorensen
\cite{HS67} long time ago
and generalized to include the particle-particle interaction by Cha \cite{Cha83}.
The use of quasiparticles in QRPA
makes it possible to include the pairing correlations in the nuclear
ground states in a simple fashion. With pairing included, the Fermi levels for protons and neutrons
not only become diffuse, but the number of nucleons in each subshell will not have a sharp
value, instead only a mean occupancy of each subshell will have a well determined value.

To include the pairing we perform first the Bogoliubov transformation relating the
particle creation and annihilation operators $a^{\dagger}_{jm}, \tilde{a}_{jm}$
with the quasiparticle creation and annihilation operators  $c^{\dagger}_{jm}, \tilde{c}_{jm}$,
\begin{equation}
\left(\begin{array}{l}
a^{\dagger}_{jm} \\  \tilde{a}_{jm}
\end{array} \right) 
= \left( \begin{array}{l l}
u_j c^{\dagger}_{jm} & + ~~ v_j  \tilde{c}_{jm} \\
-v_j c^{\dagger}_{jm} & + ~~ u_j  \tilde{c}_{jm}
\end{array} \right) ~,
\end{equation}
where  $\tilde{a}_{jm} = (-1)^{j - m} a_{j -m}$ and $u_j^2 + v_j^2 = 1$.

The amplitudes $u_j, v_j$ are determined in the standard way by solving the BCS
gap equations, separately for protons and neutrons,
\begin{equation}
~~~~\Delta_a = (2 j_a + 1)^{-1/2} \Sigma_c (2 j_c + 1)^{1/2} u_c v_c 
\langle j_a^2; 0^+ || V || j_c^2; 0^+ \rangle ~, \hspace{0.5cm} N = \Sigma_c (2 j_c + 1) v_c^2 ~.
\end{equation}   
Here $N$ is the number of neutrons or protons and
the gaps $\Delta$ are empirical quantities deduced from the usual mass differences of the
corresponding even-even and odd-A nuclei. We renormalize the strength of the pairing
interaction (the coupling constant in $\langle j_a^2; 0^+ || V || j_c^2; 0^+ \rangle$) 
slightly such that
the empirical gaps $\Delta$ are correctly reproduced.

Our goal is to evaluate the transition amplitudes associated
with charge changing one-body operator $T^{JM}$
connecting the $0^+$  BCS vacuum $| O \rangle $ of the quasiparticles $c$ and $c^{\dagger}$
in the even-even nucleus with any of the $J^{\pi}$ excited states in the neighboring
odd-odd nuclei. In the spirit of RPA we describe such states 
as harmonic oscillations 
above the BCS vacuum. Thus
\begin{equation}
| J^{\pi} M;m \rangle = \Sigma_{pn} 
\left[ X^m_{pn,J\pi} A^{\dagger} (pn;J^{\pi}M) + Y^m_{pn,J\pi} \tilde{A} (pn;J^{\pi}M) \right] 
| 0^+_{QRPA} \rangle ~,
\end{equation}
where
\begin{eqnarray}
A^{\dagger} (pn;J^{\pi}M) &  =  & \Sigma_{m_p,m_n} \langle j_p m_p, j_n m_n | J M \rangle 
c^{\dagger}_{j_p, m_p} c^{\dagger}_{j_n m_n} 
\\ \nonumber
\tilde{A} (pn;J^{\pi}M) & = & (-1)^{J-M} A (pn;J^{\pi} - M) ~,
\end{eqnarray}
and $\ 0^+_{QRPA} \rangle$ is the ``phonon vacuum" a correlated state that has the zero-point
motion corresponding to the given $J^{\pi}$ built into it. It contains, in addition to the BCS vacuum,
components with 4, 8, etc. quasiparticles.

The so-called forward- and backward-going amplitudes $X$ and $Y$ as well as the corresponding
energy eigenvalues $\omega_m$ are determined by solving the QRPA eigenvalue equations
for each $J^{\pi}$
\begin{equation}
\left( \begin{array}{cc}
A & B \\ -B & - A 
\end{array} \right)
\left( \begin{array}{c}
X \\ Y \end{array} \right)
= \omega \left(
 \begin{array}{c}
X \\ Y \end{array} \right) ~.
\label{eq_rpa}
\end{equation}

It is easy to see that  eq. (\ref{eq_rpa}) is, in fact, an eigenvalue equation for $\omega^2$
of the type $(A^2 - B^2)X = \omega^2 X$. Hence the physical solutions are such that
$\omega^2$ is positive, and we can choose $\omega$ to be positive as well. On the
other hand, there could be unphysical solutions with $\omega^2 < 0$ and hence imaginary
energies. By varying the coupling constants in the matrices $A$ and $B$ we might
trace the development of the solutions from the physical ones with $\omega > 0$
to the situation where $\omega = 0$. That point signals the onset of region where 
the original RPA (or QRPA) is no longer applicable, because the ground state
must be rearranged. Examples are the transition from a spherical to deform shape, or transition
from pairing of neutrons with neutrons (and protons with protons) to a pairing involving
neutron-proton pairs. We will see that real nuclei are rather close to that latter situation,
and we need to worry about the applicability of the method in such situations.   
   
To obtain the matrices $A$ and $B$ one needs first to rewrite the hamiltonian in the
quasiparticle representation. Then
\begin{eqnarray}
A^J_{pn,p'n'}  & = & \langle O | (c^{\dagger}_p c^{\dagger}_n )^{(JM) ^{\dagger}} \hat{H}
(c^{\dagger}_{p'} c^{\dagger}_{n'} )^{(JM)} | O \rangle \\
\nonumber
 & = &  \delta_{pn,p'n'} (E_p + E_n) \\
 \nonumber
& & + (u_p v_n u_{p'} v_{n'} + v_p u_n v_{p'} u_{n'} ) g_{ph} \langle p n^{-1} , J | V | p' n'^{-1}, J \rangle
\\ \nonumber
& & + (u_p u_n u_{p'} u_{n'} + v_p v_n v_{p'} v_{n'} ) g_{pp} \langle p n , J | V | p' n', J \rangle ~,
\label{eq_Adef}
\end{eqnarray}
and
\begin{eqnarray}
B^J_{pn,p'n'}  & = & \langle O |  \hat{H} (c^{\dagger}_p c^{\dagger}_n )^{(J - M)} (-1)^M 
(c^{\dagger}_{p'} c^{\dagger}_{n'} )^{(JM)} | O \rangle \\
 \nonumber
& & + (-1)^J (u_p v_n v_{p'} u_{n'} + v_p u_n u_{p'} v_{n'} ) g_{ph} \langle p n^{-1} , J | V | p' n'^{-1}, J \rangle
\\ \nonumber
& & - (-1)^J (u_p u_n v_{p'} v_{n'} + v_p v_n u_{p'} u_{n'} ) g_{pp} \langle p n , J | V | p' n', J \rangle ~.
\label{eq_Bdef}
\end{eqnarray}

Here $E_p, E_n$ are the quasiparticle energies. 
The particle-hole and particle-particle interaction matrix elements are related to each other by the
Pandya trasformation
\begin{equation}
 \langle p n^{-1} , J | V | p' n'^{-1}, J \rangle = - \Sigma_{J'} (2 J' + 1)
 \left\{ \begin{array}{ccc}
 p & n & J \\  p' & n' & J' 
 \end{array} \right\}  \langle p n' , J | V | p' n, J \rangle
 \end{equation}
 
Above, in eqs. (\ref{eq_Adef}, \ref{eq_Bdef}) we have introduced adjustable renormalization 
constants $g_{ph}$ and $g_{pp}$ that multiply the whole block of interaction matrix elements
for particle-hole and particle-particle configurations. Typically, one uses a realistic interaction
for $V$ (G-matrix) and thus the nominal values are $g_{ph} = g_{pp} = 1$.

The vectors $X$ and $Y$ obey the normalization and orthogonality conditions 
(for each $J^{\pi}$)
\begin{eqnarray}
& & \Sigma_{pn}  X^m_{pn} X^{m'}_{pn} - Y^m_{pn} Y^{m'}_{pn} = \delta_{m,m'}  \\
\nonumber
& & \Sigma_m X^m_{pn} X^m_{p'n'} - Y^m_{pn} Y^m_{p'n'} = \delta_{pn,p'n'}  \\
\nonumber
& & \Sigma_{pn} X^m_{pn} Y^{m'}_{pn} - Y^m_{pn} X^{m'}_{pn} = 0 \\
\nonumber
& & \Sigma_m  X^m_{pn} Y^m_{p'n'} - Y^m_{pn} X^m_{p'n'} = 0
\end{eqnarray}

For a one-body charge-changing operator $T^{JM}$ the transition amplitude 
connecting the ground state of an even-even $(N,Z)$ nucleus to the $m$th excited state
in the odd-odd $(N-1,Z+1)$ nucleus is
\begin{equation}
\langle m;JM | T^{JM} | 0^+_{QRPA} \rangle = \Sigma_{pn} \left[ t^-_{pn} X^m_{pn,J\pi}
+  t^+_{pn} Y^m_{pn,J\pi} \right] ~,
\label{eq_onebody1}
\end{equation}
where
\begin{equation}
t^-_{pn} = \frac{u_p v_n}{\sqrt{(2J + 1)}} \langle p || T^J || n \rangle ~,~~~
t^+_{pn} = (-1)^J \frac{v_p u_n}{\sqrt{(2J + 1)}} \langle p || T^J || n \rangle ~.
\label{eq_onebody2}
\end{equation}
The transition in the opposite direction, to the $(N+1,Z-1)$ is governed by analogous
formula but with $X$ and $Y$ interchanged.

Given these formulae, we are able to calculate, within QRPA, the $\beta\beta$ decay 
nuclear matrix elements, for both $2\nu$ and $0\nu$ modes. However, one needs
to make another approximation, since the initial state $| 0^+_{QRPA}; i \rangle$
and the final state $| 0^+_{QRPA}; f \rangle$ are not identical. This is a ``two vacua"
problem. The standard way of accounting for the difference in the initial and final
states is to add the overlap factor
\begin{equation}
\langle J^{\pi}_k | J^{\pi}_m \rangle = 
 \Sigma_{pn}  X^k_{pn} \tilde{X}^{m}_{pn} - Y^k_{pn} \tilde{Y}^m_{pn} ~,
 \end{equation}
 where $\tilde{X}, \tilde{Y}$ are the solutions of the QRPA equations of motion
 for the final nucleus. 

We can now write down the formula for the $2\nu\beta\beta$-decay matrix element as
\begin{eqnarray}
M^{2\nu} = \Sigma_{k,m} \frac{ \langle f || \vec{\sigma} \tau^+ || 1^+_k \rangle 
 \langle 1^+_k | 1^+_m \rangle \langle  1^+_m || \vec{\sigma} \tau^+ || i \rangle}
 { \omega_m - (M_i + M_f)/2} ~, 
 \\ 
 \nonumber
  \langle  1^+_m || \vec{\sigma} \tau^+ || i \rangle =
 \Sigma_{pn} \langle p || \vec{\sigma} || n \rangle (u_p v_n X^m_{pn} + v_p u_n Y^m_{pn} )
 \\ 
 \nonumber
  \langle f || \vec{\sigma} \tau^+ || 1^+_k \rangle =
   \Sigma_{pn} \langle p || \vec{\sigma} || n \rangle (\tilde{v}_p \tilde{u}_n \tilde{X}^k_{pn} 
   + \tilde{u}_p \tilde{v}_n \tilde{Y}^k_{pn} ) ~.
 \end{eqnarray}

For the $0\nu$ decay the corresponding formula is
\begin{eqnarray}
\label{eq:long}
&& M_K  =   \sum_{J^{\pi},k_i,k_f,\mathcal{J}} \sum_{pnp'n'}
(-1)^{j_n + j_{p'} + J + {\mathcal J}} \times\qquad\qquad
\\
&& \sqrt{ 2 {\mathcal J} + 1}
\left\{
\begin{array}{c c c}
j_p & j_n & J  \\
 j_{n'} & j_{p'} & {\mathcal J}
\end{array}
\right\}  \times \qquad\qquad\qquad\qquad\qquad\quad
\nonumber \\
&&\langle p(1), p'(2); {\mathcal J} \parallel \bar{f}(r_{12})
\tau_1^+ \tau_2^+
O_K \bar{f}(r_{12}) \parallel n(1), n'(2); {\mathcal J} \rangle \times
\qquad
\nonumber \\
&&\langle 0_f^+ ||
[ \widetilde{c_{p'}^+ \tilde{c}_{n'}}]_J || J^{\pi} k_f \rangle
\langle  J^{\pi} k_f |  J^{\pi} k_i \rangle
 \langle  J^{\pi} k_fi|| [c_p^+ \tilde{c}_n]_J || 0_i^+ \rangle\, .
\nonumber 
\end{eqnarray}
The operators $O_K, K$ = Fermi (F), Gamow-Teller (GT), and Tensor
(T) contain neutrino potentials, {see eq. (\ref{eq_poten}),
and spin and isospin operators, and
RPA energies $E^{k_i,k_f}_{J^\pi}$.
The reduced matrix elements of the one-body operators
$[ c_p^+ \tilde{c}_n ]_J $ in  eq. (\ref{eq:long})
depend on the BCS coefficients $u_i,v_j$ and on the QRPA vectors
$X,Y$; they are just the reduced matrix elements of the one-body operators
like in eqs. (\ref{eq_onebody1}, \ref{eq_onebody2}).
The function $\bar{f}(r_{12})$
in above
represents the effect of short range correlations.
These will be discussed in detail in the next Section.
Note, that the radial matrix elements in eq. (\ref{eq:long}) are
evaluated with unsymmetrized two-particle wave functions. This 
is a generic requirement of RPA-like procedures as explained in Ref.\cite{us08}

\subsection{Generalization - RQRPA}
The crucial simplifying point of QRPA (or RPA-like procedures in general) is the quasiboson
approximation, the assumption that the commutation relations of a pair of fermion operators
can be replaced by the boson commutation relation. That is a good approximation  for harmonic,
small amplitude excitations. However, when the strength of the
attractive particle-particle interaction increases, the number of quasiparticles in
the correlated ground state $| 0^+_{QRPA} \rangle$ increases and thus the method
violates the Pauli principle. This, in turn, leads to an overestimate of ground state
correlations, and too early onset of the QRPA collapse.

To cure this problem, to some extent, a simple procedure, so-called renormalized QRPA
(RQRPA) has been proposed and is widely used \cite{ToiSuh95}. In RQRPA the exact 
expectation value of a commutator of two-bifermion operators is replaced by
\begin{eqnarray}
&& \big<0^+_{QRPA}\big|\big
[A^{} (pn, JM), A^+(p'n', JM)\big]
\big|0^+_{QRPA}\big> = \delta_{pp'}\delta_{nn'}\times
\nonumber \\ 
\lefteqn{\underbrace{
\Big\{1
\,-\,\frac{1}{\hat{\jmath}_{l}}
<0^+_{QRPA}|[a^+_{p}{\tilde{a}}_{p}]_{00}|0^+_{QRPA}>
\,-\,\frac{1}{\hat{\jmath}_{k}}
<0^+_{QRPA}|[a^+_{n}{\tilde{a}}_{n}]_{00}|0^+_{QRPA}>
\Big\}
}_{
\displaystyle {\cal D}_{pn, J^\pi}
},} &&
\nonumber \\
\label{eq_rqrpa}
\end{eqnarray}
with  $\hat{\jmath}_p=\sqrt{2j_p+1}$.

To take this generalization into account, i.e. the nonvanishing values of
$ {\cal D}_{pn, J^\pi} -1$ one simply needs to use the amplitudes
\begin{equation}
{\overline{X}}^m_{(pn, J^\pi)}  =  {\cal D}^{1/2}_{pn, J^\pi}~ 
X^m_{(pn, J^\pi)}, ~~~~~
{\overline{Y}}^m_{(pn, J^\pi)}  =  {\cal D}^{1/2}_{pn, J^\pi}~ 
Y^m_{(pn, J^\pi)} ~,
\end{equation}
which are orthonormalized in the usual way instead of the standard $X$ and $Y$
everywhere also in the QRPA equation of motion (\ref{eq_rpa}).
In addition, the matrix elements of the one-body operators,
eqs. (\ref{eq_onebody1}, \ref{eq_onebody2}), must be multiplied
by ${\cal D}^{1/2}_{pn, J^\pi}$ evaluated for the initial and final nuclei.

In order to calculate the factors ${\cal D}^{1/2}_{pn, J^\pi}$ one has
to use an iterative procedure and evaluate in each iteration
\begin{eqnarray}
D_{pn}  = 1 & - & \frac{1}{2j_p + 1} \Sigma_{n'} D_{pn'} 
\left( \Sigma_{J,k} (2J+1) | \overline{Y}_{pn'}^{J,k} |^2 \right)
\\ \nonumber 
& - &  \frac{1}{2j_n + 1} \Sigma_{p'} D_{p'n} 
\left( \Sigma_{J,k} (2J+1) | \overline{Y}_{p'n}^{J,k} |^2 \right) ~.
\end{eqnarray}
Note the summation over the multipolarity $J$. Even if for the $2\nu$ decay
only $J^{\pi} = 1^+$ are seemingly needed, in RQRPA the equations need
to be solved for all multipoles.

\section{Numerical calculations in QRPA and RQRPA}

In the previous section all relevant expressions were given. But that is not
all one needs in order to evaluate
the nuclear matrix elements numerically. One has to decide, first of all,
what are the relevant input parameters, and how to choose them.

The first one to choose, as in all nuclear structure calculations, is the
mean field potential, and which orbits are 
going to be included in the corresponding
expressions. A typical choice is the calculation based on the Coulomb
corrected Woods-Saxon potential. However, sometimes it is advisable
to modify the single particle energy levels in order to better describe
certain experimental data (energies in the odd-A nuclei or occupation
numbers).

Next one needs to choose the nucleon-nucleon potential. For that one typically
uses a G-matrix based on a realistic force. For example, in \cite{us08} the
G-matrix used was derived from the Bonn-CD nucleon-nucleon force.
It turns out that the results only weakly depend on which parametrization of the
nucleon-nucleon interaction  is used.

Next, the effective coupling constants $g_{ph}$ and $g_{pp}$ in the
matrices $A$ and $B$ in equation of motion (\ref{eq_rpa}) must be determined.
The parameter $g_{ph}$ is not controversial. It is usually adjusted by requiring
that the energy of some chosen collective states, often Gamow-Teller (GT)
giant resonances, is correctly reproduced. It turns out that the calculated
energy of the giant GT state is almost independent of the size of the
single-particle basis and is well reproduced with $g_{ph} \approx 1$.
Hence, the nominal and unrenormalized value $g_{ph} = 1$ is used in  most
calculations; that reduces the number of adjustable parameters as well.

The choice of the particle-particle parameter $g_{pp}$ is, however, not only
important, but also to some extent controversial. One of the issues involved
is illustrated in Fig. \ref{fig_2nuspace}. There, one can see that the curves
of $M^{2\nu}$ versus $g_{pp}$ are rather different for different number of
included single-paricle states. Thus, the calculated magnitude of $M^{2\nu}$ will 
change dramatically if $g_{pp}$ is fixed and different number of levels is included.

\begin{figure}
\centerline{\includegraphics[width=9.0cm]{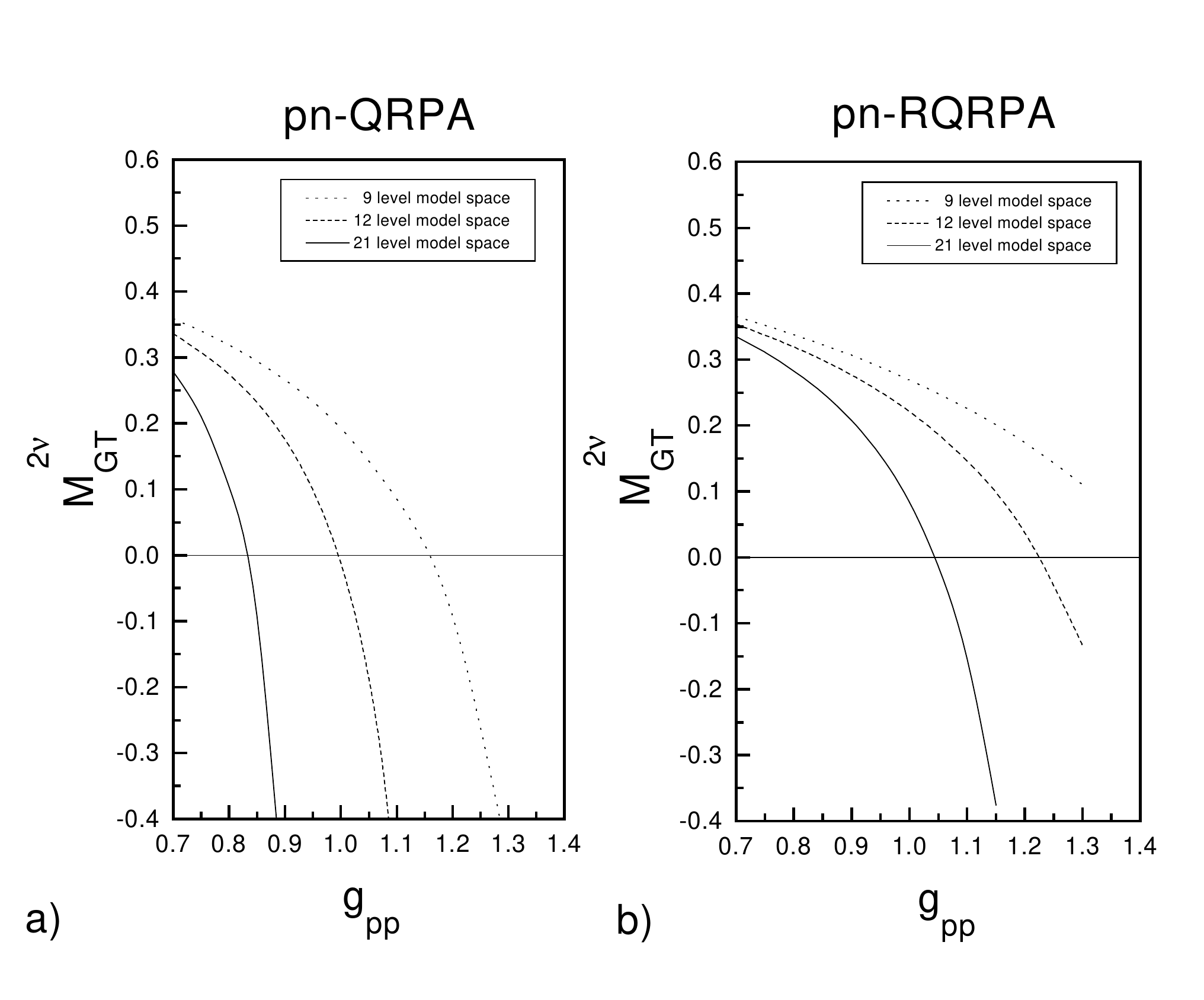}}
\caption{The matrix elements $M^{2\nu}$ for $^{76}$Ge evaluated in QRPA (panel a)
and RQRPA (panel b). The calculation was performed with the indicated number of
included single-particle states. (Adopted from \cite{FaSi98}). The experimental
value of $M^{2\nu}$ is 0.13 MeV$^{-1}$. }
\label{fig_2nuspace}
\end{figure}

The other feature, illustrated in Fig. \ref{fig_2nuspace} is the crossing of $M^{2\nu}$
of zero value for certain $g_{pp}$ that is relatively close to unity. This was first
recognized long time ago in Ref. \cite{VZ86}. 
Obviously, if $M^{2\nu} = 0$ then the $2\nu\beta\beta$-decay
lifetime is infinite, this is a absolute suppression of that mode.
Past that zero crossing the curves
become very steep and the collapse of QRPA is reached when the slope becomes vertical.
Obviously, the zero crossing is moved to larger values of $g_{pp}$ in RQRPA.

One can use all of this to an advantage by abandoning the goal of predicting the 
$M^{2\nu}$ values, but instead of using the experimental $M^{2\nu}$ and determine
the $g_{pp}$ (for a given set of s.p. states) in such a way, that the correct $M^{2\nu}$
is obtained. That is illustrated in Fig. \ref{fig_2nufix}.
The resulting $g_{pp}$ then depends on the number of included single-particle states,
and is typically in the range $0.8 \le g_{pp} \le 1.2$ when a realistic
G-matrix based hamiltonian is used.

\begin{figure}
\centerline{\includegraphics[width=10.0cm]{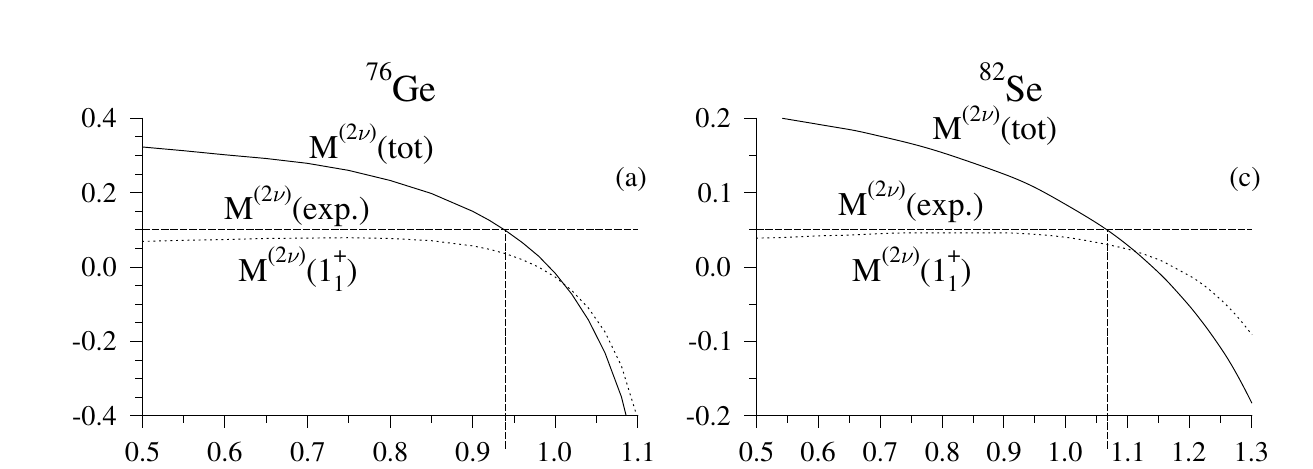}}
\caption{The calculated $2\nu$ matrix elements for $^{76}Ge$ and $^{82}$Se 
as a function of $g_{pp}$ (solid lines). The experimental values are indicated
by the horizontal dashed lines. The dotted line indicates the contribution of the
first $1^+$ state. (Adopted from\cite{Suh05}.)
 }
\label{fig_2nufix}
\end{figure}

\begin{figure}
\centerline{\includegraphics[width=5.0cm]{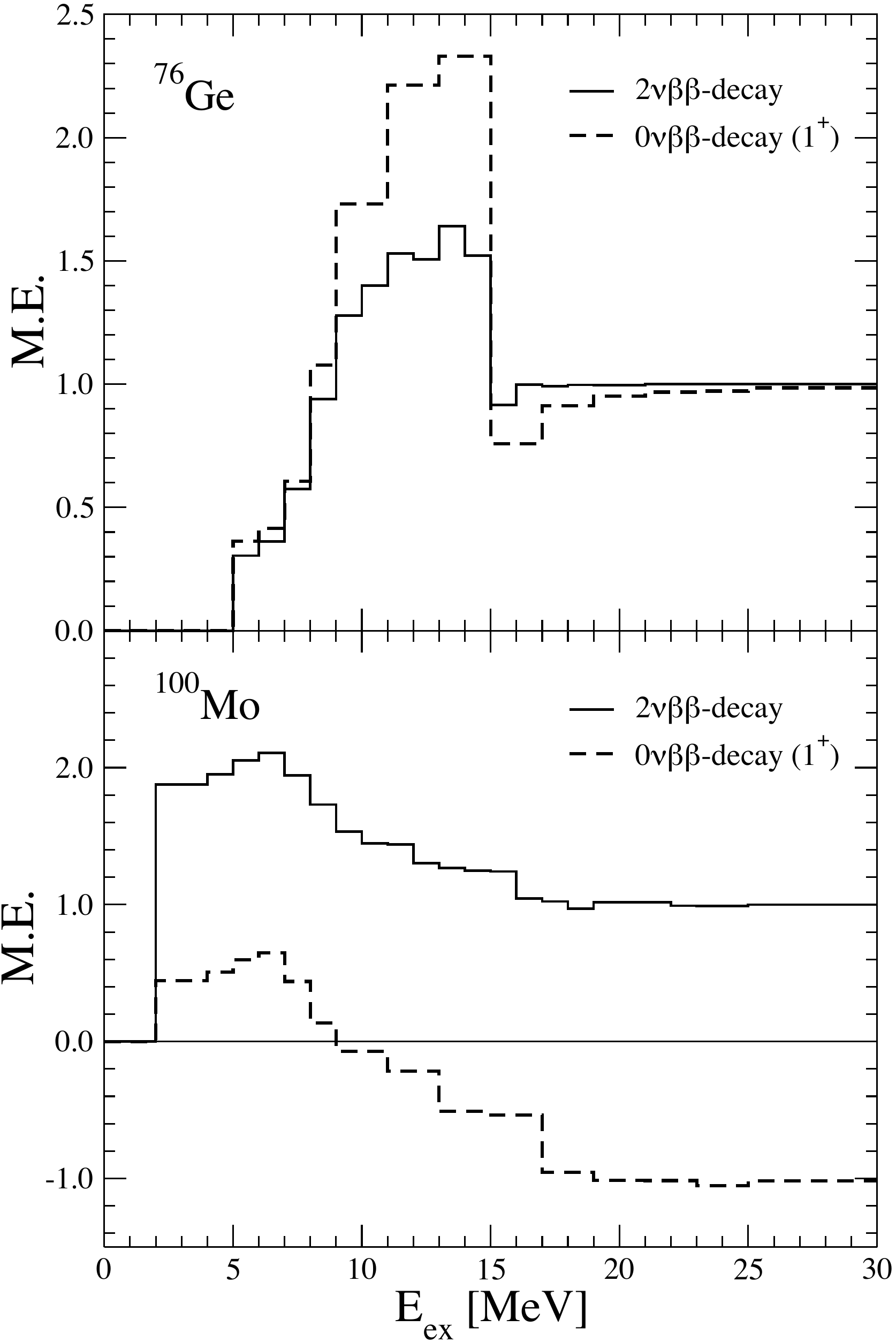}}
\caption{Running sum of the $2\nu\beta\beta$-decay  
and $0\nu\beta\beta$-decay (only $1^+$ component) 
matrix elements for $^{76}Ge$ and $^{100}Mo$ (normalized to 
unity) as a function of the  
excitation energy $E_{ex}=E_n-(E_i+E_f)/2$. }
\label{fig_1+eff}
\end{figure}

Adjusting the value of $g_{pp}$ such that the $M^{2\nu}$ is correctly reproduced
has been criticized, e.g. in Ref. \cite{Suh05}. There an alternative method, based on
the experimentally known $\beta$-decay $ft$ values connecting the ground state of the intermediate
nucleus (if that happens to be $1^+$, which is so only in $^{100}$Tc, $^{116}$In and
$^{128}$I among the $\beta\beta$-decay candidates). In QRPA the  transition
amplitudes for the EC process (decreasing nuclear charge) and $\beta^-$ (increasing
nuclear charge) move in opposite way when $g_{pp}$ is increased; the first one goes
up while the second one goes down. It is difficult, and essentially impossible to
describe all three experimental quantities with the same value of $g_{pp}$
(in the case of $^{100}$Mo this was noted already in \cite{GV92}).

\begin{figure}
\centerline{\includegraphics[width=7.0cm]{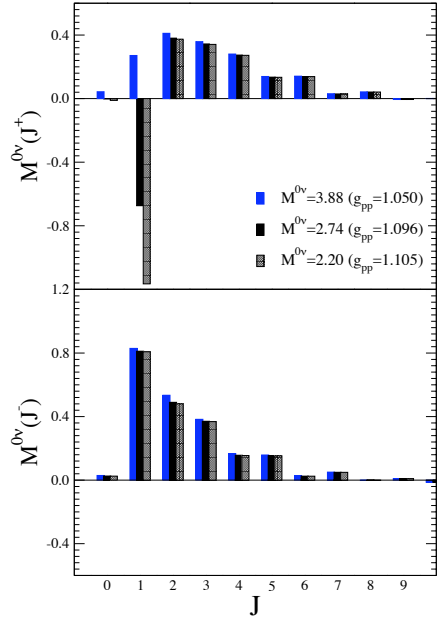}}
\caption{The contributions of different  
intermediate-state angular momenta $J^{\pi}$ 
to $M^{0\nu}$ in $^{100}$Mo (positive parities in the upper panel and negative
parities in the lower one). We show the results for several values of
$g_{pp}$. The contribution of the $1^+$ multipole changes
rapidly with $g_{pp}$, while those of the other
multipoles change slowly.}
\label{fig_multip}
\end{figure}

While the differences between these two approaches are not very large (see \cite{us06}),
there are other arguments why choosing the agreement with $M^{2\nu}$ is
preferable.
First, it is not really true that the first $1^+$ state 
is the only one responsible for the $2\nu\beta\beta$ decay. 
This is illustrated for the cases of $^{76}$Ge and $^{100}$Mo 
in Fig. \ref{fig_1+eff}. Even though for $^{100}$Mo 
the first state contributes substantially, higher lying states 
give non-negligible contribution. And in $^{76}$Ge many 
$1^+$ states give comparable contribution. Thus,  
to give preference to the lowest state is not
well justified, the sum is  
actually what matters. At the same time, the dilemma that 
the $\beta^-$ and   $\beta^+/EC$ matrix elements move with  
$g_{pp}$ in opposite directions makes it difficult to  
choose one of them.  
It seems better to use the sum of the products of the 
amplitudes, i.e. the  $2\nu\beta\beta$ decay. 

At the same time, the contribution of the $1^+$ multipole to the 
 $0\nu\beta\beta$ matrix element and the corresponding  
 $2\nu\beta\beta$ matrix element are correlated, even though they are 
not identical, as also shown in Fig. \ref{fig_1+eff}. Making sure that 
the  $2\nu\beta\beta$ matrix element agrees with its experimental value 
constrains the $1^+$ part of the  $0\nu\beta\beta$ matrix element  
as well.

Since we are really interested in the $M^{0\nu}$ matrix element, it is relevant to ask
why do we fit the important parameter $g_{pp}$ to the $2\nu$ decay lifetime.
To understand this, it is useful to point out that
two separate multipole decompositions are built into Eq.\
(\ref{eq:long}). One is in terms the
$J^{\pi}$ of the virtual states in the intermediate nucleus, the
good quantum numbers of the QRPA and RQRPA. The other decomposition
is based on the angular momenta and parities ${\mathcal J}^{\pi}$ of the
pairs of neutrons that are transformed into protons with the same
${\mathcal J}^{\pi}$.

 In Fig. \ref{fig_multip}} we show that it is essentially only the
$1^+$ multipole that is responsible for the variation of 
$M^{0\nu}$ with $g_{pp}$.  (Note that in the three variants shown
the parameter $g_{pp}$ changes only by 5\%.)
Fixing its contribution to a related
observable ($2\nu$ decay) involving the same initial and final
nuclear states appears to be an optimal procedure for determining
$g_{pp}$. Moreover, as was shown in \cite{us03,us06},
this choice, in addition,
essentially removes the dependence of $M^{0\nu}$ on
the number of the single-particle states (or oscillator shells) in
the calculations.

\subsection{Competition between `pairing' and `broken pairs'}

\begin{figure}
\centerline{\includegraphics[width=7.0cm]{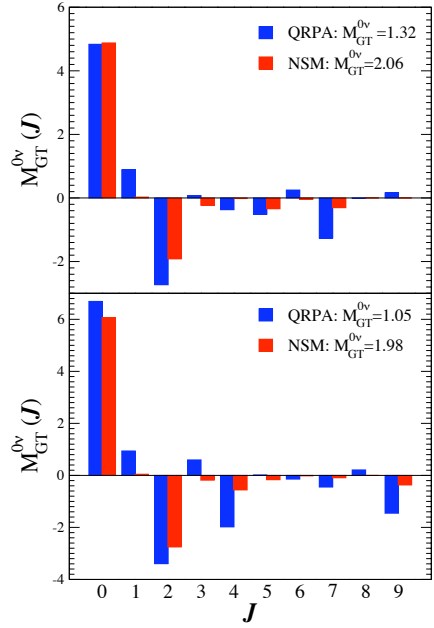}}
\caption{Contributions of different 
angular momenta ${\mathcal J}$ 
associated with the two decaying neutrons
to the Gamow-Teller part of $M^{0\nu}$ in $^{82}$Se (upper
panel) and $^{130}$Te (lower panel). The results of NSM (dark
histogram) and QRPA treatments (lighter histogram) are
compared. Both calculations use the same single-particle spaces:
($f_{5/2},p_{3/2},p_{1/2}, g_{9/2}$) for $^{82}$Se and ($g_{7/2},
d_{5/2}, d_{3/2}, s_{1/2}, h_{11/2}$) for $^{130}$Te. 
In the QRPA calculation the particle-particle interaction was
adjusted to reproduce the experimental $2\nu\beta\beta$-decay 
rate.}
\label{fig_competition}
\end{figure}

The decomposition
based on the angular momenta and parities ${\mathcal J}^{\pi}$ of the
pairs of neutrons that are transformed into protons with the same
${\mathcal J}^{\pi}$ is particularly
revealing. In Fig. \ref{fig_competition} we illustrate it both in the
NSM and QRPA, with the same single-single particle spaces in each.
These two rather different approaches agree in a semiquantitative
way, but the NSM entries for ${\mathcal J}
> 0$ are systematically smaller in absolute value.
There are two opposing tendencies in Fig.  \ref{fig_competition}.
The large positive contribution (essentialy the same in QRPA and NSM) is associated with the
so-called pairing interaction of neutrons with neutrons and protons with protons. As
the result of that interaction the nuclear ground state is mainly composed of Cooper-like
pairs of neutrons and protons coupled to ${\mathcal J} = 0$. The transformation of one 
neutron Cooper pair into one  Cooper proton pair is responsible for the ${\mathcal J} = 0$ piece in
 Fig.  \ref{fig_competition}. 

However, the nuclear hamiltonian contains, in addition, important neutron-proton interaction.
That interaction, primarily, causes presence in the nuclear ground state of ``broken pairs",
i.e. pairs of neutrons or protons coupled to ${\mathcal J} \ne 0$. Their effect, as seen in
Fig.  \ref{fig_competition}, is to reduce drastically the magnitude of $M^{0\nu}$. In treating
these terms, the agreement between QRPA and NSM is only semi-quantitative. Since the 
pieces related to the ``pairing" and ``broken pairs" contribution ale almost of the
same magnitude but of opposite signs, 
an error in one of these two competing tendencies is enhanced in the final  $M^{0\nu}$.
The competition, illustrated in Fig.  \ref{fig_competition}, is the main reason behind the
spread of the published $M^{0\nu}$
calculations. Many authors use different, and sometimes inconsistent,
treatment of the neutron-proton interaction.

\begin{figure}
\centerline{\includegraphics[width=13.0cm]{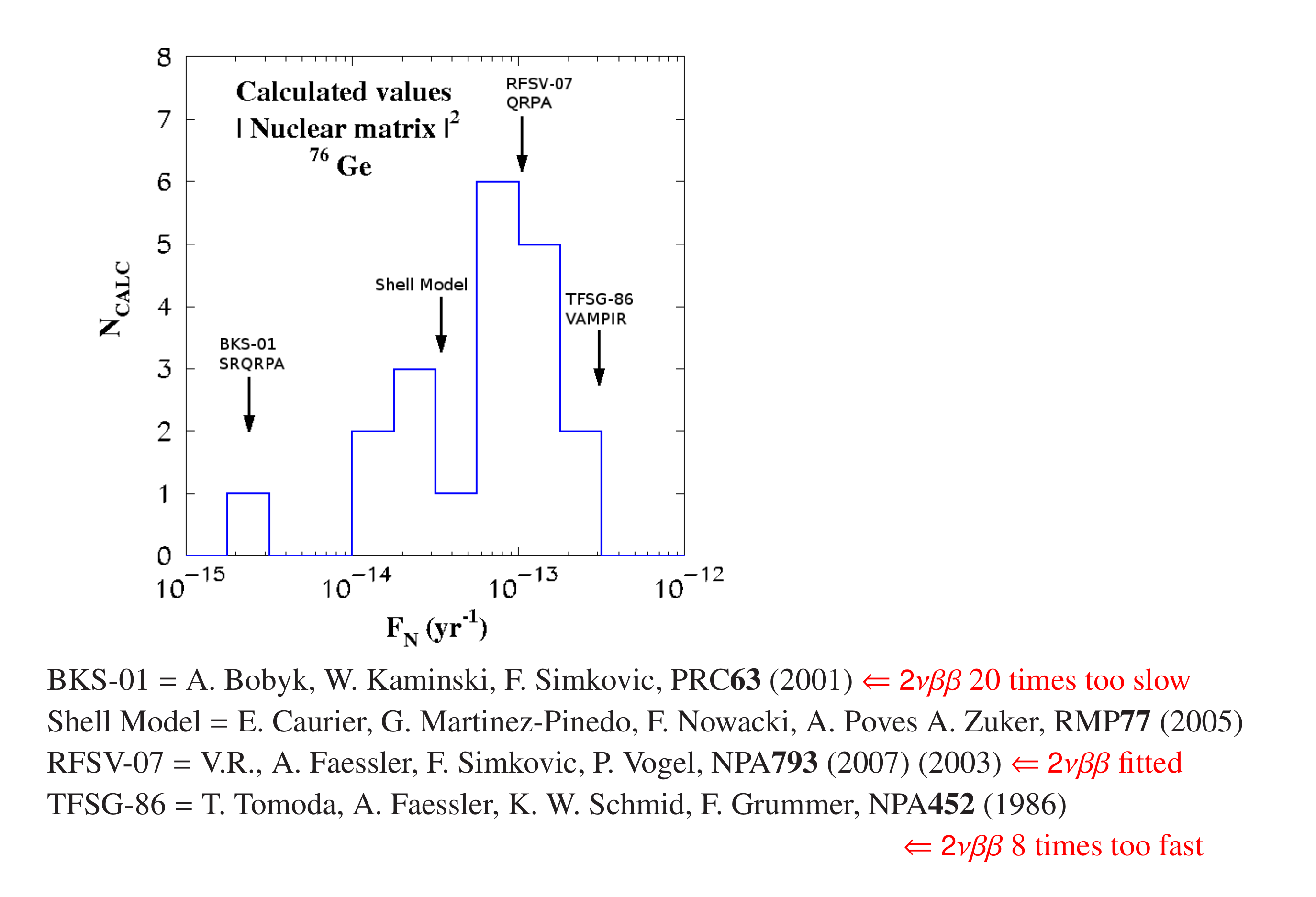}}
\caption{Histogram of older published calculated values of $(M^{0\nu})^2$ for $^{76}$Ge.
The failure of some of the calculations to reproduce the known $2\nu\beta\beta$-decay
lifetime is indicated. }
\label{fig_bah}
\end{figure}

There are many evaluations of the matrix elements $M^{0\nu}$ in the  literature
(for the latest review see \cite{AEE07}). However, the resulting matrix elements often do 
not agree with each other as mentioned above 
and it is difficult, based on the published material, to decide
who is right and who is wrong, and what is the theoretical uncertainty in $M^{0\nu}$.
That was stressed in a powerful way 
in the paper by Bahcall {\it et al.} few years ago \cite{Bahcall} where
a histogram of 20 calculated values of $(M^{0\nu})^2$ for $^{76}$Ge 
was plotted, with the implication
that the width of that histogram is a measure of uncertainty. That is clearly not a valid
conclusion as one could see in Fig. \ref{fig_bah} where the failure of the outliers
to reproduce the known $2\nu\beta\beta$-decay lifetime is indicated.

To see some additional reasons why different authors obtain in their calculations different
nuclear matrix elements we need to analyze the dependence of the  $M^{0\nu}$ on the
distance $r$ between the pair of initial neutrons (and, naturally, the pair of final protons)
that are transformed in the decay process. That analysis reveals, at the same time, 
the various physics ingredients that must be included in the calculations so that 
realistic values of the $M^{0\nu}$ can be obtained.

\subsection{Dependence on the radial distance}

The simplest and most important neutrino potential has the form
(already defined in eq. (\ref{eq_nupot}))
\begin{equation}
H(r) \sim \frac{R}{r} \Phi(\omega r) ~,
\end{equation}
where $R$ is the nuclear radius introduced here as usual to make the potential, and the resulting 
$M^{0\nu}$, dimensionless (the  $1/R^2$ in the phase space factor compensates
for this), $r$ is the distance between the transformed neutrons (or protons) and
$\Phi(\omega r) $ is a rather slowly varying function of its argument.

\begin{figure}
\centerline{\includegraphics[width=7.0cm]{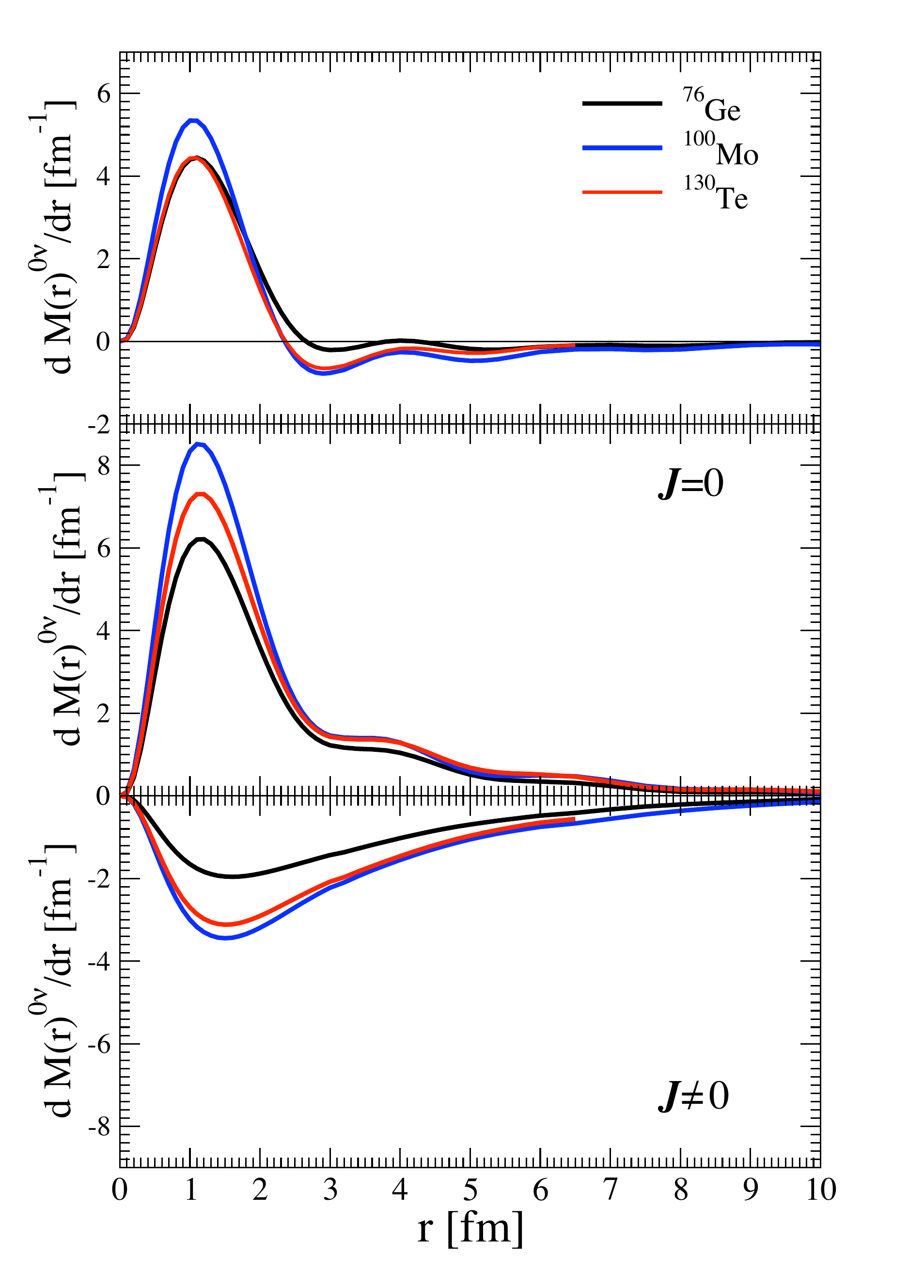}}
\caption{The dependence on $r$ of $M^{0\nu}$ for
$^{76}$Ge, $^{100}Mo$ and $^{130}$Te. The upper panel shows the full
matrix element, and the lower panel shows separately `pairing'
(${\mathcal J} = 0$ of the two participating neutrons) 
and `broken pair' (${\mathcal J} \ne 0$)
contributions.}
\label{fig_radial}
\end{figure}

From the form of the potential $H(r)$ one would, naively, expect that the 
characteristic value of $r$ is the typical distance between the nucleons in a nucleus, namely
that $\bar{r} \sim R$. However, that is not true as was demonstrated first in Ref. \cite{us08}
and illustrated in Fig. \ref{fig_radial}. One can see there that the competition between the
``pairing" and ``broken pairs" pieces essentially removes all effects of $r \ge 2-3$ fm.
Only the relatively short distances contribute significantly;
essentially only the nearest neighbor neutrons undergo the $o\nu\beta\beta$
transition. The same result was
obtained in the NSM \cite{sm08c}. (We have also shown in \cite{us08} that 
an analogous result is obtained in an exactly solvable, semirealistic model. There
we also showed that this behaviour is restricted to an interval of the parameter $g_{pp}$
that contains the realistic value near unity.)   

Once the $r$ dependence displayed in Fig. \ref{fig_radial} is accepted, several new
physics effects clearly need to be considered. 
These are not nuclear structure issue per se, since they are related more to the structure
of the nucleon.

\begin{figure}
\centerline{\includegraphics[width=9.0cm]{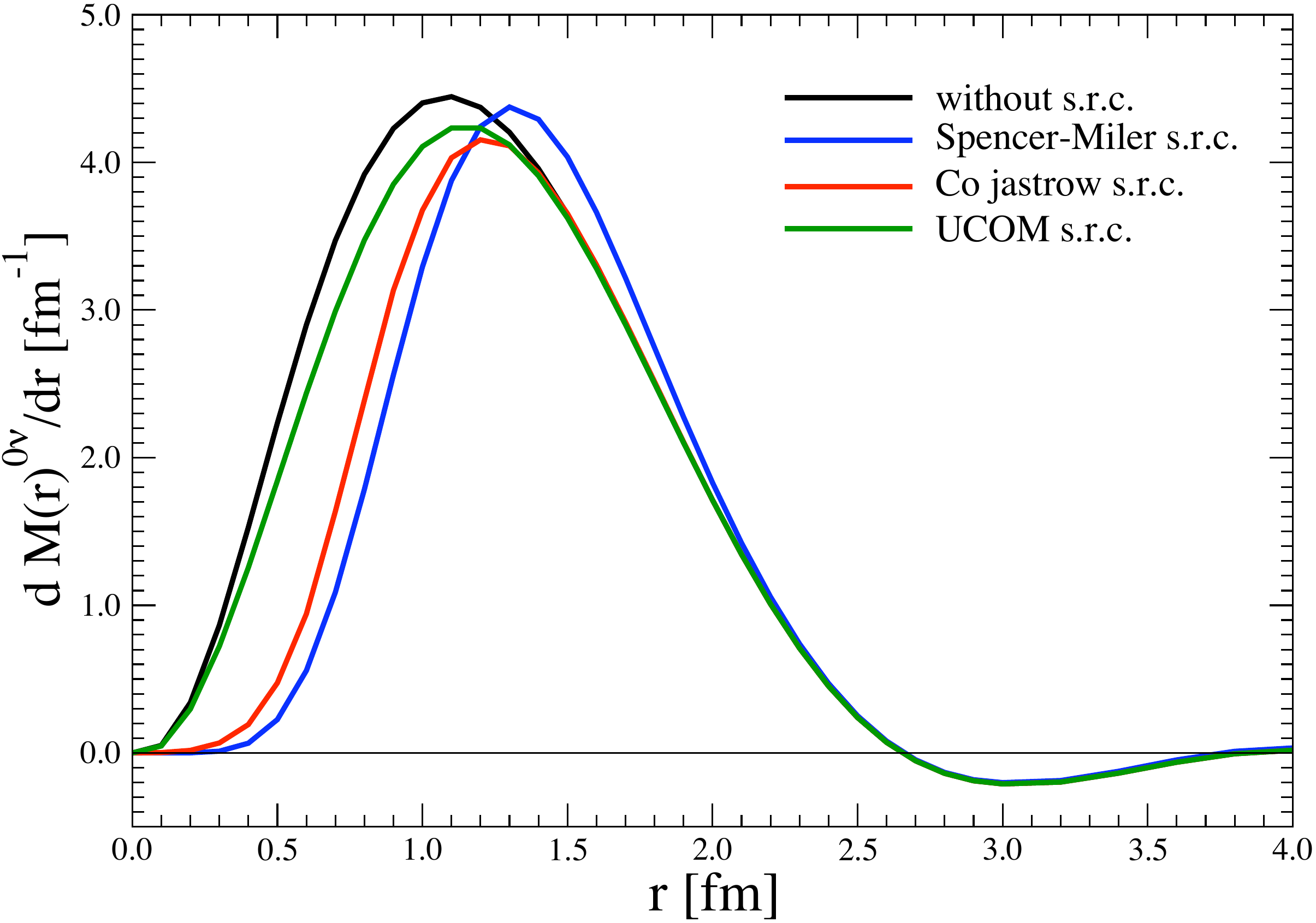}}
\caption{The $r$ dependence of $M^{0\nu}$ in
$^{76}$Ge. The four curves show the effects of different treatments
of short-range correlations. The resulting $M^{0\nu}$ values are
5.32 when the effect is ignored, 5.01 when the UCOM transformation
is applied and 4.14 when the treatment based on the
Fermi hypernetted chain 
and 3.98 when the phenomenological Jastrow 
function is used. (See the text for details.)}
\label{fig_src}
\end{figure}

One of them is the short-range nucleon-nucleon repulsion
known from scattering experiments. Two nucleon strongly repel each other at distances
$r \le 0.5-1.0$ fm, i.e. the distances very relevant to evaluation of the $M^{0\nu}$. The
nuclear wave functions used in QRPA and NSM, products of the mean field single-nucleon
wave function, do not take into account the influence of this repulsion that is irrelevant
in most standard nuclear structure theory applications. The usual and simplest way to include the
effect is to modify the radial dependence of the $0\nu\beta\beta$ operator so that the
effect of short distances (small values of $r$) is reduced. This is achieved by introducing
a phenomenological function $\bar{f}(r_{12}$ in the eq. (\ref{eq:long}). 
Examples of such Jastrow-like function were first derived in \cite{Spencer} and in
a more modern form in \cite{Co}. 
That phenomenological procedure reduces 
the magnitude of   $M^{0\nu}$ by 20-25\% as illustrated in Fig. \ref{fig_src}.
Recently, another procedure, based on the Unitary Correlation Operator Method
(UCOM) has been proposed \cite{UCOM}. That procedure, still applied not fully consistently,
reduces the $M^{0\nu}$  much less, only by about 5\% \cite{Kort}. It is prudent to include
these two possibilities as extremes and the corresponding range as systematic error.
Once a consistent procedure is developed, consisting of deriving an effective $0\nu\beta\beta$
decay operator that includes (probably perturbatively) the effect of the high momentum
(or short range) that component of the systematic error could be substantially reduced.

Another effect that needs to be taken into account is the nucleon  composite nature. 
At weak interaction reactions with higher momentum transfer the nucleon
is less likely to remain nucleon; new particles for example pions, are often
produced. That reduction is included,
usually, by introducing the dipole form of the nucleon form factor, already introduced,
\begin{equation}
f_{V,A} =\left(  \frac{1}{1 + q^2/M_{V,A}^2} \right)^2 ~,
\end{equation}
where the cut-off parameters $M_{V,A}$ have values (deduced in the reactions of
free neutrinos with free or quasifree nucleons) $\sim$ 1 GeV. This corresponds to
the nuclon size of $\sim 0.5-1.0$ fm. Note that in our case we are dealing
with neutrinos far off mass shell, and bound nucleons, hence it is not obvious
that the above form factors are applicable. It turns out, however, that
once the short range correlations are properly included (by either of the procedures 
discussed above) the $M^{0\nu}$ becomes essentially independent of the adopted
values when $M_{V,A} \ge$ 1 GeV. In the past various authors neglected the
effect of short range correlations, and in that case a proper inclusion of nucleon form factor
(or their neglect) again causes variations in the calculated $M^{0\nu}$ values.

Yet another correction that various authors neglected must be included in a correct treatment.
Since $r \le$ 2-3 fm is the relevant distances, the corresponding momentum transfer $1/r$
is of the order of $\sim$200 MeV, much larger than in the ordinary $\beta$ decay. Hence the induced
nucleon currents, in particular the pseudoscalar (since the neutrinos are far off mass shell)
give noticeable contributions \cite{us08,sim99}.

Finally, the issue of the axial current ``quenching" should be considered. As shown in a 
rather typical example in Fig. \ref{fig_sm2} calculated values of the GT $\beta$-decay 
matrix elements usually overestimate the corresponding experimental values. The
reasons for that are, at least qualitatively understood, but the explanation is beyond the
scope of these lectures. It suffice to say, that one can account for that effect phenomenologically
and conveniently, by reducing the value of the axial-vector coupling constant $g_A$ to
$\sim 1$ instead of its true value 1.25. The phenomenon of quenching has been observed only
in the GT $\beta$ decays, so it is not clear whether the same reduction of $g_A$ should be
used also in the $0\nu\beta\beta$ decay. Nevertheless, it is prudent to include that possibility
as a source of uncertainty. In anticipation , we already used the modified definition $M'^{0\nu}$,
see eq. (\ref{eq_me0nu}).

We have, therefore, identified the various physics effects that ought to be included in a
realistic evaluation of $M^{0\nu}$ values. The spread of the calculated values, noted
by Bahcall {\it et al.} \cite{Bahcall} can be often attributed to the fact that various authors either
neglect some of them, or include them inconsistently.

\section{Calculated $M'^{0\nu}$ values}

\begin{table}[htb]
  \begin{center}
\caption{\label{tab:t12}The calculated ranges of the nuclear matrix element
$M^{'0\nu}$ evaluated within both the QRPA and RQRPA and with both standard
($g_A = 1.254$) and quenched ($g_A = 1.0$) axial-vector couplings.  In each
case we adjusted  $g_{pp}$ so that the rate of the $2\nu\beta\beta$=decay is
reproduced.  Column 2 contains the ranges of $M^{'0\nu}$ with the
phenomenological Jastrow-type treatment of short range correlations (see I and
II), while column 3 shows the UCOM-based results (see Ref. \cite{UCOM}).
Columns 3 and 5 give the  $0\nu\beta\beta$-decay half-life ranges corresponding
to the matrix-element ranges in columns 2 and 4, for
$<m_{\beta\beta}>=50$~meV. Adapted from \cite{us08}. }

\vspace{0.2cm}

\begin{tabular}{lccccc}
\hline\hline
 Nuclear & \multicolumn{2}{c}{(R)QRPA (Jastrow s.r.c.)} & &
           \multicolumn{2}{c}{(R)QRPA (UCOM s.r.c.)}\\ \cline{2-3} \cline{5-6}
transition & $M^{'0\nu}$ & $T^{0\nu}_{1/2}$ ($\langle m_{\beta\beta} \rangle$ = 50 meV) &
           & $M^{'0\nu}$ & $T^{0\nu}_{1/2}$ ($\langle m_{\beta\beta} \rangle$ = 50 meV) \\\hline
$^{76}Ge\rightarrow {^{76}Se}$
  &  $(3.33,4.68)$ & $(6.01,11.9)\times 10^{26}$ &
  &  $(3.92,5.73)$ & $(4.01,8.57)\times 10^{26}$ \\
$^{82}Se\rightarrow {^{82}Kr}$
  &  $(2.82,4.17)$ & $(1.71,3.73)\times 10^{26}$ &
  &  $(3.35,5.09)$ & $(1.14,2.64)\times 10^{26}$ \\
$^{96}Zr\rightarrow {^{96}Mo}$
  &  $(1.01,1.34)$ & $(7.90,13.9)\times 10^{26}$ &
  &  $(1.31,1.79)$ & $(4.43,8.27)\times 10^{26}$ \\
$^{100}Mo\rightarrow {^{100}Ru}$
  &  $(2.22,3.53)$ & $(1.46,3.70)\times 10^{26}$ &
  &  $(2.77,4.58)$ & $(8.69,23.8)\times 10^{25}$ \\
$^{116}Cd\rightarrow {^{116}Sn}$
  &  $(1.83,2.93)$ & $(1.95,5.01)\times 10^{26}$ &
  &  $(2.18,3.54)$ & $(1.34,3.53)\times 10^{26}$ \\
$^{128}Te\rightarrow {^{128}Xe}$
  &  $(2.46,3.77)$ & $(3.33,7.81)\times 10^{27}$ &
  &  $(3.06,4.76)$ & $(2.09,5.05)\times 10^{27}$ \\
$^{130}Te\rightarrow {^{130}Xe}$
  &  $(2.27,3.38)$ & $(1.65,3.66)\times 10^{26}$ &
  &  $(2.84,4.26)$ & $(1.04,2.34)\times 10^{26}$ \\
$^{136}Xe\rightarrow {^{136}Ba}$
  &  $(1.17,2.22)$ & $(3.59,12.9)\times 10^{26}$ &
  &  $(1.49,2.76)$ & $(2.32,7.96)\times 10^{26}$ \\
\hline\hline
\end{tabular}
  \end{center}
\end{table}

\begin{figure}
\centerline{\includegraphics[width=9.0cm]{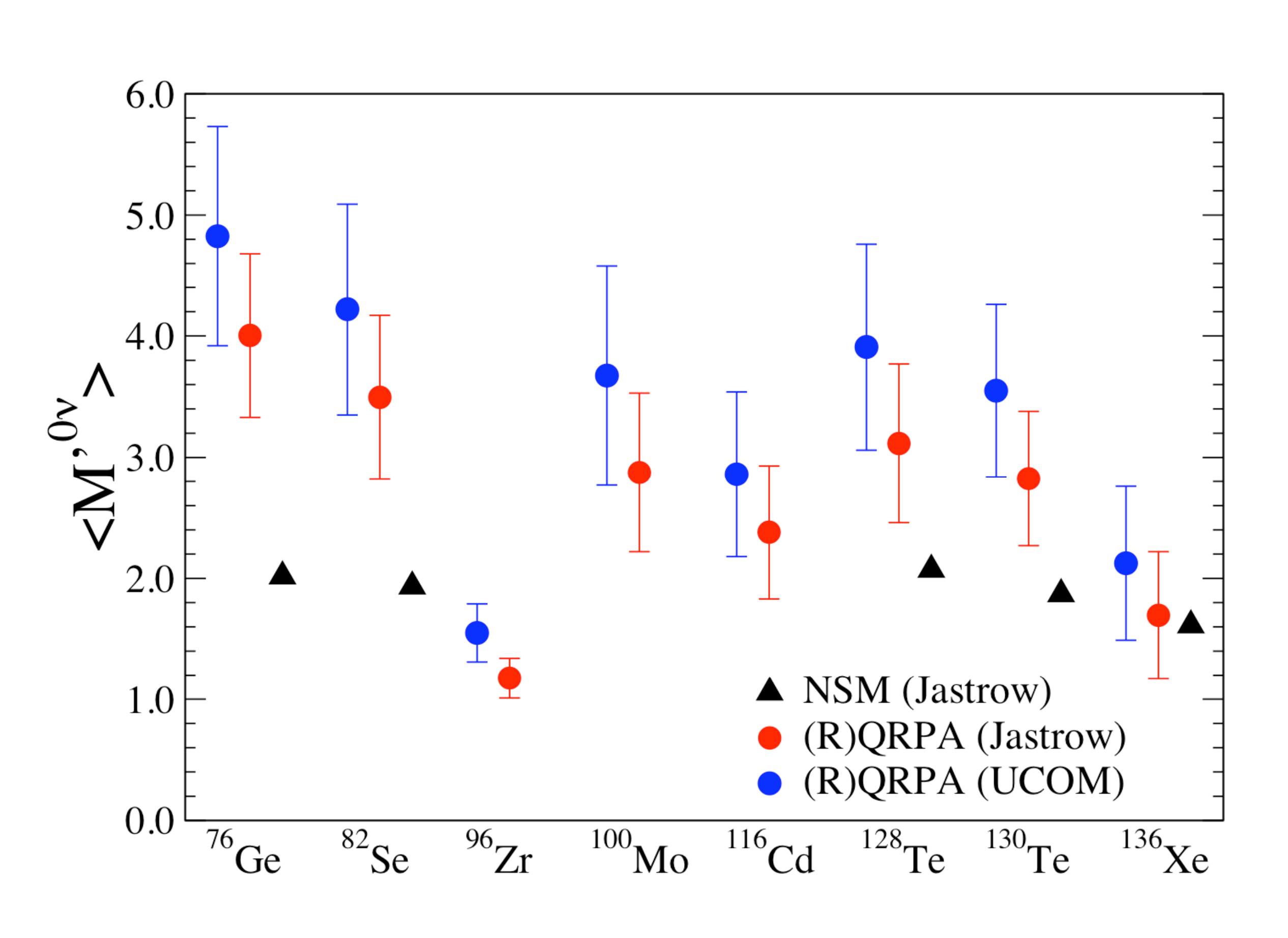}}
\caption{The full ranges of $M^{'0\nu}$ with the two alternative
treatments of the short range correlations. For comparison
the results of a recent Large Scale Shell Model evaluation of
$M^{'0\nu}$ that used the Jastrow-type treatment of short range
correlations are also shown (triangles).}
\label{fig_merange}
\end{figure}

Even though we were able to explain, or eliminate, a substantial part of the spread
of the calculated values of the nuclear matrix elements, sizeable systematic uncertainty
remains. That uncertainty, within QRPA and RQRPA, as discussed in Refs.\cite{us03,us06},
is primarily related to the difference between these two procedures, to the
size of the single-particle space included, whether the so-called quenching of the axial current
coupling constant $g_A$ is included or not, and to the systematic error in the treatment
of short range correlations \cite{us08}. In Fig. \ref{fig_merange} the full ranges of the resulting
matrix elements $M^{0\nu}$ is indicated. The indicated error bars are highly correlated; e.g.,
if true values are near the lower end in one nucleus, they would be near the lower ends in all
indicated nuclei.

The figure also shows the most recent NSM results \cite{sm08c}. Those results, obtained with
Jastrow type short range correlation corrections, are noticeably lower than the QRPA values.
That difference is particularly acute in the lighter nuclei $^{76}$Ge and $^{82}$Se. While
the QRPA and NSM agree on many aspects of the problem, in particular on the role
of the competition between ``pairing" and ``broken pairs" contributions and on the
$r$ dependence of the matrix elements, the disagreement in the actual values remains
to be explained. 

When one compares the $2\nu$ and $0\nu$ matrix elements (Figs. \ref{fig_m2nu} and 
\ref{fig_merange})
the feature to notice is the fast variation in $M^{2\nu}$ when going from one nucleus
to another while $M^{0\nu}$ change only rather smoothly, in both QRPA and NSM. This 
is presumably related to the high momentum transfer (or short range) involved in
$0\nu\beta\beta$. That property of the $M^{0\nu}$ matrix elements makes the comparison
of results obtained in different nuclei easier and more reliable.

Given the interest in the subject, we show the range of predicted half-lives
corresponding to our full range of $M'^{0\nu}$ in Table \ref{tab:t12}
(for $\langle m_{\beta\beta} \rangle$ = 50 meV).
As we argued above, this is a rather conservative range within
the QRPA and its related frameworks. One should keep in mind, however,
the discrepancy between the QRPA and NSM results as well as systematic
effects that might elude either or both calculations.
Thus, as we have seen, while a substantial progress has been achieved, we are still 
somewhat far from being able to evaluate the $0\nu\beta\beta$ nuclear matrix elements
confidently and accurately.

\appendix

\section*{}

{\it Neutrino magnetic moment and the distinction between Dirac and Majorana
neutrinos}

The topic of neutrino magnetic moment $\mu_{\nu}$ is seemingly unrelated to the
$0\nu\beta\beta$ decay. Yet, as will be shown below, experimental observation of $\mu_{\nu}$
allows one to make important conclusions about the Dirac versus Majorana nature
of neutrinos. Hence, it is worthwhile to discuss it here.

Neutrino mass and magnetic moments are intimately related. In the orthodox Standard
Model neutrinos have a vanishing mass and  magnetic moments vanish as well. 
However, in the 
minimally extended SM containing gauge-singlet right-handed neutrinos the
magnetic moment $\mu_{\nu}$ is nonvanishing, but unobservably small \cite{musm},
\begin{equation}
\mu_{\nu} = \frac{3eG_F}{\sqrt{2}8\pi^2} m_{\nu} =  3 \times 10^{-19} \mu_B \frac{m_{\nu}}{1{~\rm eV}} ~.
\label{eq_munu}
\end{equation}
An experimental observation of a magnetic moment larger than that given in
eq.(\ref{eq_munu}) would be an uneqivocal indication of physics beyond the
minimally extended Standard Model.

Laboratory searches for neutrino magnetic moments are typically based on the obsevation
of the $\nu - e$ scattering. Nonvanishing $\mu_{\nu}$ will be recognizable only if the
corresponding electromagnetic scattering cross section is at least comparable to the well
understood weak interaction cross section. The magnitude of $\mu_{\nu}$ 
(diagonal in flavor or transitional) which can 
be probed in this way  is then given by
\begin{equation}
\frac {|\mu_{\nu}|}{\mu_B} \equiv \frac{G_F m_e}{\sqrt{2} \pi \alpha} \sqrt{m_e T} \sim 10^{-10}
\left(\frac{T}{m_e} \right) ~,
\label{eq_muesc}
\end{equation}
where $T$ is the electron recoil kinetic energy. Considering realistic values of $T$, it would
be difficult to reach sensitivities below $\sim 10^{-11} \mu_B$ using the $\nu - e$ scattering.
Present limits are about
an order of magnitude larger than that.

Limits on $\mu_{\nu}$ can also be obtained from bounds on the unobserved energy loss in
astrophysical objects. For sufficiently large $\mu_{\nu}$ the rate of plasmon decay 
into the $\nu \bar{\nu}$ pairs  would conflict with such bounds. Since plasmons can also decay
weakly into the $\nu \bar{\nu}$ pairs , the sensitivity of this probe is again limited by the size
of the weak rate, leading to
\begin{equation}
\frac {|\mu_{\nu}|}{\mu_B} \equiv \frac{G_F m_e}{\sqrt{2} \pi \alpha}{\hbar \omega_P} ~,
\end{equation}
where $\omega_P$ is the plasmon frequency. Since $(\hbar \omega_P)^2 \ll m_e T$
that limit is stronger than that given in eq.(\ref{eq_muesc}). Current limits on $\mu_{\nu}$ 
based on such considerations are $\sim 10^{-12} \mu_B$.

The interest in $\mu_{\nu}$ and its relation to neutrino mass dates from $\sim$1990
when it was suggested that the chlorine data\cite{chlorine} on solar neutrinos show
an anticorrelation between the neutrino flux and the solar activity characterized by the
number of sunspots. A possible explanation was suggested in Ref.\cite{Okun} where
it was proposed that a magnetic moment $\mu_{\nu} \sim 10^{-(10-11)} \mu_B$ would
cause a precession in solar magnetic field of the neutrinos emitted initially as 
left-handed $\nu_e$ into unobservable right-handed ones. Even though
later analyses showed that the effect does not exist, the possibility of a relatively large
$\mu_{\nu}$ accompanied by a small mass $m_{\nu}$ was widely discussed and various models 
accomplishing that were suggested.

\begin{figure}
\centerline{\includegraphics[width=9.0cm]{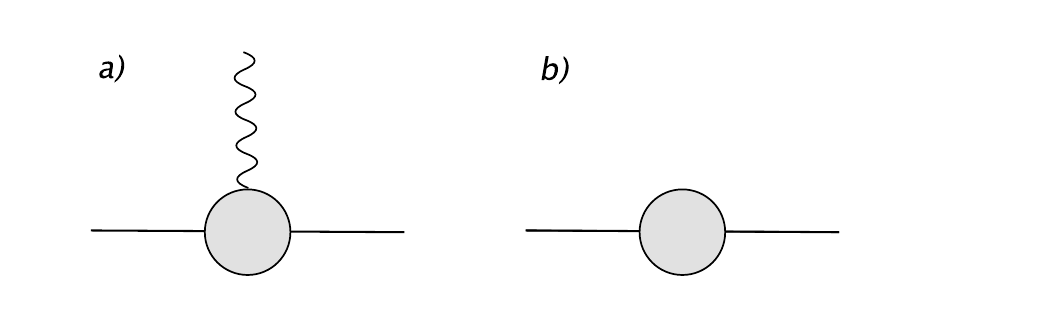}}
\caption{a) Generic contribution to the neutrino magnetic moment induced
by physics beyond the standard model. b) Corresponding contribution to the
neutrino mass. The solid and wavy lines correspond to neutrinos and photons
respectively, while the shaded circle denotes physics beyond the SM.}
\label{fig_magmom1}
\end{figure}

 If a magnetic moment is generated by physics
beyond the Standard Model (SM) at an energy scale $\Lambda$, 
we can generically express its value as
\begin{equation}
\mu_\nu \sim \frac{eG}{\Lambda},
\end{equation}
where $e$ is the electric charge and $G$ contains a combination of
coupling constants and loop factors.  Removing the photon from the 
diagram  gives a contribution to the neutrino mass of order
\begin{equation}
m_\nu \sim G \Lambda.
\end{equation}
We thus have the relationship
\begin{equation}
m_\nu \sim  \frac{\Lambda^2}{2 m_e}  \frac{\mu_\nu}{\mu_B}
~~ \sim \frac{\mu_\nu}{ 10^{-18} \mu_B}
[\Lambda({\rm TeV})]^2  \,\,\,{\rm eV},
\label{naive}
\end{equation}
which implies that it is difficult to simultaneously reconcile a small
neutrino mass and a large magnetic moment. These considerations
are schematically illustrated in Fig. \ref{fig_magmom1}.

 This na\"ive restriction given in
Eq.(\ref{naive}) can be overcome via a careful choice for the new
physics, e.g., by requiring certain additional symmetries 
\cite{Volosh,Georgi,Grimus,Babu}. Note,
however, that these symmetries are typically broken by Standard Model
interactions.
For Dirac neutrinos such symmetry (under which the left-handed neutrino 
and antineutrino $\nu$ and $\nu^c$ transform as a doublet)
is violated by SM gauge interactions. For Majorana neutrinos
analogous symmetries are not
broken by SM gauge interactions, but are instead violated by SM Yukawa
interactions, provided that the charged lepton masses are generated
via the standard mechanism through Yukawa couplings to the SM Higgs
boson. This suggests that the relation between $\mu_{\nu}$ and $m_{\nu}$
is different for Dirac and Majorana neutrinos. This distinction can be,
at least in principle, exploited experimentally, as shown below.

Earlier, I have quoted the Ref.\cite{SV82} (see Fig.\ref{fig_SV}) to stress that observation
of the $0\nu\beta\beta$ decay would necessarily imply the existence of a
novanishing neutrino Majorana mass. Analogous considerations can be
applied in this case.  By calculating neutrino magnetic moment
contributions to $m_\nu$ generated by SM radiative corrections, one may
obtain in this way general, \lq\lq naturalness" upper limits on the size of
neutrino magnetic moments by exploiting the experimental upper
limits on the neutrino mass.

In the case of Dirac neutrinos, a magnetic moment term will
generically induce a radiative correction to the neutrino mass of
order\cite{mu_D}
\begin{equation}
m_\nu \sim \frac{\alpha}{16\pi}
\frac{\Lambda^2}{m_e}  \frac{\mu_\nu}{\mu_B}
~~ \sim \frac{\mu_\nu}{3 \times 10^{-15} \mu_B}
[\Lambda({\rm TeV})]^2 \,\,\,{\rm eV}.
\end{equation}
Taking $\Lambda \simeq$ 1 TeV and $m_\nu \le$ 0.3 eV, we obtain
the limit $\mu_\nu \le 10^{-15} \mu_B$ (and a more stringent one
for larger $\Lambda)$, which is several orders of
magnitude more constraining than current experimental 
upper limits on $\mu_{\nu}$.

 The case of Majorana neutrinos is more subtle, due to the relative
flavor symmetries of $m_\nu$ and $\mu_\nu$ respectively. 
For Majorana neutrinos  the
transition magnetic moments $\left[\mu_\nu\right]_{\alpha\beta}$ are
antisymmetric in the flavor indices $\{\alpha,\beta\}$, while the mass
terms $[m_\nu]_{\alpha\beta}$ are symmetric.  These different flavor
symmetries play an important role in the limits, and are the origin of
the difference between the magnetic moment constraints for Dirac and Majorana
neutrinos.  

It has been shown in Ref.\cite{mu_M} that the constraints on Majorana
neutrinos are significantly weaker than those 
for Dirac neutrinos\cite{mu_D}, as the different flavor symmetries of $m_\nu$ and
$\mu_\nu$ lead to a mass term which is suppressed only by charged lepton
masses.  This conclusion was reached by considering one-loop mixing of the
magnetic moment and mass operators generated by Standard Model interactions.
The authors of Ref.\cite{mu_M} found that
if a magnetic moment arises through a coupling of the
neutrinos to the neutral component of the $SU(2)_L$ gauge boson, the
constraints for $\mu_{\tau e}$ and
$\mu_{\tau\mu}$ are comparable to present experiment limits, while the
constraint on $\mu_{e\mu}$ is significantly weaker.  
Thus, the analysis of  Ref.\cite{mu_M} 
lead to a model independent bound for the transition magnetic moment
of Majorana neutrinos that is less stringent than
present experimental limits.

\begin{figure}
\centerline{\includegraphics[width=11.0cm]{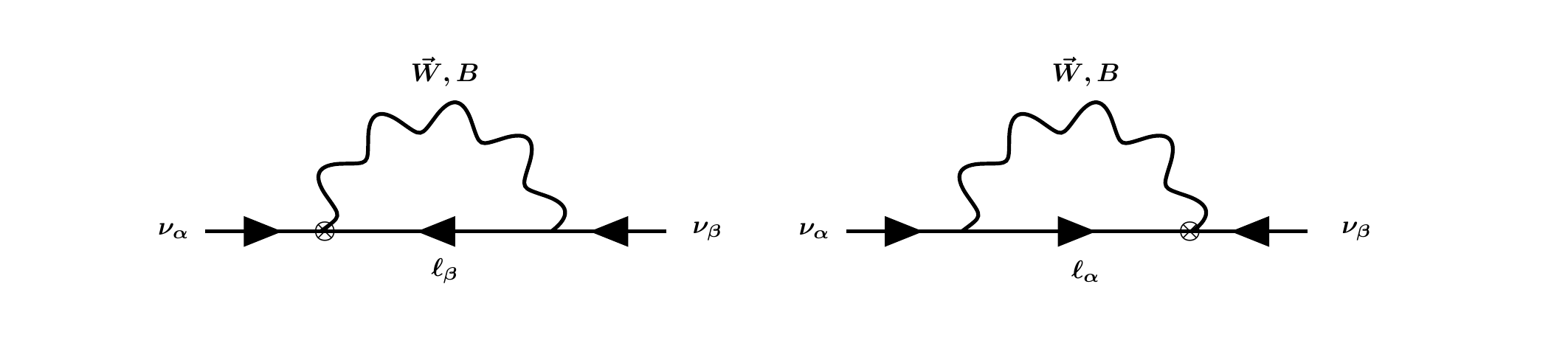}}
\caption{One loop diagram contributions to the Majorana neutrino mass associated with
the magnetic moment that sum to zero (see \cite{mu_M}). The cross indicates the magnetic moment
$[\mu]_{\alpha \beta}$ and the $l$ is the lepton doublet. }
\label{fig_magmom2}
\end{figure}

\begin{figure}
\centerline{\includegraphics[width=11.0cm]{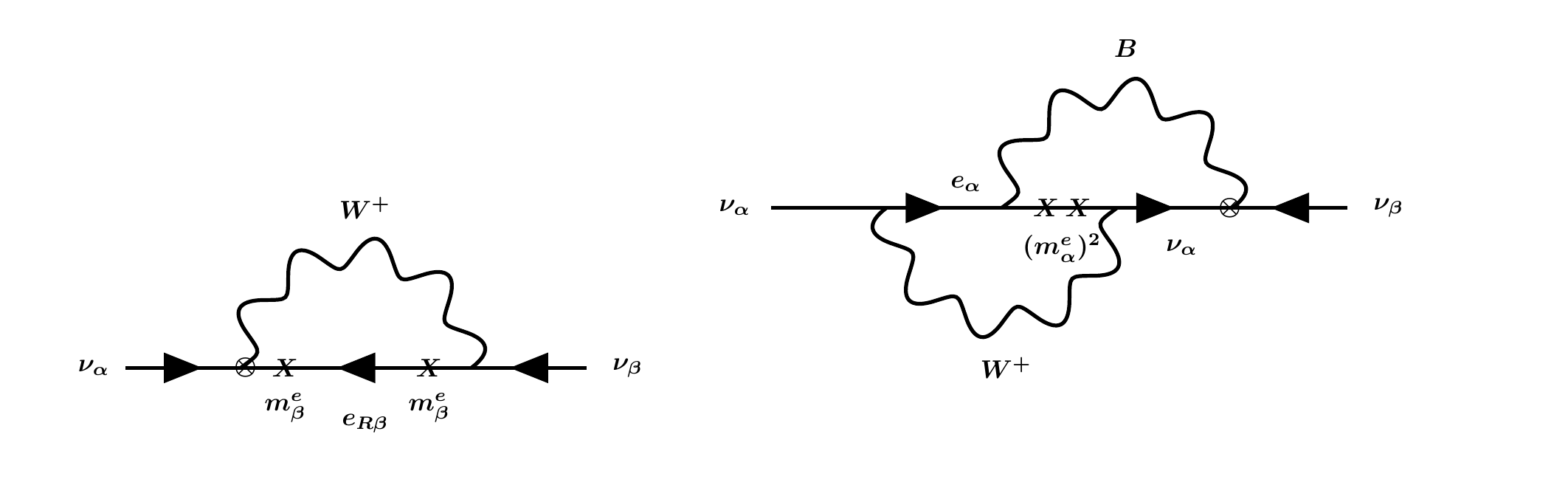}}
\caption{The one and two loop contributions  to the Majorana neutrino mass associated with
the magnetic moment. Here $X$ is the charged lepton mass insertion (see \cite{mu_M}). }
\label{fig_magmom3}
\end{figure}

Those considerations are illustrated in Figs. \ref{fig_magmom2}, \ref{fig_magmom3}. 

Even more generally it was shown in Ref.\cite{mu_last}  
that two-loop matching of mass and magnetic moment operators implies 
stronger constraints than those obtained
in\cite{mu_M} if the scale of the new physics $\Lambda \ge 10$
TeV. Moreover, these constraints apply to a magnetic moment generated
by either the hypercharge or $SU(2)_L$ gauge boson.
In arriving at these conclusions, the most general set of operators that
contribute at lowest order to the mass and magnetic moments of
Majorana neutrinos was constructed, and model independent constraints which
link the two were obtained.   Thus the results of Ref.\cite{mu_last} imply
completely model independent naturalness bound that -- for $\Lambda \ge 100$
TeV -- is stronger than present experimental limits (even for the
weakest constrained element $\mu_{e\mu}$). On the other hand, for sufficiently
low values of the scale $\Lambda$ the known small values of the neutrino
masses do not constrain the magnitude of the magnetic moment $\mu_{\nu}$
more than the present experimental limit. Thus, if these conditions are fulfilled,
the discovery of $\mu_{\nu}$ might be forthcoming any day.

The above result means that an experimental discovery of a magnetic moment
near the present limits would signify that (i) neutrinos are Majorana
fermions and (ii) new lepton number violating physics responsible for
the generation of $\mu_\nu$ arises at a scale $\Lambda$ which is well
below the see-saw scale. This would have,
among other things, implications for the mechanism of the
neutrinoless double beta
decay and lepton flavor violation as discussed above and in Ref.\cite{LNVus}.

\acknowledgments
 The original results reported here were obtained in collaboration with 
 Nicole Bell, Vincenzo Cirigliano,
 Jonathan Engel, Amand Faessler, Michail Gorchtein,
 Andryi Kurylov, Gary Prezeau,
 Michael Ramsey-Musolf, Vadim Rodin,  Fedor \v{S}imkovic,
 Peng Wang and Mark Wise. 
 The fruitful collaboration with them
 is gratefully acknowledged.

\end{document}